\documentclass[11pt,a4paper]{oxarticle}
\pdfoutput=1

\usepackage[usenames, dvipsnames]{xcolor}
\usepackage{graphicx, float, array, xspace, amscd, amsthm, amssymb, latexsym, longtable, bbm, fancyhdr,slashed,pifont}
\usepackage[letterpaper, body={6.5in, 9.25in}, top=1.0in, left=1.5in]{geometry}
\usepackage[bbgreekl]{mathbbol}
\usepackage[utf8]{inputenc}
\usepackage[sort&compress, comma, square, numbers]{natbib}
\usepackage{braket}
\usepackage{multirow}
\usepackage{subfig}
\usepackage{bm}
\usepackage{hyperref}
\usepackage{tikz}

\usepackage{tabularx, caption, boldline}
\usepackage{graphicx}
\usepackage{cellspace}
\DeclareSymbolFontAlphabet{\mathbb}{AMSb}
\DeclareSymbolFontAlphabet{\mathbbl}{bbold}

\let\SS=\S % save \S before it is redefined

\renewcommand{\b}{\beta}

\renewcommand{\th}{\theta}

\newcommand{\p}{\pi}

\renewcommand{\S}{\Sigma}
\renewcommand{\t}{\tau}

\DeclareFontFamily{OT1}{pzc}{}
\DeclareFontShape{OT1}{pzc}{m}{it}{<-> s * [1.200] pzcmi7t}{}
\DeclareMathAlphabet{\mathpzc}{OT1}{pzc}{m}{it}

\newcommand{\cC}{\mathcal{C}}

\newcommand{\cE}{\mathcal{E}}
\newcommand{\cF}{\mathcal{F}}\newcommand{\ccF}{\mathpzc F}

\newcommand{\cI}{\mathcal{I}}

\newcommand{\ccK}{\mathpzc K}

\newcommand{\cM}{\mathcal{M}}
\newcommand{\cN}{\mathcal{N}}
\newcommand{\cO}{\mathcal{O}}

\newcommand{\cS}{\mathcal{S}}

\newcommand{\cZ}{\mathcal{Z}}

\DeclareFontFamily{U}{bbold}{}
\DeclareFontShape{U}{bbold}{m}{n}
{  <-5.5> s*[1.05] bbold5
	<5.5-6.5> s*[1.05] bbold6
	<6.5-7.5> s*[1.05] bbold7
	<7.5-8.5> s*[1.05] bbold8
	<8.5-9.5> s*[1.05] bbold9
	<9.5-11.5> s*[1.05] bbold10
	<11.5-16> s*[1.05] bbold12
	<16-> s*[1.05] bbold17
}{}

\newcommand{\IC}{\mathbbl{C}}\newcommand{\Ic}{\mathbbl{c}}

\newcommand{\IQ}{\mathbbl{Q}}
\newcommand{\IR}{\mathbbl{R}}

\newcommand{\IZ}{\mathbbl{Z}}

\newcommand{\IDelta}{\mathbbl{\Delta}}
\newcommand{\one}{\mathbbm{1}}

\newcommand{\fp}{\mathfrak{p}}
\newcommand{\fq}{\mathfrak{q}}

\newcommand{\pt}{\text{pt}}

\font\eightrm=cmr8 at 8pt
\font\csc=cmcsc10

\newcommand{\beq}{\begin{equation}}
\newcommand{\eeq}{\end{equation}}
\newcommand{\beqnn}{\begin{equation*}}
\newcommand{\eeqnn}{\end{equation*}}
\newcommand{\bea}{\begin{eqnarray}}
\newcommand{\eea}{\end{eqnarray}}
\newcommand{\bean}{\begin{eqnarray*}}
	\newcommand{\eean}{\end{eqnarray*}}

\newcommand{\fref}[1]{Figure~\ref{#1}}
\newcommand{\tref}[1]{Table~\ref{#1}}
\newcommand{\sref}[1]{\SS\ref{#1}}

\newcommand{\nn}{\nonumber}

\newcommand{\defineas}{\buildrel\rm def\over =}

\newcommand{\ee}{\text{e}}
\newcommand{\ii}{\text{i}}
\newcommand{\dd}{\text{d}}

\newcommand{\place}[3]{\vbox to0pt{\kern-\parskip\kern-7pt
		\kern-#2truein\hbox{\kern#1truein #3}
		\vss}\nointerlineskip}
\newcommand{\smallfrac}[2]{\frac{\scriptstyle #1}{\scriptstyle #2}}

\DeclareFontFamily{U}{wncy}{}
\DeclareFontShape{U}{wncy}{m}{n}{<->wncyr10}{}
\DeclareSymbolFont{mcy}{U}{wncy}{m}{n}
\DeclareMathSymbol{\sha}{\mathord}{mcy}{"58}

\newcommand{\capt}[3]{\parbox{#1}{\renewcommand{\baselinestretch}{1.0}
		\caption{\label{#2}\small\it #3}}}

\newcommand{\1}{\mathbf{1}}

\renewcommand{\Re}{\text{Re~}}

\newcommand{\+}{\phantom{-}}

\renewcommand{\=}{\;=\;}

\renewcommand{\Im}{\text{Im}}
\renewcommand{\Re}{\text{Re}}
\newcommand{\wt}[1]{\widetilde{#1}}

\newcommand{\length}{\boldsymbol{\ell}}

\newcommand{\me}{\text{e}}
\newcommand{\bme}{\text{\textbf{e}}}
\newcommand{\hs}{{\circ}}
\newcommand{\sr}{\hskip1pt {*} \hskip1pt}

\newcommand{\cmark}{\ding{51}}%
\newcommand{\xmark}{\ding{55}}%

\newcommand{\Span}{\text{Span}}%

\makeatletter
\g@addto@macro\bfseries{\boldmath}

\def\blindfootnote{\xdef\@thefnmark{}\@footnotetext}
\makeatother

\newcommand{\twoFone}[4]{ {}_2F_1\left(\genfrac{}{}{0pt}{0}{#1,\,#2}{#3};~#4\right) }

\newcommand{\threeFtwo}[6]{ {}_3F_2\left(\genfrac{}{}{0pt}{0}{#1,\,#2,\,#3}{#4,~#5};~#6\right) }

\renewcommand{\baselinestretch}{1.1}
\numberwithin{equation}{section}
\begin{document}
\proofmodefalse

%%%%%%%%%%%%%%%%%%%%%%%%%%%%%%%%%%%%%%%%%%%
%%%         Title Page
%%%%%%%%%%%%%%%%%%%%%%%%%%%%%%%%%%%%%%%%%%%

\thispagestyle{empty}      
\begin{center}
\null\vskip0.1in
{\Huge Attractors with Large Complex Structure\\
for One-Parameter Families of\\[12pt]  
Calabi-Yau Manifolds}
\vskip1cm
{\csc Philip Candelas${}^1$, Pyry Kuusela${}^2$\\
and\\
Joseph McGovern${}^3$\\[2cm]}
\blindfootnote{$^1\,$candelas@maths.ox.ac.uk \hfill 
$^2\,$pyry.r.kuusela@gmail.com\hfill
$^3\,$mcgovernjv@gmail.com\kern20pt}
{\it Mathematical Institute\\
University of Oxford\\
Andrew Wiles Building\\
Radcliffe Observatory Quarter\\
Oxford, OX2 6GG, UK\\}
\vfill
{\bf Abstract}
\end{center}
\vskip-7pt
\begin{minipage}{\textwidth}
\baselineskip=15pt
\noindent 
The attractor equations for an arbitrary one-parameter family of Calabi-Yau manifolds are studied in the large complex structure region. These equations are solved iteratively, generating what we term an $N$-expansion, which is a power series in the Gromov-Witten invariants of the manifold. The coefficients of this series are associated with integer partitions. In important cases we are able to find closed-form expressions for the general term of this expansion. To our knowledge, these are the first generic solutions to attractor equations that incorporate instanton contributions. In particular, we find a simple closed-form formula for the entropy associated to rank two attractor points, including those recently discovered. The applications of our solutions are briefly discussed. Most importantly, we are able to give an expression for the Wald entropy of black holes that includes all genus 0 instanton corrections. 
\vspace*{10pt}
\end{minipage}
\clearpage

%%%%%%%%%%%%%%%%%%%%%%%%%%%%%%%%%%%%%%%%%%%
%%%          Contents
%%%%%%%%%%%%%%%%%%%%%%%%%%%%%%%%%%%%%%%%%%%

\thispagestyle{empty}  
{\baselineskip=17pt
	\tableofcontents} 
\clearpage

%%%%%%%%%%%%%%%%%%%%%%%%%%%%%%%%%%%%%%%%%%%
%%%        MAIN TEXT BEGINS HERE
%%%%%%%%%%%%%%%%%%%%%%%%%%%%%%%%%%%%%%%%%%%

\setcounter{page}{1}
\renewcommand{\baselinestretch}{1.1}
\section{Introduction} \label{sect:Introduction}
\vskip-10pt
\subsection{Preamble} \label{sect:Preamble}
\vskip-10pt
Ever since the discovery of the attractor mechanism \cite{Ferrara:1996dd}, the study of supersymmetric black holes in $\cN=2$ theories in four dimensions has been intimately linked to the study of attractor points on Calabi-Yau manifolds. This has led to many interesting conjectures and results, such as the intricate connections to number theory, discussed by Moore in \cite{Moore:1998zu,Moore:1998pn} and the BPS state existence conjecture~\cite{Denef:2007vg}, which links the existence of certain type of attractor points, in the $d=4, \, \cN=2$ supergravity theories, to the existence of BPS states in the quantum theory of gravity. 

While the attractor mechanism applies to any four-dimensional $\cN=2$ supergravity theory coupled to vector multiplets, an especially interesting class of examples is provided by compactifying 10-dimensional type IIA and IIB supergravities on Calabi-Yau threefolds (see \cite{Sen:2005wa,Mohaupt:2000mj} and references cited therein). In this case, special geometry is used to describe the complex structure or complexified K\"ahler moduli space. The scalars in $\cN=2$ vector multiplets can be interpreted as coordinates on the moduli space, and the attractor mechanism can be viewed as specifying, on the black hole horizon, a particular Calabi-Yau manifold in the moduli space as a function of the charges.

In this paper, we study attractor equations in $d=4, \, \cN=2$ supergravity theories that have been obtained by compactifying a type IIA or IIB supergravity on a one-parameter family of Calabi-Yau manifolds. We concentrate on the region of the moduli space that is near the large complex structure point. In this region, it is possible to solve the instanton-corrected attractor equations using perturbative methods.

What is even more surprising than the existence of this kind of solution is the fact that for some important special cases, we have been able to find a closed-form for the instanton corrections. With the help of these solutions, we are able to give interesting formulae for locations of attractor points, among other physical quantities of interest. For example, we are able to give a formula for the Wald entropy of a black hole that, in the large charge limit, is fully quantum- and instanton-corrected. 

Our aim in this paper is to develop what we term \textit{$N$-expansions}, by which we mean expansions in terms of the Gromov-Witten invariants $N^{\mbox{\tiny GW}}_k$. These include all perturbative and non-perturbative terms. The prototypical quantity that exhibits such an $N$-expansion is the prepotential, which is a key ingredient in formulating the attractor equations. Recall that the prepotential is introduced by first choosing a homology basis $A^a, B_b \in H_3(X, \IZ)$ and a dual cohomology basis $\alpha_a, \beta^b \in H^3(X,\IZ)$, for a Calabi-Yau manifold $X$. The holomorphic three-form $\Omega$ can be expressed in terms of the cohomology basis
\vskip-35pt
\begin{align}
\Omega \= z^a \alpha_a - \cF_b \beta^b~, \qquad a,b\=0,\dots,h^{2,1}(X)~, \label{eq:Omega_in_terms_of_Prepotential}
\end{align}
where \cite{Candelas:1990pi} the $\cF_b$ are the derivatives $\frac{\partial \cF}{\partial z^b}$ of a prepotential $\cF$. In the body of this paper, we will concentrate, for simplicity, on Calabi-Yau manifolds $X$ that lie in one-parameter families.

It is believed that near the large complex structure point there is a choice of a symplectic basis such that, in the one-parameter case, the prepotential $\cF(z^0,z^1)$ takes the form
\begin{align}
\begin{split}
\cF \,= -\frac{1}{3!} Y_{abc} \frac{z^a z^b z^c}{z^0} - (z^0)^2 \cI \left( \frac{z^1}{z^0} \right) \, = -(z^0)^2 \left(\frac{1}{6} Y_{111} t^3 + \frac{1}{2} Y_{110} t^2 + \frac{1}{2} Y_{100} t + \frac{1}{6} Y_{000} + \cI \left( t \right) \! \right)\!, \label{eq:defn_Calabi-Yau_prepotential}
\end{split}
\end{align}
where in the second relation $t=x+\ii y=z^1/z^0$. 

The terms proportional to $Y_{100},\,Y_{110},$ and $Y_{000}$ come from perturbative worldsheet corrections in $\alpha'$ to the string sigma model. The term $\mathcal{I}$ incorporates the nonperturbative genus-0 instanton contributions. Apart from the imaginary part of $Y_{000}$, all of these perturbative corrections can be added or removed by performing a Peccei-Quinn transformation and in this manner their contribution to attractor point locations was analysed in \cite{Bellucci:2010zd}. The $Y_{100}$ terms, in particular, were included in \cite{Shmakova:1996nz}.

The Yukawa coupling $Y_{111}$ is related to the topological quantity
\begin{equation}\label{eq:yukawa_definition}
Y_{111}\=\int_{\wt{X}}e_{1}\wedge e_{1}\wedge e_{1}~,
\end{equation}
where $e_{1}$ is a basis vector of the one-dimensional space $H^{2}(\wt{X},\IZ)$, and $\wt{X}$ is the mirror manifold of $X$. It is believed that a basis $(\alpha_a, \beta^b)$ in \eqref{eq:Omega_in_terms_of_Prepotential} can be chosen such that \cite{Candelas:1990pi,Candelas:1990rm,Hosono:1994ax,Halverson:2013qca}
\begin{align}\notag
\begin{split}
Y_{110}\; \in \; \left\{0,\frac{1}{2}\right\}~,\ \ \ \
Y_{100}\=-\frac{1}{12}\int_{\wt{X}}c_{2}\wedge e_{1}~,\ \ \ \
Y_{000}\=-3\frac{\zeta(3)}{(2\pi i)^{3}}\chi(\wt{X})~.
\end{split}
\end{align}
In the one-parameter case the component $Y_{110}$ can be taken to be zero if the integer $Y_{111}$ is even, and $1/2$ if $Y_{111}$ is odd. The quantity $\cI(t)$ corresponds to the exponentially small terms, and has the form
\begin{align}
\cI(t) \= \frac{1}{(2\pi \ii)^3} \sum_{k=1}^\infty n_k \text{Li}_3(\me^{2 \pi \ii k t}) \= \frac{1}{(2\pi \ii)^3} \sum_{k=1}^\infty \frac{N_k}{k^3} \me^{2\pi \ii k t}~, \label{eq:Instanton_Sum_Definition}
\end{align}
where the integral coefficients $n_k$ are the instanton numbers or Gopakumar-Vafa invariants \cite{Gopakumar:1998ii,Gopakumar:1998jq} and in the second expression we have written the instanton sum in terms of what we shall call the \emph{scaled Gromov-Witten invariants} $N_k$, which are given by
\begin{align}
N_k \= \sum_{d|k} d^3 n_d~, \qquad N_k \= k^3 N_k^{GW}~, \label{eq:defn_N_k_invariants}
\end{align}
with $N_{k}^{GW}$ the usual Gromov-Witten invariants. The form of the prepotential has long been known. However, we have recalled this to emphasise that $\cF$ has a \emph{universal} form. Given the topological quantities $Y_{ijk}$ and the instanton numbers $n_k$ corresponding to a given manifold $X$, one can write down the prepotential and this $N$-expansion converges in a neighbourhood of the large complex structure point. What we seek here are similar universal expansions for quantities of interest such as the coordinate $t$ and the central charge corresponding to an attractor point. In the general case, the expressions for these quantities are somewhat detailed. However, in simple cases, they can be concisely stated. For example, for the case of a D6-D0 brane system with attractor point on the imaginary axis and nonvanishing central charge, which we specify completely in \sref{sect:Solutions}, we have 
\begin{align} \label{eq:y_solution_introduction}
\begin{split}
t = \ii y, \qquad \text{ where } \quad
\frac{y}{y_0} = 1 - \sum_{m=1}^\infty \; \sum_{\fp \in \pt(m)} a_\fp N_\fp \left( \frac{m}{2 \pi y_0 Y_{111}} \right)^{\length(\fp)} \bm{k}_{\length(\fp)-1}(2 \pi m y_0).
\end{split}
\end{align}
In this expression $y_0$ is the perturbative solution for $y$, obtained by solving the perturbative form attractor equations, where the non-perturbative terms $\cI$ are set to zero. The quantity $m$ is an integer while $\fp$ runs over the partitions of $m$. If $\fp = \{j^{\mu_j}\}$, with $\sum_{j} \mu_j j = m$, is such a partition, then  $\length(\fp)=\sum_j \mu_j$ is the length of the partition. Surprisingly, the quantity $\bm{k}_n$ that arises here is the modified spherical Bessel function $\bm{k}_n(z) = \smash{\sqrt{\smallfrac{2}{\pi z}} K_{n+\frac{1}{2}}(z)}$. The coefficient $a_\fp$ is a combinatorial factor related to the partition, and the quantity $N_\fp$ is a combination of scaled Gromov-Witten invariants
\begin{align} \label{eq:a_p_N_p_definition}
a_\fp \= \prod_j \frac{1}{\mu_j! \, j^{2\mu_j}}~, \qquad N_\fp \= \prod_{j} N_j^{\mu_j}~.
\end{align}
Our results apply, in particular, to one of the rank two attractor points of \cite{Candelas:2019llw}. In that case, $y$ is known explicitly to high precision, and this formula has been checked for integers $m$ up to 75 and so for over 80 million partitions, giving an accuracy of over 50 decimal places. The sum also responds well to the techniques of accelerated convergence and using an iterated Shanks transformation of order 11, the agreement can be extended to 80 decimal places. The difficulty in checking this to a higher degree of accuracy lies in the need to store large numbers of partitions.

A similarly concise expression can be given for $|\cZ|^2$, where $\cZ$ is the central charge associated to a certain class of rank two attractor points,
\begin{align}
|\cZ(Q_1)|^2 \= \frac{\kappa^2 y}{2} |\Upsilon{-}\Lambda| ~, \label{eq:Central_Charge_Introduction}
\end{align}
where $y$ is as in the expression given above and the rank two attractor point corresponds to the two charge vectors
\begin{align} \notag
Q_1 \=
\kappa \begin{pmatrix}
0\\
\Upsilon\\
1 \\
0
\end{pmatrix}, \qquad Q_2 \= \lambda \begin{pmatrix}
\Lambda\\
0\\
0 \\
1
\end{pmatrix}.
\end{align}
In \eqref{eq:Central_Charge_Introduction}, we have given an expression for $|\cZ(Q_1)|^2$. The corresponding expression for $|\cZ(Q_2)|^2$ is
\begin{align} \notag
|\cZ(Q_2)|^2 \= \frac{\lambda^2}{2y}|\Upsilon{-}\Lambda|~.
\end{align}
Though the expression \eqref{eq:Central_Charge_Introduction} appears simpler, it is clear that the quantity $1/y$ can be re-expressed as an $N$-expansion.

These $N$-expansions have a compelling interpretation in terms of microstate counting, and this is particularly clear in virtue of the appearance of the instanton numbers. The $\cN{\=}4$ case was famously studied by Strominger and Vafa \cite{Strominger:1996sh}.
Important early papers on microstate counting of black holes in $\cN{\=}2$ theories are Maldacena, Strominger and Witten \cite{Maldacena:1997de}, the papers \cite{LopesCardoso:1998tkj,LopesCardoso:1999cv,LopesCardoso:1999fsj,LopesCardoso:1999xn,Mohaupt:2000mj,Behrndt:1996jn,Behrndt:1996mm,Behrndt:1998eq} and the extensive research of Sen (see for example \cite{Sen:2007qy} and references therein). These calculations take into account the leading term with the coefficient $Y_{111}$ of the prepotential and also, in the case of \cite{Maldacena:1997de}, the term with coefficient $Y_{100}$, which is important owing to its relation to the genus one correction to the prepotential. What we wish to do here is to take into account also the remaining terms, namely the perturbative $Y_{000}$ term, and the instanton corrections, whose importance for the application to counting is manifest.\newpage

The utility of the prepotential for us is that, as in \eqref{eq:Omega_in_terms_of_Prepotential}, we can write the holomorphic 3-form in terms of its derivatives. This immediately implies that the period vector $\Pi$ can also be written in this form. 
\begin{align}
\Pi(t)=\begin{pmatrix}
\frac{\partial}{\partial z^0} \cF\\[3pt]
\frac{\partial}{\partial z^1} \cF\\[3pt]
z^0\\
z^1
\end{pmatrix} = z^0 \begin{pmatrix}
\frac{1}{6} Y t^3 - \frac{1}{2} Y_{100} t - \frac{1}{3}\ii Y\sigma  - 2 \cI (t) + t \cI'(t)\\[3pt]
-\frac{1}{2} Y t^2 - Y_{110} t - \frac{1}{2} Y_{100}- \cI'(t) \\[3pt]
1 \\[3pt]
t
\end{pmatrix}. \label{eq:defn_Period_Vector}
\end{align}
Here we have used the fact that the period vector is only defined up to a scale, and is a homogeneous of degree one as a function of $z_0,z_1$. For brevity, we shall write in the following
\begin{align} \notag
Y_{111} \= Y \qquad \text{and} \qquad Y_{000} \= \ii Y\sigma~.
\end{align}
Black holes in type II supergravity theories compactified on Calabi-Yau manifolds can be viewed as consisting of D0-, D2-, D4-, and D6-branes wrapped on 0-, 2-, 4-, and 6-cycles in type IIA theory, or alternatively as D3-branes wrapping 3-cycles in type IIB theory. This implies that the charge vector of the black hole can be viewed as an element of \hbox{$H^0(X,\IZ) {\oplus} H^2(X,\IZ) {\oplus} H^4(X,\IZ) {\oplus} H^6(X,\IZ)$} or $H^3(X,\IZ)$. We will concentrate on one-parameter families of Calabi-Yau manifolds and usually choose a basis for these vector spaces so that we can think of the charge vector $Q$ as an element of $\IZ^4$ and the period vector $\Pi$ as an element of $\IC^4$. 

A generic element $\Gamma \in H^0(X,\IZ) {\oplus} H^2(X,\IZ) {\oplus} H^4(X,\IZ) {\oplus} H^6(X,\IZ)$ can be written as
\begin{align} \notag
\Gamma \= p^0 + p \, e_1 + q \, \frac{e_1^2}{Y} + q_0 \, \frac{e_1^3}{Y}~.
\end{align}
The corresponding charge vector can be written as
\begin{equation}\label{eq:defn_Q}
Q\=\left(\begin{matrix}_{\phantom{0}}q_0\\_{\phantom{0}}q_{\phantom{0}}\\_{\phantom{0}}p^{0}\\_{\phantom{0}}p^{\phantom{0}}\end{matrix}\right) \= \left(\begin{matrix}q_{\text{\tiny D0}}\\q_{\text{\tiny D2}}\\q_{\text{\tiny D6}}\\q_{\text{\tiny D4}}\end{matrix}\right),
\end{equation}
where the last equality reminds us of the relation of the components to the D-brane charges\footnote{In these conventions the D-brane charges do not include the charges induced on various branes by the coupling to the world-volume curvature. Often it is useful to switch to conventions where these charges are included. For the relation between the two conventions, see appendix \ref{app:D-Brane_Interpretation_of_Q}.}. 
This choice of basis coincides with the choice that resulted in the expression \eqref{eq:defn_Period_Vector} for the period vector, so that we are allowed to think of the charge and period vectors both as elements of same vector space isomorphic to $\IC^4$, although due to charge quantization the elements of $Q$ are integral. 

Mirror symmetry implies that the IIB theory compactified on a Calabi-Yau manifold $X$ is equivalent to IIA theory compactified on $\wt{X}$, the mirror of $X$. The mirror symmetry exchanges the complex structure moduli space $\cM_{\IC}$ with coordinate $\varphi$ and complexified K\"ahler moduli space $\cM_{\IC K}$ with coordinate $t$ of the two manifolds with each other. At the level of supergravity, the equivalence is that the vector multiplet scalars in IIB theories take values in $\cM_{\IC}$, while the hypermultiplet scalars take values in $\cM_{\IC K}$, with the opposite correspondence in IIA theories. In the following, we will focus mostly on the IIA perspective.

\subsubsection*{The large complex structure region}
\vskip-10pt
We use the term \textit{large complex structure region} to refer to a region in the Calabi-Yau moduli space where our solutions, such as \eqref{eq:y_solution_introduction}, converge. We will later in appendix \ref{sect:Asymptotics_Convergence} argue that such a region does exist and contains an open neighbourhood of the point of maximal unipotent monodromy which we will refer to simply as the \textit{large complex structure point}.

There are also at least two other closely related, but ultimately different, regions in the moduli space that could reasonably be called large complex structure regions. One is the region where the instanton sum $\cI(t)$ converges. This is determined by the asymptotic growth of the Gromov-Witten invariants (see for example \cite{Candelas:1990rm,Bershadsky:1993cx,Klemm:1999gm,Couso-Santamaria:2016vcc})
\begin{align} \notag
\log N^{\text{\tiny{GW}}}_k \; \sim \; 2\pi y_{\text{sing}} k~,
\end{align}
where $y_\text{sing}$ is the coordinate of the singularity nearest to the large complex structure point. Thus the instanton sum $\cI(t)$ converges if and only if $\Im \, t > y_\text{sing}$. 

It is also possible to study this region using the mirror symmetry relation to the complex structure moduli space on the mirror manifold. On the complex structure moduli space, the coordinates can be chosen so that the point of maximal unipotent monodromy is located at $\varphi=0$. The period vector is most conveniently expressed in terms of a basis of solutions $\widehat{\varpi}_i$ to a Picard-Fuchs equation around $\varphi=0$, where close to the large complex structure point the vector with components these solutions, $\widehat{\varpi}$, has asymptotics
\begin{align} \notag
\widehat{\varpi}\; \sim \; \left(\begin{matrix}1\\t\\t^2\\t^3\end{matrix}\right)~.
\end{align}
After identifying $(z_{0},z_{1}){\=}(\widehat{\varpi}_{0},\widehat{\varpi}_{1})$, one expresses the complexified K\"ahler moduli space coordinate as follows:
\begin{align} \label{eq:t_mirror_map}
t \= \frac{\widehat{\varpi}_1(\varphi)}{\widehat{\varpi}_0(\varphi)} \= \frac{1}{2\pi \ii} \log \varphi + \cO(\varphi)~,
\end{align}
with the last equality following from the specifics of the Picard-Fuchs equation \cite{Braun2015TwoOS,Candelas:2019llw}. One should note the slightly different notation of each reference, and that we use the same notation as \cite{Candelas:2019llw}. The change of basis from this vector to the period vector $\Pi$ that we work with is given by
\begin{align} \notag
\Pi \= -\left(\begin{matrix}\frac{1}{3}Y_{000} & \frac{1}{2}Y_{100} & 0 &-\frac{1}{6}Y_{111}\\[2pt]
\frac{1}{2}Y_{100} & Y_{110} & \frac{1}{2}Y_{111} & 0\\
-1\+&0&0&0\\0&-1\+&0&0\end{matrix}\right) \widehat{\varpi}~.
\end{align}
The solutions $\widehat{\varpi}_i$ have a radius of convergence determined by the singularities of the Picard-Fuchs operator. This determines yet another region that could be reasonably called the large complex structure region. Applying the mirror map \eqref{eq:t_mirror_map}, this region gets mapped to a neighbourhood of $t \to \infty$, bounded by a circle around the nearest singularity, unless the Picard-Fuchs operator has apparent singularities. However, this image does not in general coincide with the $\Im \, t > y_\text{sing}$ region, as the correction terms of order $\cO(\varphi)$ affect the relation. In \fref{fig:LCS_regions}, we have sketched the region around $\varphi=0$, its image under the mirror map, and image of the resulting region under the map $t \to 1/t$. 
\begin{figure}[!htb]
	\centering
	\framebox[230pt]{
		\begin{minipage}{.45\textwidth}
			\vskip10pt
			\centering
			\includegraphics[width=0.9\linewidth, height=0.3\textheight]{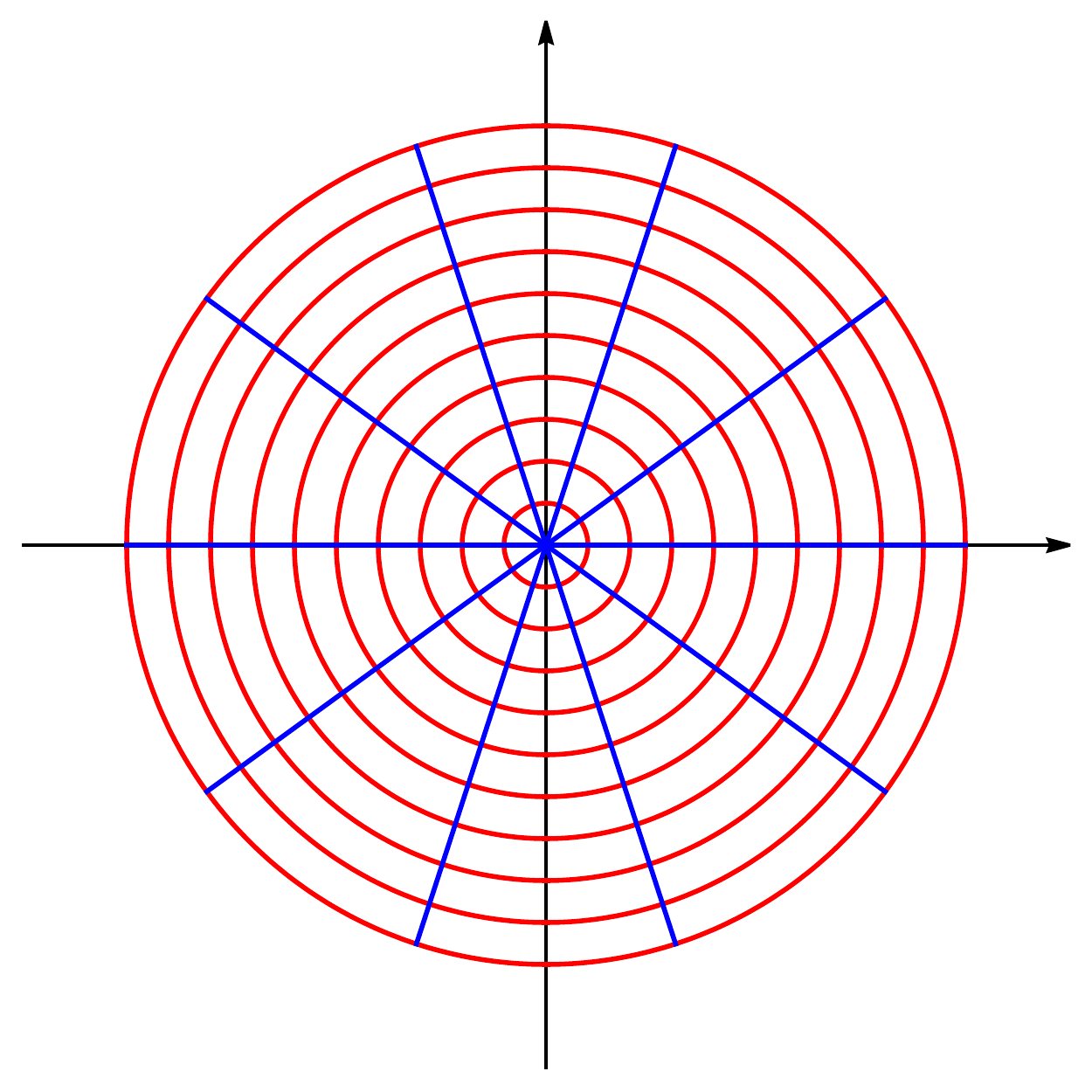}
			\caption*{i}
			\label{fig:prob1_6_2}
	\end{minipage}}
	\framebox[230pt]{
		\begin{minipage}{.45\textwidth}
			\vskip10pt		
			\centering
			\includegraphics[width=0.9\linewidth, height=0.3\textheight]{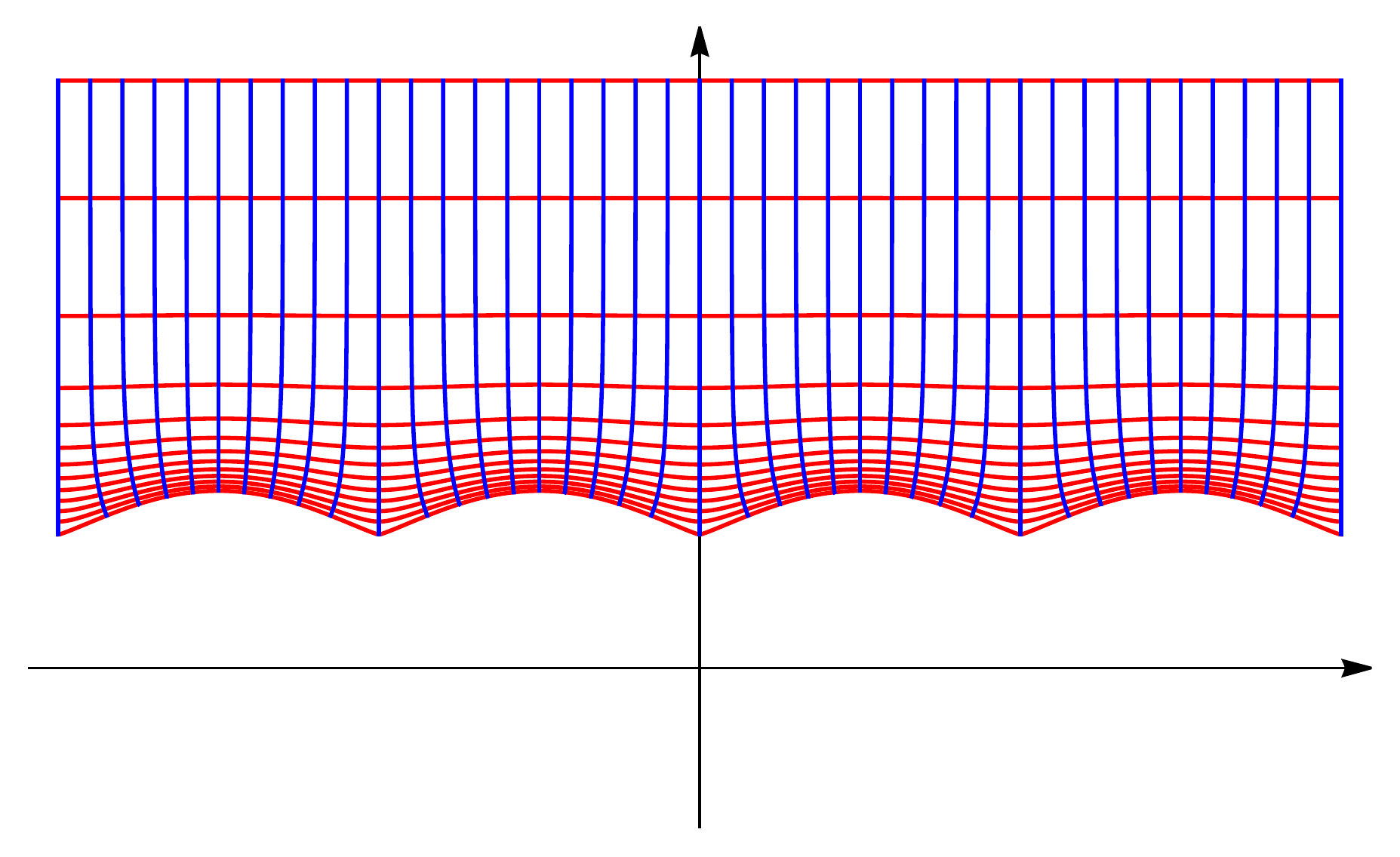}
			\caption*{ii}
			\label{fig:prob1_6_1}
	\end{minipage}}\vskip10pt
	\framebox[230pt]{
		\begin{minipage}{0.45\textwidth}
			\vskip10pt		
			\centering
			\includegraphics[width=.9\linewidth, height=0.3\textheight]{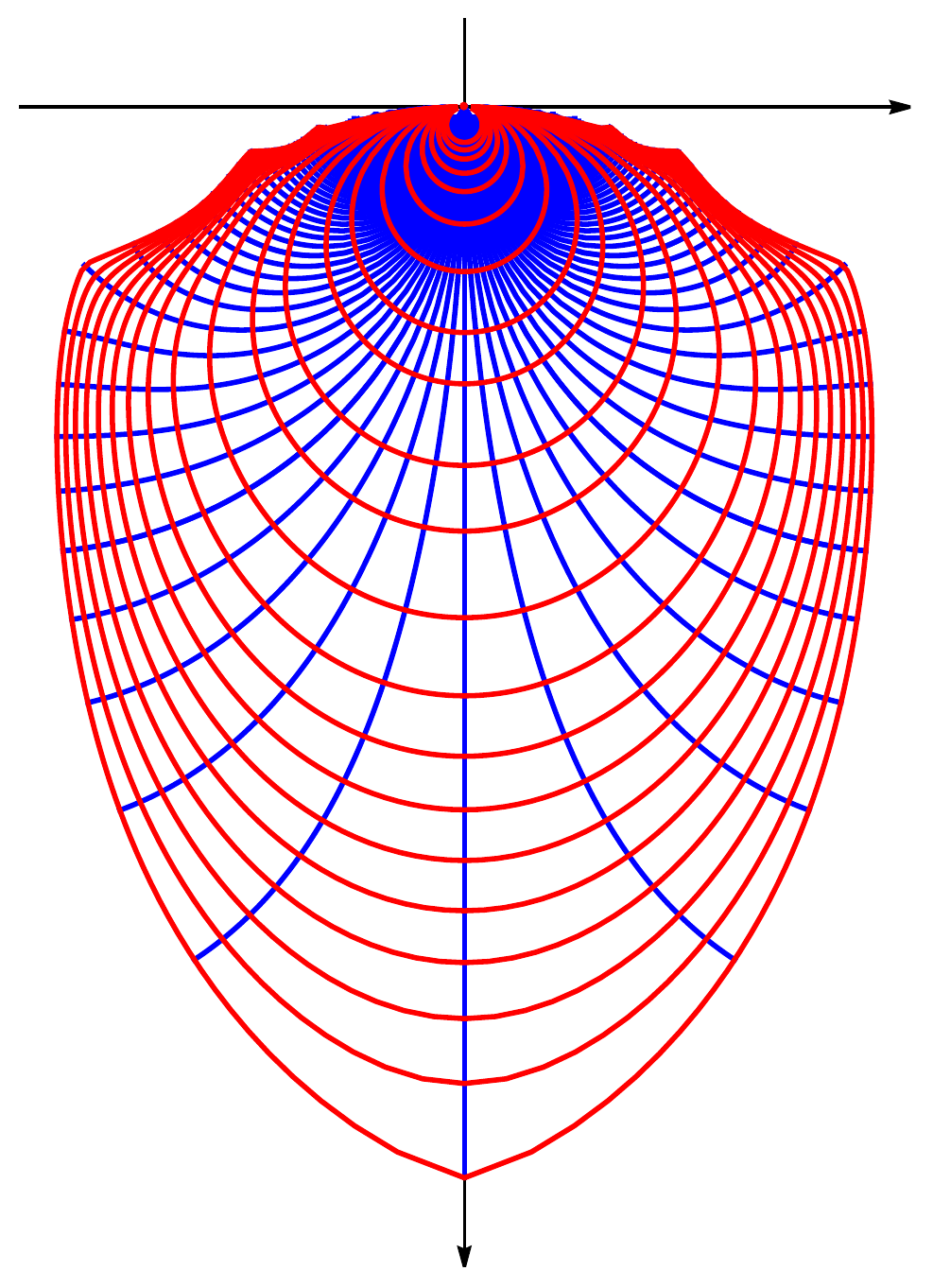}
			\caption*{iii}
	\end{minipage}}
	\vskip10pt		
	\place{3.03}{6.91}{$\boxed{\varphi}$}
	\place{6.31}{6.91}{$\boxed{t}$}	
	\place{4.64}{3.41}{$\boxed{\frac{1}{t}}$}	
	\capt{6in}{fig:LCS_regions}{Sketch of three different views on the large complex structure region. Figure i displays the large complex structure region on the mirror manifold. The region is centred around the large complex structure singularity at $\varphi = 0$ and bounded by a circle through the singularity nearest to the $\varphi=0$ singularity. Under the mirror map, this gets mapped to the subset of the $t$ upper half plane displayed in Figure ii, bounded from below and extending to imaginary infinity. This figure contains multiple copies of the fundamental domain, corresponding to the multiples branches of the mirror map. This can be more conveniently displayed by making a transformation $t \to 1/t$, resulting in the region displayed in Figure iii. Note that the exact shapes of these regions depend on the manifold in question, but the qualitative features of these regions are similar for any one-parameter family of Calabi-Yau manifolds.}
\end{figure}

\subsubsection*{The attractor mechanism}
\vskip-10pt
In \cite{Ferrara:1995ih,Ferrara:1996dd}, the authors studied a four-dimensional $\cN{\=}2$ supergravity coupled to $\cN{\=}2$ vector multiplets, which can be conveniently formulated in terms of special geometry \cite{Cremmer:1984hj}. For special geometry itself, see \cite{Strominger:1990pd,Candelas:1990pi}, and for a short summary in the style of the present paper, see \cite{Candelas:2019llw}. For a pedagogical introduction to the attractor mechanism, see for example \cite{Becker:2007zj,Freedman:2012zz,supersymmetric_mechanics_2}.  It was noticed that, for the supersymmetric black hole solutions in these theories, the special geometry moduli have to satisfy a pair of relatively simple differential equations. These equations can be interpreted as requiring that the central charge of the supergravity theory has a critical point at the horizon of the black hole. In this section, we will look at things from the IIB point-of-view.

If the black hole solution in question is spherically symmetric and static, we can write the line element 
\begin{equation*}\notag
\dd s^{2}\= e^{2U(r)} \dd t^{2}-e^{-2U(r)}(\dd r^{2}+r^{2}\dd \Omega^{2}),
\end{equation*}
where the radial coordinate $r$ is zero on the horizon. A key quantity is the central charge $\cZ$ of the superymmetry algebra, which is related to the mass and entropy of the black hole.
\begin{align}
\cZ(Q) \= \frac{Q^{T}\,\Sigma\,\Pi}{\sqrt{-\ii \Pi^{\dagger}\,\Sigma\, \Pi}}~, \label{eq:Central_Charge_Definition}
\end{align}
The intersection matrix $\Sigma$ appearing here is given in our chosen basis by
\begin{align} \notag
\Sigma \= \begin{pmatrix} 
\+0 & \one\\
- \one & 0
\end{pmatrix}.
\end{align}
The dynamics of the scalars $z^{a}$ and the function $U$ can be written in terms of a new coordinate $\rho=1/r$:
\begin{align}
\frac{dU(\rho)}{d\rho}&\=-e^{U(\rho)}|\cZ|~, \qquad
\frac{dz^{a}(\rho)}{d\rho}\=-2e^{U(\rho)}g^{a \, \bar{b}}\partial_{\bar{b}}|\cZ|\label{eq:fermionsAM2}~,
\end{align}
where $g_{a \, \bar{b}}$ is the special geometry metric on the moduli space $\mathcal{M}_{\IC}$ and $z^a$ are the scalars of the $\cN{\=}2$ vector multiplets, which are viewed as coordinates on the moduli space $\cM_{\IC}$. These equations describe flows of scalar field values from $\rho=0$ at spatial infinity to $\rho=\infty$ at the black hole horizon. The terminology ``attractor mechanism'' owes to the fact that the values of $z^a$ are fixed at the black hole horizon at $r=0$ ($\rho = \infty$) by the black hole charges, and so are independent of the values they take at $r=\infty$ ($\rho = 0$), at least for sufficiently small variations of boundary values.

The second equation, (\ref{eq:fermionsAM2}), can be used to find a gradient flow equation:
\begin{equation*}\notag
\frac{d|\cZ|}{d\rho}\=-4e^{U(\rho)}g^{a\bar{b}}\partial_{a}|\cZ|\partial_{\bar{b}}|\cZ|~,
\end{equation*}
revealing that the end points of attractor flows are critical values of $|\cZ|$, since the metric $g_{a\bar{b}}$ is positive definite. One can also see that $|\cZ|$ decreases along attractor flows.

It was pointed out early on by Strominger \cite{Strominger:1996kf} that at an attractor point where the central charge is not vanishing, the period vector of the Calabi-Yau manifold satisfies a vector relation that we shall refer to as the \textit{alignment equations}
\begin{align}
\Im \left[ C \Pi \right] &\= Q~, \label{eq:Strominger_Equations_Original}
\end{align}
where $\Pi$ is the period vector of the Calabi-Yau manifold, $Q$ is the charge vector of the corresponding black hole, and 
\begin{align}\notag
C \; = \; 2\me^{-K/2} \overline{\cZ}, \qquad \text{where} \qquad \me^{-K} \; \defineas \; \ii \int_X \Omega \wedge \overline{\Omega}~.
\end{align} 
In the following, we often choose a gauge in which $z^0 = 1$. The equation \eqref{eq:Strominger_Equations_Original} determines the relation between the charges and the prepotential. 

If, on the other hand, the central charge vanishes, then the corresponding relation is
\begin{align}
Q^{T}\,\Sigma\,\Pi  \= 0~, \label{eq:Orthogonality_Equations_Original}
\end{align}
which we will refer to as the \textit{orthogonality equation}. This determines the prepotential in such a~case. 

The one-parameter attractor equations for a certain classes of Calabi-Yau manifolds were studied for example in \cite{Bellucci:2006ib,Bellucci:2007eh,Bellucci:2008tx}.

\subsubsection*{The alignment equations} \label{sect:Strominger_Equations}
\vskip-10pt
It is convenient to write $C = a + \ii b$ and express the alignment equation \eqref{eq:Strominger_Equations_Original} in the form
\begin{align}
a \, \Im \Pi + b \, \Re \Pi \= Q~. \label{eq:Strominger_Equations}
\end{align}
This equation can also be understood in terms of cohomology, and amounts to the condition that
\begin{align} \notag
\Gamma \in H^{(3,0)}(X,\IZ) \oplus H^{(0,3)}(X,\IZ)~.
\end{align}
We can think of the period vector $\Pi$ as defining a 2-plane over the reals, spanned by $\Re \Pi$ and $\Im \Pi$. This two plane varies as we move in the moduli space giving a family of 2-planes in $\IR^4$, parametrised by the moduli-space coordinate $t$.

In order to satisfy the alignment equations, the plane
\begin{align} \notag
V \= \Span(\Re \Pi, \Im \Pi)
\end{align}
must contain a non-zero point of the charge lattice $H^3(X,\IZ) \simeq \IZ^4$, and so a lattice line consisting of integer multiples of smallest such point. If the intersection is a lattice line, the attractor is said to be of rank one. Exceptionally, the plane $V$ will intersect the charge lattice in two linearly independent points, and so in a lattice plane. Then the attractor point is said to be of rank two. If this is so, the alignment equations are satisfied for two linearly independent vectors $Q$.

\subsubsection*{The orthogonality equation}
\vskip-10pt
In addition to the attractor points solving the alignment equations, there are attractor points where the central charge vanishes, since these points are necessarily local minima of the nonnegative quantity $|\cZ(Q)|$. In this case
\begin{align} \notag
\Gamma \in H^{(2,1)}(X,\IZ) \oplus H^{(1,2)}(X,\IZ). 
\end{align}
There is again an interpretation of this equation in terms of planes and lattices. The charge vector $\Gamma$ lies in the plane $V^\perp$ that is orthogonal to $V$ under the symplectic inner product, corresponding to $\Sigma$. For there to be an attractor point, the plane $V^\perp$ must also intersect the charge lattice $H^3(X,\IZ)$, and this intersection can also be of rank one, or exceptionally of rank two.

Solutions to the alignment equations and the orthogonality equation are related. The fact that $V$ is to some extent rational with respect to the charge lattice has consequences for the rationality of the orthogonal plane $V^\perp$. In particular, if $V$ intersects the lattice in a lattice plane, then $V^\perp$ does so also. To see this, note that in this case $V {\=} \text{Span}(Q_1,Q_2)$ for some independent lattice vectors $Q_1$ and $Q_2$. Then the space $V^\perp$ that is orthogonal to $V$ under the symplectic inner product, can also be taken to be spanned by two lattice vectors $V {\=} \text{Span}(Q_3,Q_4)$. The alignment equations at rank two attractor points can be then written as 
\begin{align} \nn
\text{Span}(\Re \Pi, \Im \Pi) \= \text{Span}(Q_1,Q_2)^{\phantom{\perp}} =\; \text{Span}(Q_3,Q_4)^\perp~.
\end{align} 
Likewise, the orthogonality equation can be written as 
\begin{align} \nn
\text{Span}(\Re \Pi, \Im \Pi) \= \text{Span}(Q_1,Q_2)^\perp = \;  \text{Span}(Q_3,Q_4)^{\phantom{\perp}}~.
\end{align} 
Thus alignment equations can be, in this case, written as an orthogonality equation and vice versa. In the case that $V$ intersects the lattice in a lattice line, the situation is more complicated, and is explored in more detail in \sref{sect:Solutions}.

We wish to be clear about an issue of terminology. The term ``attractor point" is often used to refer exclusively with regards to  massive black holes, and so to solutions of the alignment equations, ignoring possible solutions to the orthogonality equation. In the following, we will understand an attractor point to be a point that is either a solution of the alignment equations or a solution of the orthogonality equation. By a rank two attractor point, we will continue to mean that there are two independent solutions to either the alignment or the orthogonality equation, and hence that either $V$ or $V^\perp$ intersects the charge lattice in a plane.
\begin{figure}[H]
	\centering
	\includegraphics[width=0.6\textwidth]{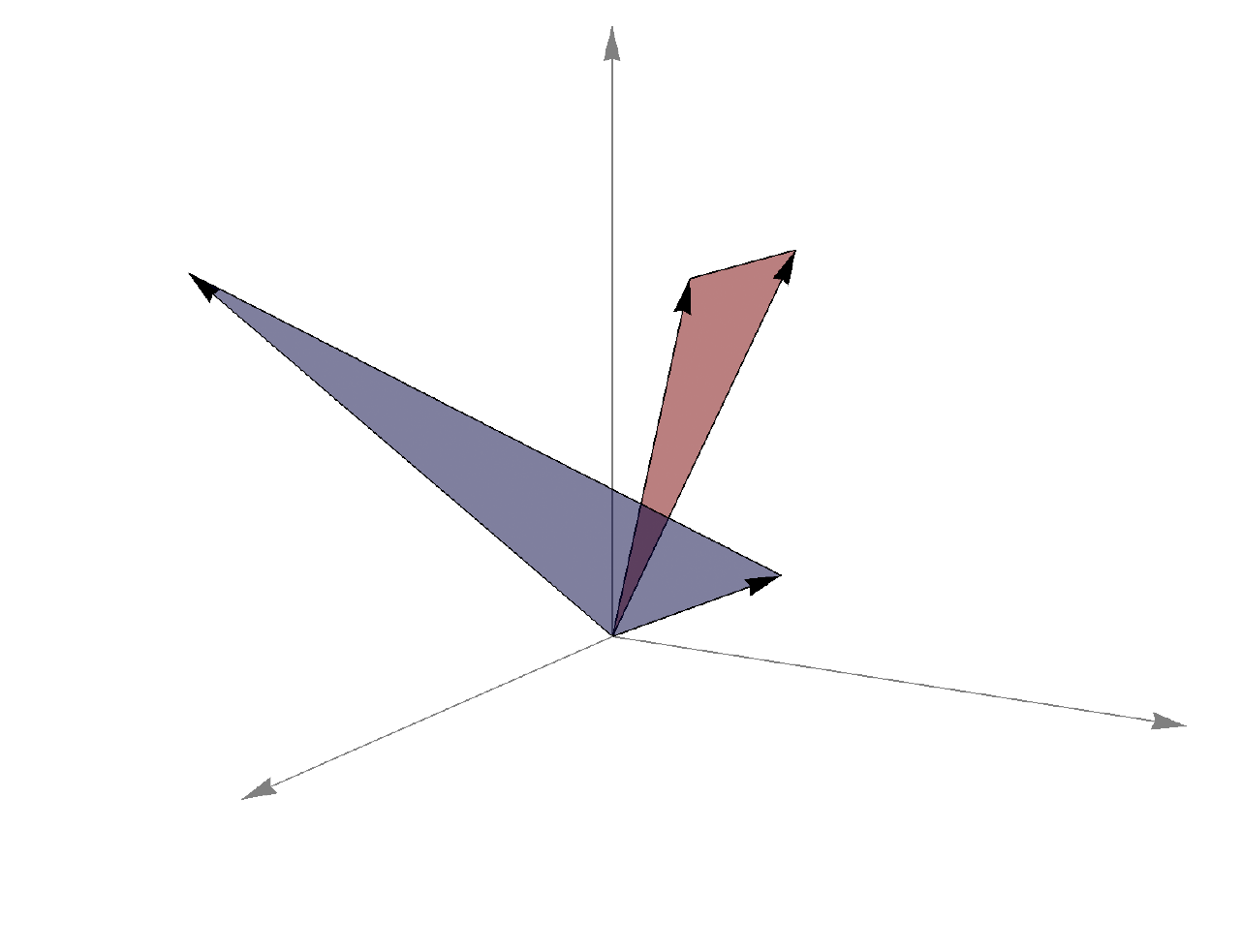}
	\vskip0pt
	\place{3.35}{2.45}{$Q_1$}
	\place{3.8}{2.5}{$Q_2$}
	\place{3.8}{1.4}{$\Im \Pi$}	
	\place{1.8}{2.4}{$\Re \Pi$}
	\vskip-30pt	
	\capt{6in}{fig:Lattice_Planes}{A schematic figure of the $H^3(X,\IR)$, which we identify with $\IR^4$. Here we show the real and imaginary parts of $\Pi$ as vectors in this space, along with the charge vectors $Q_1$, $Q_2$. As the moduli space coordinate $t$ varies, the vectors $\Re \, \Pi$ and $\Im \, \Pi$, as well as the vectors orthogonal to these, move. When we have a solution to the alignment equations, at least one of the vectors $Q_1$ and $Q_2$ will align with the plane generated by $\Re \, \Pi$ and $\Im \, \Pi$.}	
\end{figure}
\subsubsection*{Interpretation}
\vskip-10pt
The solutions to the alignment and orthogonality equations can be given very different physical interpretations, owing to their relation to the BPS state existence conjecture and the nature of the corresponding attractor flows. The BPS state existence conjecture \cite{Denef:2007vg,Denef:2000nb,Denef:2001xn,Moore:1998pn,Moore:1998zu} asserts that, in the full quantum theory of gravity, the existence of BPS states with certain charge vectors is determined by the nature of the attractor point where the flow ends. If we leave aside the possibility of a limiting cycle, there are three different types of flows:
\begin{enumerate}
	\item The attractor flow ends at a point where $|\cZ| {\,\neq\,} 0$.
	\item The attractor flow ends at a point where $|\cZ| {\,=\,} 0$, and the point is a regular point on the moduli space of the Calabi-Yau manifold.
	\item The attractor flow ends at a point where $|\cZ| {\,=\,} 0$, and the point is a singular point on the moduli space of the Calabi-Yau manifold.
\end{enumerate}
The BPS state existence conjecture asserts that BPS states always exist in the first case, and never in the second case, while in the third case BPS states may or may not exist, and this question cannot be determined from the attractor flow data alone \cite{Moore:1998pn,Moore:1998zu}. Thus the conjecture asserts that the solutions to the alignment equations should have an associated BPS state in the full theory, while the solutions to the orthogonality equation do not correspond to such states.

Even though the BPS state existence conjecture asserts that solutions to the orthogonality equation do not correspond to BPS states, this does not mean that the solutions are not significant. First of all, as we have seen, the alignment and orthogonality equations are intimately connected: in some cases, solutions to the alignment equations can be found from solutions to the orthogonality equation. In addition, the orthogonality equation has a relation to flux compactifications \cite{Kachru:2020abh,Kachru:2020sio} (see also section 4 of \cite{Candelas:2019llw}). Finally, it has been suggested \cite{Candelas:2019llw} that, at least at rank two attractor points, the existence of solutions to the orthogonality equation may indicate supersymmetry enhancement.

\subsubsection*{The central charge and entropy}
\vskip-10pt
In the case of Calabi-Yau compactifications, the attractor equations not only make it possible to find the complex or complexified K\"ahler structure of the internal manifold, but also can be used to compute the central charge \eqref{eq:Central_Charge_Definition} of the supergravity theory, which in turn is related to the Bekenstein-Hawking entropy of the black hole associated to the attractor point.
\begin{align} \notag
S_{\mbox{\tiny BH}} \= \frac{A}{4} \= \pi |\cZ|^2. 
\end{align}
Solutions for the attractor equations in the approximation where the instanton corrections to the prepotential are ignored, but the genus 1 perturbative contribution is included, were found in \cite{Shmakova:1996nz,Behrndt:1996jn} and used to give an approximate formula for the Bekenstein-Hawking entropy in any Calabi-Yau compactification, where the internal manifold has a large complex structure. For black holes for which $p^{0}=0$ but the other charges are large,
\begin{align}\label{eq:Shmakova_Entropy}
S_{\mbox{\tiny BH}} = \pi \sqrt{-\frac{2}{3}Y(\wt{p}^3+c_2 \wt{p})\left(\wt{q}_0-\frac{1}{2\wt{p}Y}\wt{q}^2\right)}~.
\end{align}
Here $\wt{p}$, $\wt{q}$ and $\wt{q}_0$ are the microscopic charges, that is numbers of D4-, D2-, and D0-branes. The relation to the macroscopic charges $p$, $q$ and $q_0$ can be found in appendix \ref{app:D-Brane_Interpretation_of_Q}.

In the calculation of this formula $Y_{000}$ and $Y_{110}$ cancel out, and $Y_{100}$ and $Y_{111}$ appear via $c_2$ and~$Y$, respectively. The existence of such a result, which bears a clear resemblance to Cardy growth~\cite{Cardy:1986ie}, raises the question whether there is a modular form related to the microstate counting, whose asymptotics reproduce the above formula in a manner similar to the celebrated results of Strominger and Vafa \cite{Strominger:1996sh}. This question was first tackled in \cite{Maldacena:1997de}, where the authors showed that this entropy can be reproduced, under some mild assumptions, by counting massless degrees of freedom in an M-theory scenario with fivebranes wrapping a four-cycle on $X$ and the M-theory circle, together with momentum $q_0$ around the circle. This setup gives rise to a two-dimensional $\cN = (4,0)$ conformal field theory on the fivebrane world-volume. It is possible to define a \emph{modified elliptic genus}, which is a supersymmetric index that it is invariant under continuous deformations of the theory, at least for small enough deformations. This modified elliptic genus can be expressed in terms of a vector-valued modular form.  The coefficients of this modular form contain information about the number of microstates. The asymptotic growth of the number of microstates is then given by the Cardy formula \cite{Cardy:1986ie} with the central charges
\begin{align} \notag
\begin{split}
c_R &= Y \left( \wt{p}^3 + c_2 \wt{p}\right), \qquad c_L = Y \left( \wt{p}^3 + \frac{1}{2}c_2 \wt{p}\right). 
\end{split}
\end{align}
In the large charge limit $\wt{p} \to \infty$, this correctly reproduces \eqref{eq:Shmakova_Entropy}. 

For cases where the charge $\wt{p}$ is small, the corresponding modified elliptic genera have been found explicitly, for some simple but interesting Calabi-Yau manifolds in \cite{Gaiotto:2006wm,Gaiotto:2007cd}. To do this, one can consider the geometry of divisors, curves, and points on the manifold. The fivebranes are understood as divisors, and the lower-dimensional branes as curves or points on these. The coefficients in the modular form that appears in the modified elliptic genus can then be found by studying the moduli spaces of various such configurations of branes and computing their Witten indices.

There is, however, an alternative method, which uses \emph{split attractor flows} \cite{Denef:2000nb,Denef:2000ar,Denef:2001xn} to compute elliptic genera. This was used in \cite{Collinucci:2008ht} to repeat the calculation for the Calabi-Yau manifolds referred to above. This technique is based on the observation that at least some of the coefficients count microscopic degrees of freedom in situations where the black hole can be viewed as a bound state of a D6 and an anti-D6 brane \cite{Denef:2007vg}. Such microstate indices can be expressed in terms of Donaldson-Thomas invariants \cite{Donaldson:1996kp,Thomas:1997}, which in turn can be expressed in terms of the Gromov-Witten invariants. Interestingly, these invariants also appear in our macroscopic expressions, such as \eqref{eq:y_solution_introduction}. Unfortunately, the modified elliptic genus is not currently known for the manifold AESZ34, which we use as the main example of a one-parameter Calabi-Yau manifold in this paper owing to that fact that it has rank two attractor points that are known explicitly. We hope to revisit this issue elsewhere.

It can also be shown, as we will review in \sref{sect:Microstate_Counting}, that the central charge can be viewed as the leading-order contribution to the Wald entropy \cite{Wald:1993nt} of the black hole in the large charge limit~\cite{LopesCardoso:1999cv}. The Wald entropy can be obtained from an analogue of the attractor equations, known as the \textit{stabilisation equations}, where the period vector $\Pi$ is replaced by a more general vector that includes the higher-genus prepotentials. However, as we will see in \sref{sect:Microstate_Counting}, in the large charge limit the stabilisation equations reduce to the alignment equations \eqref{eq:Strominger_Equations_Original}. 

One of the aims of the present work is to extend these results by including the contributions of instantons that have not been included in the approximation used in \cite{Shmakova:1996nz,Behrndt:1998eq,Maldacena:1997de,Bellucci:2010zd}. We set up an iterative procedure that generates the $N$-expansions, which have a universal form that depends on the manifold through the perturbative coefficients of the prepotential together with the instanton numbers. A simple example of such an expansion has been given in~\eqref{eq:y_solution_introduction}.

The black hole charge determines the location of the attractor point, and so also the period vector. The value of the central charge therefore depends in a detailed way on the particular brane system under consideration. Nevertheless, we close this subsection with some general remarks, which we will later refine in order to obtain explicit expressions for different brane systems. 

From the alignment equations it follows that at an attractor point the charge vector can be expressed as a linear combination of real and imaginary parts of the period vector. Substituting \eqref{eq:Strominger_Equations} into the definition of the central charge, the square of the central charge can be simplified
\begin{align} \notag
|\cZ|^2 &\= \frac{a^2+b^2}{2} \left|\Re \Pi^T \, \Sigma \; \Im \Pi \right|.
\end{align}
Without yet specialising to any particular case, we can expand this near the large complex structure limit, so for large $y$, in the form
\begin{align} \notag
\frac{|\cZ|^2}{a^2+b^2} &\= \frac{1}{3}  y^3 Y + \frac{1}{6}Y\sigma + \sum_{k=1}^\infty \frac{(2 \pi  k y+1) \cos (2 \pi k  x)}{(2 \pi k)^3} N^{\text{GW}}_k \me^{-2\pi k y}~.
\end{align}
This formula, however, is deceptively simple. It depends on the location of the attractor point $x+\ii y$, which in turn depends on the charge vector $Q$. Our aim is to solve this dependence explicitly and so find a formula for the central charge that depends only on the black hole charges and the internal geometry of the manifold. An interesting aspect of this expression is the appearance of the Gromov-Witten invariants, or equivalently the instanton numbers, since these hint at counting of microstate contributions. Since $x$ and $y$ also have expansions in terms of the Gromov-Witten invariants, this feature is preserved when we eliminate $x$ and $y$ from the expression.

\subsubsection*{Rank two attractor points}
\vskip-10pt
\label{sect:Entropy_Rank-2_Attractor_Points}
A particularly interesting examples of attractor points are provided by \textit{rank two attractor points} where two linearly independent charges solve the alignment equation \eqref{eq:Strominger_Equations_Original} at the same point in the moduli space. These are much rarer than rank one attractor points, as the period vector has to satisfy much more stringent conditions. However, precisely due to these restrictions, we also have more control over various quantities associated to these~points.

As first emphasised by Moore \cite{Moore:1998pn,Moore:1998zu}, rank two attractor point also have interesting connections to number theory. The explicit examples of rank two attractor points found for the manifold AESZ34 in \cite{Candelas:2019llw} display some intriguing connections between attractor point coordinates, central charge and L-functions. For example, it was conjectured (supported by extensive numerical evidence) that there is a rank two attractor point at $\varphi = 33 - 8 \sqrt{17}$ whose location in the complexified K\"ahler moduli space is
\vskip-20pt
\begin{align} \notag
t \= \ii \frac{5}{16 \cdot 17} (9 + \sqrt{17}) \frac{\pi \lambda_4(1)}{\lambda_4(2)}~,
\end{align}
where $\lambda_4(s)$ are the real parts of an L-function $L(s)$ for the group $\Gamma_1(34)$ with the LMFDB designation \textbf{34.4.b.a} \cite{LMFDB}. In this paper, we will be able to rederive some of these results using a complementary point of view that emphasises the relation of the attractor point coordinate and the charge vector, whereas the method utilised in \cite{Candelas:2019llw} uses zeta-function factorisation to find the location attractor point in terms of the mirror coordinate $\varphi$. As a result of our different approach, we are able to derive some interesting identities between the L-function values $\lambda_4(1)$, $\lambda_4(2)$ and Gromov-Witten invariants. Identities like this were already anticipated by Moore \cite{Moore:1998pn}. 

\subsection{Outline of the paper}
\vskip-10pt
In this paper, our aim is to solve the attractor equations for an arbitrary one-parameter family of Calabi-Yau manifolds by developing $N$-expansions for the attractor point coordinate $t$. Although we consider an arbitrary manifold, we specialise the charge vector for simplicity and concentrate on two important special cases that correspond to the D6-D2-D0 and D4-D2-D0 brane systems. The main reason for this choice is that restricting to a case where one of the charges vanishes makes the resulting equations much more tractable. In addition, these systems are of particular interest with microstate counting in mind, as their microstate indices can be viewed either as Donaldson-Thomas invariants or as Witten indices of divisor moduli spaces~\cite{Collinucci:2008ht,Gaiotto:2006wm,Gaiotto:2007cd}.

In \sref{sect:Method_Overview}, prior to studying the two special cases separately, we discuss the general method that will be used to derive the $N$-expansions. We use this section to set notation and find expressions that we will later specialise.

In \sref{sect:Partitions}, we review some basic facts concerning partitions. While this material is standard, we take the opportunity to set the notation that is used extensively in the rest of the paper. We also define operations on partitions that are useful for expressing various functions as perturbative expansions.

Following these preliminaries, in \sref{sect:D6-system} and \sref{sect:D4-system} we are able to calculate fully instanton-corrected solutions to the attractor equations for the general D6-D2-D0 and D4-D2-D0 configurations. The alignment and orthogonality equations are considered separately. The solutions are separated into a perturbative part and instanton corrections.

Our strategy is to express the instanton corrections as series in $q_y = \me^{-2\pi y_0}$. These series have coefficients which are polynomials in the perturbative coordinates $x_0$, $y_0$, as well as the related quantities $\sin 2\pi x_0$ and $\cos2 \pi x_0$. The polynomials are most conveniently expressed as solutions to recurrence relations. These relations can be quite complicated, but in \sref{sect:Solutions} we see that in some important cases there exist closed-form expressions. In particular, the polynomials appearing in the solutions to the D4-D2-D0 orthogonality equation and a special case of the D6-D2-D0 alignment equations can be written in terms of modified spherical Bessel functions.

This special case of the D6-D2-D0 alignment equations is of particular interest since the rank two attractor points of AESZ34, found in \cite{Candelas:2019llw}, are included in this case. The highly constrained nature of the rank two attractor points allows one, for example, to write the entropy in the particularly simple form~\eqref{eq:Central_Charge_Introduction}. Combining the solutions obtained here to the formulae in \cite{Candelas:2019llw}, where the entropy at rank two attractor points was related to L-function values, we are able to derive an intriguing formula relating L-function values to Gromov-Witten invariants and Bessel functions. 

In \sref{sect:Applications} we discuss some applications of our expressions. First the attractor points in the large complex structure region are divided according to whether or not the black hole central charge vanishes: an important distinction for the physical interpretation of the solutions. We are able, based on consideration of the black hole charge only, to determine whether solutions to the alignment or orthogonality equations exist near the large complex structure point.

To conclude this section, we consider our entropy formulae from the perspective of microstate counting and verify that in leading order our expressions agree with those in \cite{Maldacena:1997de} and with modular form asymptotics. We are, however, considering a limit that is different from that in~\cite{Maldacena:1997de} and so differences appear in subleading orders. Building on the work of \cite{Behrndt:1996jn,Behrndt:1996mm,Behrndt:1998eq,LopesCardoso:1998tkj,LopesCardoso:1999cv,LopesCardoso:1999fsj,LopesCardoso:1999xn} (for a review, see \cite{Mohaupt:2000mj}), we show that the proper interpretation of our entropy formula is that it gives the leading-order Wald entropy of the black hole in the large charge limit. In this sense, our results extend those in~\cite{Shmakova:1996nz,Maldacena:1997de,LopesCardoso:1998tkj}.

Matters that would otherwise disrupt the narrative are relegated to four appendices. In  appendix~\ref{app:Symplectic_Transformations}, we discuss the action on our formulae of a $\IZ^3$ subgroup of $\text{Sp}(4,\IZ)$, and identify certain combinations of the Yukawa couplings and the charges which are invariant under this group. In appendix \ref{app:Monodromy}, we examine the role of the monodromy transformations around the large complex structure limit, showing that these can either be interpreted as a symmetry of the solutions or used to extend our solutions to cases where the D4 brane charge divides the D6 charge. Appendix \ref{app:D-Brane_Interpretation_of_Q} sets out the relation of the charge vector $Q$ to the various D-brane charges. In appendices \ref{app:D6_Orth_Polynomials} and \ref{app:D6_Polynomials}, we list a set of polynomials which can be used to express the perturbative solutions to the D6-D2-D0 attractor equations. Finally, in appendix \ref{sect:Asymptotics_Convergence}, we discuss the asymptotics and convergence of solutions we find.

\subsection{Notation}
\vskip-10pt
We introduce notation in an attempt to streamline otherwise complicated equations, and collect the notation which is used throughout this paper, in the tables below. 
\begin{table}[H]
	\renewcommand{\arraystretch}{1.7}
	\begin{center}
		\begin{tabular}{ |c|c|c|c|}
			\hline
		 $Y_{111} = Y$ &  $\text{Im} \, Y_{000} =Y\sigma$ & $\zeta=\frac{\sigma}{y_{0}^{3}}$  & $\widehat{\sigma}=\frac{\sigma}{\beta^{3}}$\\[5pt]
			\hline
	   \quad $\alpha =-\frac{Y_{110}+\Upsilon}{\wt\beta Y}$ \quad \null	& \quad $\beta=\sqrt{\frac{Y_{100}+2\Upsilon}{Y}}$ \quad \null & \quad $\wt\beta=\sqrt{\frac{Y_{100}+2\Lambda}{Y}}$ \quad \null&  \qquad $\gamma=-\frac{3\Lambda}{\beta^{3}Y}$ \qquad \null\\[5pt]
	   \hline
		\end{tabular}
		\vskip10pt
		\capt{4.5in}{tab:shorthand}{Shorthand for $Y_{ijk}$- and charge-related quantities. We take $\beta$ and $\wt\beta$ to have argument $0$ or $\frac{\pi}{2}$, depending on the argument of the square root.}
	\end{center}
\end{table}
\vskip-33pt
In the main body of this paper, we choose a symplectic basis for $H^3(X,\IZ)$ in which $Y_{000}$ has no real part. When attention is restricted to a system with no D6 charge, the $\IZ^{3}$ symmetry discussed in appendix \ref{app:Symplectic_Transformations} leaves the combinations $\alpha$ and $\wt\beta$ invariant. In the setup where the D4 charge vanishes, this symmetry leaves $\beta$ and the quantity \hbox{$\beta^{3} \gamma Y-\Re \, Y_{000}$} invariant.
\begin{table}[H]
	\renewcommand{\arraystretch}{1.49}
	\centering
	\capt{6in}{tab:notation2}{Quantities that are used throughout the paper with references to where they are defined. }
	\begin{tabularx}{16.5cm}{|>{\hsize=.125\hsize\linewidth=\hsize}X|
			>{\hsize=0.4\hsize\linewidth=\hsize}X|>{\hsize=0.4\hsize\linewidth=\hsize}X|  >{\hsize=0.07\hsize\linewidth=\hsize}X|}
		\hline
		\textbf{Notation} & \hfil \textbf{Definition} & \hfil \textbf{Description} & \textbf{Ref} \\[2pt]
		\hline	\hline
		$\Pi$ & $(\partial_0 \cF, \partial_1 \cF,	z^0, z^1)^T$ & period vector of $X$ & \eqref{eq:defn_Period_Vector} \\[4pt] \hline
		$Q$ & $(q_0,q,p_0,p)^T$ & charge vector & \eqref{eq:defn_Q} \\[4pt] \hline	
		$\cF$ & $-\frac{1}{3!} Y_{abc} \frac{z^a z^b z^c}{z^0} - (z^0)^2 \cI \left( \frac{z^1}{z^0} \right)$ & Calabi-Yau prepotential & \eqref{eq:defn_Calabi-Yau_prepotential}\\[4pt] \hline
		$t=x+\ii y$ & $z^1/z^0$ & attractor point coordinate & \sref{sect:Preamble}\\[4pt] \hline
		$t_0, x_{0}, y_{0}$ & $t_0 = x_0 + \ii y_0$ & perturbative coordinates & \sref{sect:period_vector_decomposition} \\[4pt] \hline		
		$x_{00}, y_{00}$ & $x_0 |_{\sigma = 0}$, $y_0 |_{\sigma = 0}$  & coordinates at $\sigma=0$ & \sref{sect:period_vector_decomposition} \\[4pt] \hline			
		$\cI$ & $\frac{1}{(2\pi \ii)^3} \sum_{k=1}^\infty n_k \text{Li}_3(\me^{2 \pi \ii k t})$ & instanton sum & \eqref{eq:Instanton_Sum_Definition}\\[4pt] \hline
		$\cI_1,\,\cI_2$ & $\Re[\cI], \, \Im[\cI]$ & real and imaginary parts & \sref{sect:Third_and_Fourth_Alignmnent} 
		\\[4pt] \hline							
		$Y_{ijk}$ & $\int_{\wt{X}}e_{i}\wedge e_{j}\wedge e_{k}$ & $Y_{ijk}$ coefficients & \eqref{eq:yukawa_definition}\\[4pt] \hline	
		$\cZ$ & $\frac{Q^{T}\Sigma \Pi}{\sqrt{-\ii \Pi^{\dagger} \Sigma \Pi }}$ & central charge & \eqref{eq:Central_Charge_Definition} \\[4pt] \hline	
		
		$n_k$ & & instanton numbers & \eqref{eq:Instanton_Sum_Definition} \\[4pt] \hline	
		$N^{\mbox{\tiny GW}}_k$ & & Gromov-Witten invariants & \eqref{eq:defn_N_k_invariants} \\[4pt] \hline
		$N_k$ & $\sum_{d|k} d^3 n_d = k^3 N^{GW}_k$ & scaled GW invariants & \eqref{eq:defn_N_k_invariants} \\[4pt] \hline
		$q_t$ & $\me^{2\pi \ii t_0}$ & small expansion parameter & \eqref{eq:txy_ansatz} \\[4pt] \hline
		$q_x,q_y$ & $q_x = \ee^{2\pi\ii x_{0}},\, q_y=\ee^{-2\pi y_{0}}$ & phase and modulus of $q_t$ & \eqref{eq:txy_ansatz} \\[4pt] \hline
		$t_i$ & $t = \sum_{j=0}^\infty t_j q_t^j$ & $q_t$-expansion of $t$ & \eqref{eq:txy_ansatz} \\[4pt] \hline			
		%$t_i$ & $t = \sum_{j=0}^\infty t_j q_t^j$ & $q_t$-expansion of $t$ & \eqref{eq:txy_ansatz} \\[4pt] \hline						
	\end{tabularx}\\
	\vskip10pt
\end{table}

\begin{table}[H]
	\renewcommand{\arraystretch}{1.5}
	\centering
	\begin{tabularx}{16.5cm}{|>{\hsize=.125\hsize\linewidth=\hsize}X|
			>{\hsize=0.4\hsize\linewidth=\hsize}X|>{\hsize=0.4\hsize\linewidth=\hsize}X|>{\hsize=0.07\hsize\linewidth=\hsize}X|}
		\hline
		\multicolumn{4}{|>{\hsize=15cm}X|}{ \sffamily \tref{tab:notation2} continued. }\\\hline		
		\hline
		\textbf{Notation} & \hfil \textbf{Definition} & \hfil \textbf{Description} & \textbf{Ref} \\
		\hline	\hline	
		$x_i,y_i$ & $x = \sum_{j=0}^\infty x_j q_y^j$, $y = \sum_{j=0}^\infty y_j q_y^j$ & real and imaginary parts & \eqref{eq:xy_q_j_expansion} \\[4pt] \hline				
		$R^{t}_\fp$ & $t_{j} = \sum_{\fp \in \pt(j)} N_\fp R^{(t)}_\fp$ & $N_k$-expansion of $t_j$ & \eqref{eq:x_y_ansatse} \\[4pt] \hline
		$R^{x}_\fp,\, R^{y}_\fp$ & $\Re \left[R^t_\fp \right], \, \Im \left[R_\fp^t \right]$ & real and imaginary parts & \eqref{eq:x_y_ansatse} \\[4pt] \hline	
		$\fp,\fq$ & $\sum_{j} \mu_j j, \; \sum_{j} \nu_j j$ & partitions & \sref{sect:Partitions} \\[4pt] \hline	
		$c_\fp$ & $\frac{\left(\sum_{j=1}^\infty \mu_j \right)!}{\prod_{j=1}^\infty \mu_j !}; \qquad \fp = \sum_{j} \mu_j j$ & multinomial coefficient & \eqref{eq:defn_c_fp} \\[4pt] \hline	
		$a_\fp$ & $\prod_{j=1}^{\infty} \frac{1}{j^{2 \mu_j} \mu_j!}, \, \text{when} \, \, \fp = \sum_{j=1}^\infty \mu_j j$ & coeffs. in the solutions in \sref{sect:Solutions} & \eqref{eq:defn_a_fp} \\[4pt] \hline	
				
		$(F \sr G)_\fp$ & $\sum_{\fq \subseteq \fp} F_{\fp - \fq}\, G_{\fq}$ & natural convolution-like product & \eqref{eq:defn_sr_product} \\[4pt] \hline
		$(F {\circ} G)_\fp$ & $\sum_{\fq \subset \fp} F_{\fp - \fq}\, G_{\fq}$ & as above, empty partitions excluded & \eqref{eq:defn_sr_product} \\[4pt] \hline		
		$\Lambda, \Upsilon$ & varies according to context & charge ratios & \eqref{eq:chargevectors} \\[4pt] \hline
		$|\fp|$ & the integer with partition $\fp$ &  & \sref{sect:Partitions} \\[4pt] \hline
		$\length(\fp)$ & number of parts of a partition $\fp$ & length of $\fp$ & \sref{sect:Partitions} \\[4pt] \hline
		$\pt(k)$ & $\left\{\,\fp~\big|~ |\fp| = k \right\}$ & set of all partitions of $k$ & \eqref{eq:defn_pt(k)} \\[4pt] \hline
		$\pt_s(k)$ & $\{\,\fp~\big|~|\fp|=k~~\text{and}~~ \length(\fp) = s \}$ & all partitions of length $s$ & \eqref{eq:defn_pt_s(k)} \\[4pt] \hline
		$\eta_k$ & $\sum_{j=0}^k \frac{(-1)^j}{y_0^{j+1}} \sum_{\fp \in \pt_j(k)} c_\fp y_\fp$ & coeffs. in the $q_y$-expansion of $1/y$ & \eqref{eq:xy_Series_q_Ansatz} \\[4pt] \hline
		$\bm{\eta}_\fp$ & $\frac{1}{y_0} \sum_{j=0}^{l(\fp)} \frac{(-1)^j}{y_0^j} r^{y_{\,}}_{(j,j),\fp}$ & coeffs. in the expansion of $1/y$ & \eqref{eq:defn_N_p_expansions} \\[4pt] \hline						
		$\me^{y}(\alpha)$ & $\sum_{j=0}^k \frac{(2\pi m)^j}{j!} \sum_{\fp \in \pt_j(k)} c_\fp y_\fp$ & coeffs. in the $q_y$-expansion of $\me^{2\pi y}$ & \eqref{eq:xy_Series_q_Ansatz} \\[4pt] \hline
		$\bme^{y}_\fp$ & $\sum_{j=0}^{l(\fp)} \frac{(2\pi \alpha)^j}{j!} r^{y_{\,}}_{(j,j),\fp}$ & coeffs. in the expansion of $\me^{2\pi y}$ & \eqref{eq:defn_N_p_expansions} \\[4pt] \hline
		$r^{x}_{(m,n),\mathfrak{p}}$ & $\displaystyle \sum_{\substack{\sum_{j=1}^n \fp_j = \mathfrak{p} \\ \fp_1,...,\fp_m \neq \emptyset \\[3pt]}} c\big( \hskip-2pt \left \{\mathfrak{p}_j \right\}_{j=1}^i \big) \prod_{j=1}^i R^{x}_{\mathfrak{p}_j}$ & combinations of functions $R^{x}$ & \eqref{eq:defn_r^(x)} \\[4pt] \hline			
		$\bm{k}_n(z)$ & $\sqrt{\frac{2}{\pi z}} K_{n+\frac{1}{2}}(z)$ & spherical $k$ Bessel functions & \eqref{eq:D6-D0_P_s_M} \\[4pt] \hline	
		$f_n(y)$ & $\me^{y} y^{n+1} \bm{k}_{n}(y)$ &polynomials associated to $\bm{k}_n$ & \eqref{eq:D6-D0_P_s_M} \\[4pt] \hline	
		$\cF^{(g)}(X)$ &  & genus $g$ Calabi-Yau prepotential & \sref{sect:Microstate_Counting} \\[4pt] \hline
		$\ccF(X,W^2)$ & $\sum_{g=0}^\infty \cF^{(g)}(X) W^{2g}$ & all-genus CY prepotential & \eqref{eq:All-Genus_Prepotential_Expansion} \\[4pt] \hline																	
	\end{tabularx}\\
	
\end{table}
\clearpage

\newpage

\section{Solving the Attractor Equations: a General Overview} \label{sect:Method_Overview}
\vskip-10pt
Our strategy for solving the attractor equations relies on using the expression \eqref{eq:defn_Period_Vector} for the period vector $\Pi$ to express both the alignment equations \eqref{eq:Strominger_Equations} and the orthogonality equation \eqref{eq:Orthogonality_Equations_Original} explicitly as a set of equalities involving geometric data (the topological quantities $Y_{abc}$ and Gromov-Witten invariants of the Calabi-Yau manifold) and the black hole charges. 

In the following sections \sref{sect:D4-system} and \sref{sect:D6-system}, we solve these equations perturbatively in the instanton corrections. We first find the zeroth-order solution corresponding to what we shall refer to as the \textit{perturbative} part of the prepotential, meaning the $Y_{abc}$-terms, corresponding to ignoring the contributions of the instanton sums. This gives us a pair of polynomial equations of degree at most four. Explicit solutions of such polynomial equations are of course known since the Renaissance, but we dedicate some discussion to giving the most concise solutions. We show that the solutions can, in most cases, be given by simple trigonometric expressions.

Near the large complex structure point the instanton corrections are exponentially small and we expand in terms of the quantity $q_y {\=} \me^{-2\pi y_0}$, where $y_0$ is the perturbative value of the attractor coordinate, found by using the perturbative equations as outlined above. We find that the instanton corrections take the form of a series in $q_y$,
\begin{align}
x \= \sum_{j=0}^\infty x_j q_y^j~, \qquad y \= \sum_{j=0}^\infty y_j q_y^j~, \label{eq:xy_q_j_expansion}
\end{align}
with coefficients $x_j$ and $y_j$ that are rational functions of $x_0$, $y_0$, $\sin 2 \pi x_0$, and $\cos 2 \pi x_0$.
\begin{align}
x_{j} &\= \sum_{\fp \in \pt(j)} \hskip-5pt N_\fp R^{x_{\,}}_\fp(x_0,y_0), \qquad y_{j} \= \sum_{\fp \in \pt(j)} \hskip-5pt N_\fp R^{y_{\,}}_\fp(x_0,y_0), \label{eq:x_y_ansatse}
\end{align}
where $\fp$ is a partition of $j {\,\in\,} \IZ^{+}$, $R^{x_{\,}}_\fp(x_0,y_0)$ and $R^{y_{\,}}_\fp(x_0,y_0)$ are the rational functions of $x_0$, $y_0$, $\sin(2\pi x_0)$ and $\cos(2\pi x_0)$ and are independent of the Gromov-Witten invariants. The $N_\fp$ are combinations of the scaled Gromov-Witten invariants as in \eqref{eq:a_p_N_p_definition}. We give recurrence relations for the $R_\fp^x$ and $R_\fp^y$, which are amenable to rapid computer calculation. Since the individual terms in the sum are exponentially small the first few terms suffice for many practical purposes. 

A pleasant surprise is that even though the recurrence relations are somewhat complicated in the general case, we are able to give simple closed-form solutions in important special cases. In particular, for the D4-D2-D0 orthogonality equation and a special case of D6-D2-D0 alignment equations, we will find in \sref{sect:Solutions} interesting expressions for the location of the attractor point which curiously involve spherical Bessel functions. 

\subsection{Particular brane systems} \label{sect:Particular_Brane_Systems}
\vskip-10pt
We study extremal charged black holes that, in the string theory picture, arise from wrappings of \mbox{D-branes} on cycles on the internal Calabi-Yau manifold. We restrict our attention to arrangements containing either D6- or D4-branes but not both. We attach to these some D2- and D0-branes. Recalling the form of a generic charge vector \eqref{eq:defn_Q}, this means that we consider charge vectors of the form 
\begin{equation}\label{eq:chargevectors}
Q_{\text{D4}}=\kappa\left(\begin{matrix}\Lambda\\\Upsilon\\0\\1\end{matrix}\right), \hspace{2cm} Q_{\text{D6}}=\kappa\left(\begin{matrix}\Lambda\\\Upsilon\\1\\0\end{matrix}\right).
\end{equation}
The physical charge vectors have integral components, so $\Lambda$ and $\Upsilon$ are rational numbers, and $\kappa$ is a suitable integer. The charge vectors are written in this way partly to simplify later expressions since the locations of attractor points depend only on the \textit{charge ratios} $\Upsilon$ and $\Lambda$. Furthermore, this decomposition allows us to make an important distinction between what we call the \textit{large charge limit}, where ${|\kappa| \to \infty}$ and the \textit{large charge ratio limits}, where either $|\Lambda|$ or $|\Upsilon|$ is large. 

In~\cite{Candelas:2019llw}, where rank two attractor points for one-parameter families were first explicitly identified, the associated charge vectors were of this type, and charge vectors of this form have received much attention in the literature. Microscopic state counting has been performed both for black holes of $Q_{\text{D4}}$ charge using elliptic genera \cite{Gaiotto:2006wm,Gaiotto:2007cd} and D6-D0 bound states using Donaldson-Thomas invariants \cite{Denef:2007vg,Collinucci:2008ht}.

By utilising the monodromy around the large complex structure point we can extend our solutions to any brane system where the D6-charge divides the D4-charge, as we will show in detail in appendix~\ref{app:Monodromy}.

\subsection{General comments concerning the alignment equations} \label{sect:Third_and_Fourth_Alignmnent}
\vskip-10pt
The alignment equations \eqref{eq:Strominger_Equations_Original} depend on the quantity $\Im[C\Pi]$, which is manifestly not holomorphic in $t$, so it becomes useful to decompose the coordinate $t$ into its real and imaginary parts, $t = x +\ii y$. Similarly, the constant $C$ is written as $a+\ii b$. All of the $Y_{abc}$ can be chosen to be real, apart from $Y_{000}=\ii Y\sigma$ which becomes pure imaginary. The instanton sum $\mathcal{I}(t)$ and its derivative $\mathcal{I}'(t)$ can be decomposed as $\mathcal{I}(t)=\mathcal{I}_{1}(x,y)+\ii\mathcal{I}_{2}(x,y)$ and $\mathcal{I}'(t)=\mathcal{I}_{1}'(x,y)+\ii\mathcal{I}_{2}'(x,y)$.

In terms of the components of the period vector \eqref{eq:defn_Period_Vector} the alignment equations \eqref{eq:Strominger_Equations} are
\beq\begin{split}
	&Q \= \\
	&a \hskip-3pt
	\begin{bmatrix}
		\vrule height13pt width0pt \hskip-2pt \frac{1}{6} Y \hskip-2pt y \hskip-2pt \left( 3x^2 {-} y^2 \right) {-} \frac{1}{2} Y_{100} y {-} \frac{1}{3}Y\hskip-2pt \sigma  {+} 
		y \cI_1' {+} x \cI_2' {-} 2 \cI_2\!{} \\[3pt]
		-Yxy {-} Y_{110} y {-} \cI_2' \\[3pt]
		0 \\[3pt]
		y
	\end{bmatrix} \hskip-2pt	
	+b \hskip-3pt
	\begin{bmatrix}
		\vrule height13pt width0pt \hskip-2pt \frac{1}{6} Y \hskip-2pt x \hskip-2pt \left( x^2 {-} 3y^2 \right) {-} \frac{1}{2} Y_{100} x {+} x \cI_1'{-} y \cI_2' {-}
		2 \cI_1\!{} \\[5pt]
		\frac{1}{2} Y (y^2{-}x^2) {-} Y_{110}x {-} \frac{1}{2} Y_{100} {-} \cI_1' \\[3pt]
		1 \\[3pt]
		x
	\end{bmatrix}
\end{split}\label{eq:Strominger_Equation_Full}\eeq
The lower two components of this vector equation determine $a$ and $b$
\begin{equation}\label{eq:aandb}
(a,b)=\begin{cases}\frac{\kappa}{y}\left(-x\, ,y\right), & \text{D6 system}\\ & \\\frac{\kappa}{y}\left(\phantom{-}1\, ,0\right), & \text{D4 system.}\end{cases}
\end{equation}
Substituting these two values into the upper two components of equation \eqref{eq:Strominger_Equation_Full} leaves two equations for $x$ and $y$. This offers another perspective on the ubiquity of rank one attractor points and rarity of those of rank two: the alignment equations written in the form \eqref{eq:Strominger_Equation_Full} can be solved very often, but satisfying this equation for two linearly independent charge vectors involves four equations and only two unknowns. This latter system is overdetermined and usually insoluble.

After substituting in \eqref{eq:aandb}, there is an overall factor of $\kappa$ on both sides of \eqref{eq:Strominger_Equation_Full} which cancels. This leaves the equations with which we begin the discussion of the alignment equations in \sref{sect:D4-system} and \sref{sect:D6-system}.
\subsection{Setting up the iteration}\label{sect:period_vector_decomposition}
\vskip-10pt
To set up the perturbative problem, we will separate, in the period vector, those terms that do not contain instanton corrections from those that do. \vskip-3pt
\begin{equation*}\notag
\begin{gathered}
\Pi=\Pi_{0}+\Delta_{\mathcal{I}}\Pi,\\[15pt]
\Pi_{0}=\left(\begin{matrix}\+ \frac{1}{6}Yt^{3}-\frac{1}{2}Y_{100}t-\frac{1}{3} \ii Y\sigma\\[3pt]
-\frac{1}{2}Yt^2-Y_{110}t-\frac{1}{2}Y_{100}\\1\\t\end{matrix}\right),\quad \quad \Delta_{\mathcal{I}}\Pi=\left(\begin{matrix}t\mathcal{I}'-2\mathcal{I}\\-\mathcal{I}'\\0\\0\end{matrix}\right).
\end{gathered}
\end{equation*} 
As stated above, we wish to treat the instanton terms as small corrections, and so in all cases we first solve the zeroth-order \textit{perturbative equations}. These are obtained from the full alignment or orthogonality equations by simply replacing the full period vector $\Pi$ by $\Pi_0$,
\begin{align} \label{eq:perturbative_eqs}
\Im[C \Pi_0] \= Q~, \qquad\qquad \Pi_0^T \Sigma Q \= 0~.
\end{align} 
We denote the solutions to the perturbative alignment equations by $t_0 = x_0 + \ii y_0$, and largely work directly with the real and imaginary parts. We can find a closed-form solution for $x_0$ and $y_0$, which depends on $\sigma$. This solution is usually most conveniently expressed in terms of trigonometric functions. We can also give $x_{0}$ and $y_{0}$ as a power series in $\sigma$. Thus, in addition to the perturbative coordinates $x_0$ and $y_0$, we define quantities $x_{00}$ and $y_{00}$ as the $\sigma=0$ limits of these.
\begin{align} \notag
x_{00} \= x_0 \big|_{\sigma = 0}~, \qquad y_{00} \= y_0 \big|_{\sigma = 0}~. 
\end{align}
The utility of this expansion will be seen in \sref{sect:Applications}, where it is a natural expansion to use for the large charge ratio limits, as well as for comparing supergravity results to microscopic formulae.
 
Solving the orthogonality equation proceeds in a very similar manner. The principal difference is that, as the orthogonality equation \eqref{eq:Orthogonality_Equations_Original} depends on $t$ holomorphically, it is most convenient to work directly with $t$ itself instead of splitting it into real and imaginary parts. Similarly, we do not split the instanton sums $\cI$ and $\cI'$ into their real and imaginary parts. We denote the solution to the perturbative equations \eqref{eq:perturbative_eqs} by $t_0$ and express this both as a closed-form expression involving trigonometric functions, as well as a Taylor series expansion around $\sigma=0$. To include the instanton corrections we make the ansatz
\begin{align}\label{eq:txy_ansatz}
t \= \sum_{j=0}^\infty t_j q_t^j~.
\end{align}
The small parameter we use here is $q_t = \me^{2\pi \ii t_0}$. Note however that $|q_t| = q_y$, so we can expect similar convergence. As before, we are able to derive recurrence relations for the coefficients $t_j$. In \eqref{eq:txy_ansatz} the $t_j$ are also expressed in terms of rational functions and Gromov-Witten invariants in the~form
\begin{align} \label{eq:t_j_R^t_ansatz}
t_j \= \sum_{\fp \in \pt(j)} \hskip-5pt N_\fp R^t_\fp(t_0)~.
\end{align}

\newpage
\section{Partitions} \label{sect:Partitions}
\vskip-10pt
Since many of the formulae we will come across while solving the attractor equations involve partitions, we find it useful to introduce some terminology and notation from the domain of enumerative combinatorics and provide a rudimentary introduction to the theory of partitions. This serves to fix our notation and give some of the basic results that will be extensively used in the following. A more thorough treatment can be found in, for example, \cite{Stanley}, whose notation we largely follow. 

\subsection{Basic definitions and notation}
\vskip-10pt
A \emph{partition} $\fp$ of a positive integer $k \in \IZ_{>0}$ is a sequence of positive integers 
\beq 
0 < p_1 \leqslant p_2 \leqslant \ldots \leqslant p_m \quad \text{such that} \quad \sum_{i=1}^m p_i \= k~.
\notag\eeq 
If $\fp$ is a partition of $k$, we write $|\fp| {\=} k$ and refer to the terms $p_i$ as the \emph{parts} of $\fp$. The number of parts is called the \emph{length} of the partition, which is denoted by $\length(\fp)$. If $\mu_i$ denotes the number of parts equal to $i$, we can write partitions by means of a multiset notation $\fp {\=} \{1^{\mu_1},2^{\mu_2},...\}$, where only a finite number of $\mu_i$ are non-zero.

The partitions of an integer $k$ can be also viewed as the different ways of expressing $k$ as a sum of positive integers without distinguishing the order of the summands. In virtue of this, partitions are often also written using a summation notation:
\beq
\fp \= \sum_{i=1}^\infty \mu_i i ~.
\notag\eeq
Every integer $k$ is a partition of itself, so we can regard integers as special cases of partitions.

Although partitions are defined as sequences of positive numbers, it is sometimes convenient to generalise the notion of partition to include \emph{partitions with zeros}. A partition with zeros $\fp$ of $k {\,\in\,} \IZ_{>0}$ is a sequence of non-negative integers
\beq
0 \leqslant p_1 \leqslant p_2 \leqslant ... \leqslant p_m \quad \text{such that} \quad
\sum_{i=1}^m p_i \= k~ . 
\notag\eeq
The definitions of $\length(\fp)$ and $|\fp|$ are as before.

There is an important special case where the partition corresponds to an empty multiset $\fp = \emptyset$. This is the \emph{empty partition}, and we take $\length(\emptyset)=0$ and $|\emptyset|=0$. It is important to distinguish the empty partition $\emptyset$ from the partition with zeros $\{0^n\}$. For example, we have $\length(\emptyset)=0$, while $\length(\{0^n\})=n$.
\subsection{Operations on partitions}
\vskip-10pt
In this subsection 
\begin{align*} \nonumber
\fp \= \sum_{i=0}^\infty \mu_i i \= \{0^{\mu_0},1^{\mu_1},2^{\mu_2},...\} \qquad \text{and} \qquad \fq \= \sum_{i=0}^\infty \nu_i i \= \{0^{\nu_0},1^{\nu_1},2^{\nu_2},...\}
\end{align*} 
are either partitions or partitions with zeros, for which we use the multiset or sum notation, as convenient. 

We write  $\fp \subseteq \fq$ if, as multisets, $\fp \subseteq \fq$, which means that $\mu_i \leqslant \nu_i$ for all $i$. We use $\subset$ when~$\mu_i < \nu_i$.

We can always define a sum of two partitions as a union of the partitions: $\fp + \fq {\=} \fp \cup \fq$, where in writing $\fp \cup \fq$ we understand $\fp$ and $\fq$ as multisets. Alternatively, in the sum notation, we can write 
\beq \notag
\fp + \fq \= \sum_{i=0}^\infty (\mu_i + \nu_i) \, i~. 
\eeq 
It is clear that this is always a partition, and that $|\fp + \fq| {\=} |\fp| + |\fq|$, and $\length(\fp + \fq) {\=} \length(\fp) + \length(\fq)$.

When $\fp \subseteq \fq$, we can define also the difference of these partitions as $\fq - \fp {\=} \fq \setminus \fp$ or, slightly informally, using the sum notation, as 
\beq \notag
\fq - \fp \= \sum_{i=0}^\infty (\nu_i - \mu_i) \, i~. 
\eeq 
Note that there is a potential source of confusion here with the treatment of zeros. We emphasise that this sum is just an informal way of writing the difference, and the unambiguous definition is given via the multiset difference. Similarly to the sum, we have that \hbox{$|\fq - \fp| {\=} |\fq| - |\fp|$} and $\length(\fq - \fp) {\=} \length(\fq) - \length(\fp)$. In the special case $\fp = \fq$, we have 
\begin{align} \nonumber
\fp - \fp \= \emptyset,
\end{align}
which follows unambiguously by taking the multiset difference.

It is useful also to set notation for some important sets of partitions. We denote the set of all partitions of $k$, of any length, by 
\beq
\pt(k) \; \defineas \; \left\{\,\fp~\big|~ |\fp| = k \right\}~. \label{eq:defn_pt(k)}
\eeq 
The set of all partitions of $k$ with length $s$ is denoted by
\beq
\pt_s(k) \; \defineas \; \{\,\fp~\big|~|\fp|=k~~\text{and}~~ \length(\fp) = s \}~. \label{eq:defn_pt_s(k)}
\eeq
In addition, we define some unions of the above sets: 
\beq \notag
\pt_{{\scriptscriptstyle\leqslant}\, s}(k) \; \defineas \; \bigcup_{r=1}^s \pt_r(k)\,,~~\text{and}~~ 
\pt_{{\scriptscriptstyle <}\, s}(k) \; \defineas \; \bigcup_{r=1}^{s-1} \pt_r(k)~,
\notag\eeq 
with the sets $\pt_{{\scriptscriptstyle\geqslant}\, s}(k)$ and $\pt_{{\scriptscriptstyle >}\, s}(k)$ defined similarly. 

The set of partitions of $k$ with zeros is denoted by $\pt^0(k)$, and the definitions of the variants such as $\pt^0_{{\scriptscriptstyle\leqslant}\, s}(k)$ are given straightforwardly, as in the above. To avoid any confusion arising from the somewhat subtle issue of the empty partition, we take $\pt(0)=\{\emptyset\}$, and $\emptyset \notin \pt^0_i(k)$ for any $i$ or $k$. 

It is sometimes convenient to use a partition $\fp$ as a multi-index, specifying a product. This is understood exclusively for the following five quantities:
\beq
x_\mathfrak{p} \= \prod_{j=0}^\infty x_{j}^{\mu_j}~, \qquad y_\fp \= \prod_{j=0}^\infty y_{j}^{\mu_j}~, \qquad t_\fp \= \prod_{j=0}^\infty t_{j}^{\mu_j}~, \qquad N_\fp \= \prod_{j=0}^\infty N_{j}^{\mu_j}~, \qquad \bm{m}_\fp \= \prod_{j=0}^\infty \bm{m}_{j}^{\mu_j}~.
\notag
\eeq
In all other cases where $\fp$ appears in a subscript, such as the later $\cE_{j,\fp}$, it is meant as an index that associates a quantity to a partition in some other way specific to that symbol. In other words, said symbol is a different function of $\fp$. 

Functions arise that depend on partitions and also Gromov-Witten invariants. We often expand these functions as power series in the rescaled Gromov-Witten invariants $N_k$. Using the partition notation, the resulting series can be written as
\beq
F \= \sum_{\fp} N_\fp F_\fp~.
\notag\eeq
We shall refer to such an expansion as an \emph{$N$-expansion}. 

We can think of the coefficients $F_\fp$ as functions of the partitions $\fp$. For two such functions $F_\fp$ and $G_\fp$, there is a natural convolution-like product
\beq
(F \sr G)_\fp \, \defineas \, \sum_{\fq \subseteq \fp} F_{\fp - \fq}\, G_{\fq}~. \label{eq:defn_sr_product}
\eeq
This product is clearly commutative and associative, so we write 
\beq
F \sr (G \sr H) \= (F \sr G) \sr H \= F \sr G \sr H~.
\notag\eeq 
The main utility of this product is that if 
\begin{align} \notag
F(j) \= \hskip-3pt \sum_{\fp \in \pt(j)} \hskip-5pt F_\fp \, N_\fp~, \qquad \text{and} \qquad G(j) \= \hskip-3pt \sum_{\fp \in \pt(j)} \hskip-5pt G_\fp \, N_\fp~,
\end{align}
then we have the following identity
\beq
\sum_{j=0}^k F(j)\, G(k-j) \;= \sum_{\fp \in \pt(k)} (F \sr G)_\fp \, N_\fp~.
\notag\eeq
We will find frequent use for this identity in formulating different recurrence relations.

The sum in the convolution product includes the cases $\fq{\=}\fp$ and $\fq{\=}\emptyset$, which both correspond to the empty partition. It is sometimes useful to consider a convolution product which does not include these two terms. We denote this product by $\circ$.
\beq \notag
(F {\circ} G)_\fp \, \defineas \hskip-3pt \sum_{\substack{\fq \subset \fp, \fq \neq \emptyset}} F_{\fp - \fq}\, G_{\fq} \= (F \sr G)_\fp - F_\emptyset G_\fp - F_\fp G_\emptyset~.
\eeq

\subsection{Series expansions}
\vskip-10pt
The main utility of the theory of partitions to the discussion of the attractor equations is that the generating functions of many series that arise may be expressed in terms of the quantities defined above. The prototypical quantities where sums or series can be written with the help of partitions are the powers of sums. For example
\beq
\left( \sum_{j=1}^\infty x_j q^j \right)^n \=
\sum_{k=1}^\infty q^k \sum_{\fp \in \pt_n(k)} \hskip-5pt c_\fp x_\fp~, 
\label{eq:Partitions_x^n}\eeq
where $c_\fp$ is the multinomial coefficient associated to the partition $\fp {\=} \sum_{j=1}^\infty \mu_j j$
\beq
c_\fp \= \frac{\left(\sum_{j=1}^\infty \mu_j \right)!}{\prod_{j=1}^\infty \mu_j !}~. \label{eq:defn_c_fp}
\eeq

Expressions like this will be appear repeatedly in what follows. In this way, we are led to series expansions for 
\begin{align} \nn
x \= \sum_{i=0}^\infty x_i \, q_y^i~, \quad y \= \sum_{i=0}^\infty y_i \, q_y^i~, \quad t\= \sum_{i=0}^\infty t_i \, q_t^i~,
\end{align}
and the following closely related quantities. These expressions follow straightforwardly from \eqref{eq:Partitions_x^n}.
\begin{equation}\label{eq:xy_Series_q_Ansatz}
\begin{aligned}
x^n &\= \sum_{k=0}^\infty q_y^k \sum_{\fp \in \pt^0_{n}(k)} \hskip-5pt c_\fp x_\fp~, \qquad n > 0~,\\[5pt]
\frac{1}{y} &\= \sum_{k=0}^{\infty} \eta_k \, q_y^k~, \qquad\qquad\qquad\text{where}  \;\qquad\qquad\eta_k \; \defineas \; \sum_{j=0}^k \frac{(-1)^j}{y_0^{j+1}} \sum_{\fp \in \pt_j(k)} \hskip-5pt c_\fp y_\fp~,  \\[5pt]
\ee^{-2\pi m y} &\= \sum_{k=0}^\infty \ee_{k}^{y_{\,}}(-m) \, q_y^{k+m}~, \;\,\qquad\text{where}\qquad\;\, \ee_{k}^{y_{\,}}(m) \; \defineas \; \sum_{j=0}^k \frac{(2\pi m)^j}{j!} \sum_{\fp \in \pt_j(k)} \hskip-5pt c_\fp y_\fp~,  \\[5pt]
\me^{2\pi \ii m x} &\= \sum_{k=0}^\infty \me_{k}^{x_{\,}}(\ii m) \, q_y^k q_x^{m}~, \;\;\;\, \qquad\text{where}\qquad\;\, \ee_{k}^{x_{\,}}(m) \; \defineas \; \sum_{j=0}^k \frac{(2\pi m)^j}{j!} \sum_{\fp \in \pt_j(k)} \hskip-5pt c_\fp x_\fp~,\\[5pt]
\me^{2\pi \ii m t} &\= \sum_{k=0}^\infty \me_{k}^{t_{\,}}(\ii m) \, q_t^{k+m} ~, \;\;\;\, \qquad\text{where}\qquad\;\, \ee_{k}^{t_{\,}}(m) \; \defineas \; \sum_{j=0}^k \frac{(2\pi m)^j}{j!} \sum_{\fp \in \pt_j(k)} \hskip-5pt c_\fp t_\fp~.
\end{aligned}
\end{equation}

It is also useful to express these in terms of the quantities in \eqref{eq:x_y_ansatse} and \eqref{eq:t_j_R^t_ansatz}. Substituting these for $x$ and $y$, we find a set of expressions listed below. In the intermediate expressions, the following quantity appears often:
\vskip-27pt
\begin{align}
r^{x_{\,}}_{(m,n),\mathfrak{p}} &\; \defineas \sum_{\substack{\sum_{j=1}^n \fp_j = \mathfrak{p} \\ \fp_1,...,\fp_m \neq \emptyset}} c\Bigl( \hskip-2pt \left \{\mathfrak{p}_j \right\}_{j=1}^n \Bigr) \prod_{j=1}^n R^{x_{\,}}_{\mathfrak{p}_j}~. \label{eq:defn_r^(x)}
\end{align}
The sum here is over all partitions such that their sum is equal to $\fp$, with the condition that the partitions $\fp_1,...,\fp_m$ are non-empty (equivalently, there are between 0 and $n-m$ empty partitions). The coefficients $c(\{\fp_j\})$ are given by $c_\fq$, where $\fq$ is the partition $\sum_{j=1}^n q_j$ formed out of the multiplicities $q_j$ of distinct partitions in the set $\{\fp_j\}_{j=1}^n$. We define functions $r^{y}_{(m,n),\fp}$ and $r^{t}_{(m,n),\fp}$ similarly, with the only difference being that $R^{x_{\,}}$ in replaced by $R^{y_{\,}}$ or $R^{t_{\,}}$.

With the help of these quantities, we can express various quantities appearing in \eqref{eq:xy_Series_q_Ansatz}, in a form that is convenient for later computations, for example:
\begingroup
\allowdisplaybreaks
\begin{alignat}{2}
\eta_k &\= \sum_{\fp \in \pt(k)} \hskip-3pt N_\fp \sum_{j=1}^{l(\fp)} \frac{(-1)^j}{y_0^{j+1}} r^{y_{\,}}_{(j,j),\fp} \, &&\defineas \, \sum_{\fp \in \pt(k)} \bm{\eta}_\fp N_\fp~,\notag \\[5pt] 
\eta'_k &\= \sum_{\fp \in \pt(k)} \hskip-3pt N_\fp \sum_{j=2}^{l(\fp)} \frac{(-1)^j}{y_0^{j+1}} r^{y_{\,}}_{(j,j),\fp} \, &&\defineas \, \sum_{\fp \in \pt(k)} \bm{\eta'}_\fp N_\fp~, \label{eq:defn_N_p_expansions}\\[5pt] 
\me_k^{x_{\,}}(\alpha) &\= \sum_{\fp \in \pt(k)} \hskip-3pt N_\fp \sum_{j=0}^{l(\fp)} \frac{(2\pi \alpha)^j}{j!} r^{x_{\,}}_{(j,j),\fp} \, &&\defineas \, \sum_{\fp \in \pt(k)} \bme_\fp^{x_{\,}}(\alpha) \, N_\fp~, \notag
\end{alignat}
\endgroup
with similar expressions for $\me^{y}_k(\alpha)$ and $\me^{t}_k(\alpha)$. When $\fp=\emptyset$, it is convenient to define exceptional values
\begin{align}
\bm{\eta}_\emptyset \= \frac{1}{y_0}~, \qquad \bm{\eta}'_\emptyset \= 1~.
\end{align}
We give some combinations of these functions their own names due to the fact that these will appear in the instanton sums.
\begin{align} \label{eq:defn_C_S}
\begin{split}
\cC_j(k) &\= \hskip-3pt \sum_{\fp \in \pt(k)} N_\fp \, \cC_{j,\fp}~, \qquad
\cS_j(k) \= \hskip-3pt \sum_{\fp \in \pt(k)} N_\fp \cS_{j,\fp}~, \qquad \cE_j(k) \= \hskip-3pt \sum_{\fp \in \pt(k)} N_\fp \cE_{j,\fp}~, \\[5pt]
\cC_{j,\fp} &\= \frac{1}{Y (2\pi)^j} \sum_{m \in \fp} \frac{1}{2 m^j} \Big( q_x^m \big(\bme^{x_{\,}}(\ii m) \sr \bme^{y_{\,}}(-m)\big)_{\fp-m}+q_x^{-m} \big(\bme^{x_{\,}}(-\ii m) \sr \bme^{y_{\,}}(-m)\big)_{\fp-m} \Big)~,\\[5pt]
\cS_{j,\fp} &\= \frac{1}{Y (2\pi)^j} \sum_{m \in \fp} \frac{1}{2 \ii m^j} \Big( q_x^m \big( \bme^{x_{\,}}(\ii m) \sr \bme^{y_{\,}}(-m)\big)_{\fp-m}-q_x^{-m} \big(\bme^{x_{\,}}(-\ii m) \sr \bme^{y_{\,}}(-m)\big)_{\fp-m}  \Big)~,\\[5pt]
\cE_{j,\fp} &\= \frac{1}{Y (2\pi \ii)^j} \sum_{m \in \fp} \frac{1}{m^j} \bme^{t_{\,}}(\ii m)_{\fp-m}~,
\end{split}
\end{align}
where the sums over $m \in \fp$ denote the sum where every $i$ with $\mu_i \neq 0$ is counted once. Now we can write the instanton sums and their derivatives compactly as
\begin{align} \notag
\begin{split}
\cI_1 &\= -Y \sum_{k=1}^\infty q_y^k \, \cS_3(k)~, \qquad \cI_2 \= \+ Y \sum_{k=1}^\infty q_y^k \, \cC_3(k)~,\\[5pt]
\cI'_1 &\= -Y \sum_{k=1}^\infty q_y^k \, \cC_2(k)~, \qquad \cI'_2 \= - Y\sum_{k=1}^\infty q_y^k \, \cS_2(k)~,\\[5pt]
\cI &\= \+ Y \sum_{k=1}^\infty q_t^k \cE_3(k)~, \qquad \;\cI' \= \+ Y \sum_{k=1}^\infty q_t^k \cE_2(k)~.
\end{split}
\end{align}
\newpage

\section{The D4-D2-D0 System}\label{sect:D4-system}
\vskip-10pt
In this and the following section, we use the methods outlined previously to find the perturbative solutions to the zeroth-order attractor equations and the recurrence relations for the instanton corrections. We will consider the orthogonality and alignment equations separately. 

We start with the case where the charge vector takes the following form, corresponding to the D4-D2-D0 system:
\begin{align} \notag
Q_{\text{D4}} = \kappa\begin{pmatrix}
\Lambda\\
\Upsilon\\
0\\
1
\end{pmatrix}.
\end{align}
The orthogonality equation for a charge vector of this form turns out to be the simplest among the four different cases we study in this paper. Consequently, this is one of the cases where we are able to find a closed-form solution for the instanton corrections, as will be shown in detail in \sref{sect:Solutions}. 

With regards to microstate counting, the alignment equations in this case are particularly interesting. For example in \cite{Gaiotto:2006wm,Gaiotto:2007cd} the microstates of black holes with these charges are counted by the coefficients of modular forms. Using these results, it is possible to express the supergravity entropy in terms of the perturbative corrections corresponding to $\sigma$ as well as the instanton corrections. This can be used to compare the macroscopic expression to the asymptotics of modular forms, a subject to which we intend to return elsewhere. We discuss this in a little more detail in \sref{sect:Microstate_Counting}, where we give the relation of our results to the Wald entropy of the black hole associated to the attractor point we find here. We show that the leading-order asymptotics of the coefficients of the modular forms agree with the formulae in this section.
\subsection{The orthogonality equation}
\vskip-10pt
The orthogonality equation depends only on the complex period ${\Pi}$ and not on the real and imaginary parts separately. Therefore it is most natural to write the orthogonality equation as a condition on the complex coordinate $t$ instead of separating this into the real and imaginary parts $x$ and $y$ as we are obliged to do for the non-holomorphic alignment equations later. 

The orthogonality equation $\Pi^T \, \Sigma \, Q_{\text{D4}} =0$ can be written as
\begin{align}
t^2 - 2 t \alpha \wt{\beta} + \wt{\beta}^2  + \frac{2}{Y}\, \cI' \= 0~.
\label{eq:D4_Zero_Equations}
\end{align}
\subsubsection*{The perturbative equations}
\vskip-10pt
Ignoring the instanton corrections, the equation is just a quadratic. Moreover, it is independent of~$\sigma$. Of the two solutions, the solution that can, for suitable values of $\Lambda$ and $\Upsilon$, lie near the large complex structure point is
\begin{align}
t_0 \= \wt{\beta}\left(\alpha+\sqrt{\alpha^2-1} \right). \label{eq:D4zero}
\end{align}
For this to correspond to a solution that lies in the large complex structure region, $t_{0}$ needs to have a large imaginary part. Noting that $\alpha \wt{\beta}=-\frac{Y_{110}+\Upsilon}{Y}$ is always real and $\wt\beta$ has a phase $0$ or $\frac{\pi}{2}$, we have that the pair $\alpha$ and $\wt\beta$ are either both real or both purely imaginary. In fact, in the large complex structure region they cannot both be imaginary, for then $t_{0}$ would be real. Thus we require that $\wt\beta>0$ and $-1 <\alpha < 1$, in order for $t_{0}$ to have a positive imaginary part. We will discuss the conditions for existence of perturbative solutions near the large complex structure point in more detail in \sref{sect:Classification_of_Attractor_Points}. This said, the real and imaginary parts $x_0$ and $y_0$ can be written as
\begin{align}
x_{0}\=\wt\beta\alpha~, \qquad y_{0}\=\wt\beta\sqrt{1-\alpha^{2}}~.\label{eq:d4d2d0zqsols}
\end{align}
\subsubsection*{Instanton corrections}
\vskip-10pt
The instanton corrections to the solutions of the orthogonality equation are included as a series in~$q_t$ by writing $t = \sum_{j} t_j q_t^j$. To solve the attractor equations, we require that the coefficient of every power of $q_t$ vanishes. This requirement can be written as
\begin{align} \notag
0 &\= \frac{1}{2} \sum_{\fp \in \pt^0_2(k)} \hskip-6pt c_\fp t_\fp - \alpha \wt{\beta} t_k + \cE_{2}(k)~.
\end{align}
It is simple to solve this equation for $t_k$ in terms of $t_i$ with $i<k$. As $\cE_2(k)$ is independent of $t_k$, the above equation contains $t_k$ multiplied by the factor $(t_0-\alpha \wt{\beta})$. Thus $t_k$ is given by
\begin{align} \notag
t_k &\= \frac{1}{\alpha \wt{\beta}-t_0} \left( \frac{1}{2} \sum_{\fp \in \pt_2(k)} \hskip-6pt c_\fp t_\fp +  \cE_2(k) \right).
\end{align}
We now express $t_k$ in terms of the Gromov-Witten invariants in the form \eqref{eq:t_j_R^t_ansatz}. The rational functions $R^t_\fp$ are chosen precisely to be themselves independent of the scaled Gromov-Witten invariants $N_k$. This leads to a recurrence relation
\begin{align}
R^{t}_\fp(t_0) &\= \frac{1}{\alpha \wt{\beta} - t_0} \left( \frac{1}{2} \left(R^t \hs R^t \right)_\fp + \cE_{2,\fp} \right). \label{eq:D4_R^t_Recurrence}
\end{align}
From this we see that the $R^t_\fp$ are rational functions of $t_0$. The denominators of these rational functions are due to the overall prefactor $1/(\alpha \wt{\beta} - t_0)$, and can be written explicitly. We can concentrate on the numerators by introducing polynomials~$P^t_\fp$
\begin{align} \notag
P^t_\fp(w) \= -\frac{\ii \pi \IDelta(w)^{\length(\fp)}}{w} R^t_\fp \left(\alpha \wt{\beta} + \ii \frac{w}{\pi} \right),
\end{align} 
where in this subsection we set
\begin{align} \notag
\IDelta(w) &\= - 4 Y w^2~.
\end{align}
This formula differs somewhat from the relation between the polynomials and rational functions in the other cases considered later, since here the argument of $R^t_\fp$ is shifted by $\alpha \wt{\beta}$, while in the other cases we just multiply the argument $v$ by an overall factor. The reason for the shift in this case, as will soon become apparent, is that with this definition we can later substitute $v = y_0$ to re-obtain the rational functions $R_\fp^t(t_0)$ satisfying the recurrence relation \eqref{eq:D4_R^t_Recurrence}. This is the consequence of the fact that the instanton corrections depend essentially on $y_0$ only.

It is easy to write down the first few polynomials. These have a simple form, and will later play an important role in finding solutions to the attractor equations in closed form.
\begin{align}
\begin{split}
P_{\{1\}}^{t}(w) &\= \+ \hskip2pt 1~,\\[3pt]
P_{\{2\}}^{t}(w) &\= \+ \hskip2pt \frac{1}{4}~,\\[3pt]
P_{\{1^2\}}^{t}(w) &\= -\hskip2pt \frac{1}{2}   \hskip3pt \left(1+4 w\right),\\[3pt]
P_{\{1,2\}}^{t}(w) &\= - \hskip2pt \frac{1}{4}  \hskip3pt \left(1+6 w\right),\\[3pt]
P_{\{1^3\}}^{t}(w) &\= \+ \hskip2pt \frac{1}{2} \hskip3pt \left(1+6 w + 12 w^2\right),\\[3pt]
P_{\{1^2,2\}}^{t}(w) &\= \+ \hskip2pt \frac{1}{8} \hskip3pt \left(3+24 w + 64 w^2\right),\\[3pt]
P_{\{1^4\}}^{t}(w) &\= -\frac{1}{24} \left(15+120 w + 384 w^2 +512 w^3\right).
\end{split}\nonumber
\end{align}
Combining these observations, and recalling from \eqref{eq:d4d2d0zqsols} that $\alpha \wt{\beta} - t_0 = - \ii y_0$, we are able to write the coordinate~$t$ in the form
\begin{align} \notag
\begin{split}
t&\=t_0 + \ii y_0 \sum_{k=1}^\infty \me^{2\pi \ii k t_0} \! \sum_{\fp \in \pt(k)} \frac{ N_\fp}{\IDelta \left(\pi y_0 \right)^{\length(\fp)}} P^t_\fp\left(\pi y_0 \right).
\end{split}
\end{align}
Let us also consider the real and imaginary parts of $t$. The polynomials $P^t_\fp(v)$ are real for real $v$, as can be seen from the recurrence \eqref{eq:D4_R^t_Recurrence}, so the real and imaginary parts of $t$ are given by
\begin{align}\label{eq:D4_orthogonality_equations_solution_x_y}
\begin{split}
x \= x_0 - y_0 \sum_{k=0}^\infty \me^{-2\pi k y_0} \sin (2\pi k x_0) \sum_{\fp \in \pt(k)} \frac{ N_\fp}{\IDelta \left(\pi y_0 \right)^{\length(\fp)}} P^t_\fp\left(\pi y_0 \right), \\[5pt]
y \= y_0 + y_0 \sum_{k=0}^\infty \me^{-2\pi k y_0} \cos (2\pi k x_0) \sum_{\fp \in \pt(k)}  \frac{ N_\fp}{\IDelta \left(\pi y_0 \right)^{\length(\fp)}} P^t_\fp\left(\pi y_0 \right).
\end{split}
\end{align}

\subsection{The alignment equations} \label{sect:D4_Strominger_Equations}
\vskip-10pt
The alignment equations are most conveniently expressed in terms of the real coordinates $x$ and $y$. After replacing $a$ and $b$ in the equations \eqref{eq:Strominger_Equation_Full} via \eqref{eq:aandb}, we are left with two nontrivial conditions:
\begin{align}
\begin{split}
(3x^2 - y^2)y - 2\sigma +\frac{6}{Y}\left(x \cI_2'- 2\cI_2 + y \cI_1'  \right)-3\wt\beta^{2}y&\=0~,\\[10pt]
x  + \frac{1}{Yy} \cI_2' - \alpha \wt\beta&\=0~. 
\end{split} \label{eq:D4_Strominger_Equations}
\end{align}
\subsubsection*{The perturbative equations}
\vskip-10pt 
Neglecting the instanton contributions, the attractor equations become
\begin{align}
\begin{split}\label{eq:D4quantum}
\left(3x_0^{2}-y_0^{2}\right)y_0-2\sigma-3\wt\beta^{2} y_0&\=0~, \qquad x_0-\alpha\wt\beta\=0~.
\end{split}
\end{align}
The second of these equations is solved trivially, and substituting this solution into the first equation yields a cubic for~$y_0$:
\begin{align} \notag
y_0^3 - 3\wt{\beta}^2 (\alpha^2-1) y_0 + 2 \sigma \= 0~.
\end{align} This makes it possible to give a simple solution for $y_0$ in terms of trigonometric functions. When $\wt\beta\neq0$ and $|\alpha|\neq1$, we make the substitution 
\begin{align}
y_0 \= 2 \wt{\beta} \sqrt{\alpha^2 - 1} \sin \theta~, \label{eq:D4_y_0_Solution}
\end{align}
which brings the equation to the form
\begin{align} \notag
-4 \sin^3 \th + 3 \sin \th \= \frac{\sigma}{\wt{\beta}^3 (\alpha^2-1)^{3/2}}~,
\end{align}
and we recognise the left-hand side as $\sin 3\th$, with $\theta$ defined by
\begin{align}
\sin 3 \theta \= \frac{\sigma}{\wt{\beta}^3 (\alpha^2-1)^{3/2}}~. \label{eq:D4_theta_defn}
\end{align}
A cubic has three solutions, corresponding here to the freedom to replace $\theta$ by $\theta + \frac{2\pi}{3}$ or $\theta + \frac{4\pi}{3}$. We want to choose a branch so that $y_0$ is real and $x_0 + \ii y_0$ lies in the large complex structure region. This choice depends on the values of the charges and Yukawa couplings, and so on $\alpha,\wt{\beta},$ and $\sigma$, and on the size of the large complex structure region. To better describe these various possibilities we tabulate different relevant cases in \tref{tab:D4_Alignment_x00} below.
\begin{table}[H]
	\renewcommand{\arraystretch}{1.5}
	\begin{center}
		\begin{tabular}{ |l|l|c|}
			\hline
			\hfil \textbf{Charge data}  &  \hfil \textbf{Positive solutions for $y_{0}$} & \textbf{Definition of $\th$} \\
			\hline
			\hline			
			$0 {\leqslant} \sigma {\leqslant} \wt{\beta}^3 (\alpha^2{-}1)^{3/2} $ & $2 \wt{\beta} \sqrt{\alpha^2{-}1} \sin \th$, $2 \wt{\beta} \sqrt{\alpha^2{-}1} \sin \left(\th {+} \frac{2\pi}{3}\right)\!$ & $\th = \frac{1}{3} \sin^{-1} \frac{\sigma}{(\wt{\beta}\sqrt{\alpha^2-1})^3}$\\[7pt]
			\hline
			$\wt{\beta}^3 (\alpha^2{-}1)^{3/2} {\leqslant} \sigma {\leqslant} 0$ & $2 \wt{\beta} \sqrt{\alpha^2{-}1} \sin \left(\th {+} \frac{4\pi}{3}\right)$ & $\th =  \frac{1}{3} \sin^{-1} \frac{\sigma}{(\wt{\beta}\sqrt{\alpha^2-1})^3}$\\[7pt]
			\hline
			$-\wt{\beta}^3 (\alpha^2{-}1)^{3/2} {\leqslant} \sigma {\leqslant} 0$& $2 \wt{\beta} \sqrt{\alpha^2{-}1} \sin \left(\th {+} \frac{2\pi}{3}\right)$ & $\th =  \frac{1}{3} \sin^{-1} \frac{\sigma}{(\wt{\beta}\sqrt{\alpha^2-1})^3}$ \\[7pt]
			\hline
			$0 {\leqslant} \sigma {\leqslant} -\wt{\beta}^3 (\alpha^2{-}1)^{3/2}$  & $2 \wt{\beta} \sqrt{\alpha^2{-}1} \sin \th$, $2 \wt{\beta} \sqrt{\alpha^2{-}1} \sin \left(\th {+} \frac{4\pi}{3}\right)\!$ & $\th =  \frac{1}{3} \sin^{-1} \frac{\sigma}{(\wt{\beta}\sqrt{\alpha^2-1})^3}$\\[7pt]
			\hline
			$\sigma {<} \wt{\beta}^3 (\alpha^2{-}1)^{3/2} {<} 0$ & $-2 \wt{\beta} \sqrt{\alpha^2{-}1} \cosh \nu$ &$\th {=} \frac{3 \pi}{2} {+} \ii \nu$, $\nu {=} \frac{1}{3} \cosh^{-1} \frac{\sigma}{(\wt{\beta}\sqrt{\alpha^2-1})^3}$\\[7pt]
			\hline
			$\sigma {<\!} -\wt{\beta}^3 (\alpha^2{-}1)^{3/2} {<} 0$ & $2 \wt{\beta} \sqrt{\alpha^2-1} \sin \th$ & $\theta = \frac{5\pi}{6} {+} \ii \nu$ $\nu {=} \frac{1}{3} \cosh^{-1} \frac{\sigma}{(\wt{\beta}\sqrt{\alpha^2-1})^3}$\\[7pt]\hline
			$\sigma {<} 0$, $\ii \wt{\beta} \sqrt{\alpha^2{-}1} \in \IR$ & $2 |\wt{\beta} \sqrt{\alpha^2{-}1} \sinh \nu |$ & $\th {=} \ii \nu$, $ \nu {=} \frac{1}{3} \sinh^{-1} \frac{\sigma}{|\wt{\beta}\sqrt{\alpha^2-1}|^3}$\\[7pt]\hline		
		\end{tabular}
		\vskip10pt
		\capt{6in}{tab:D4_Alignment_x00}{Solutions to the perturbative attractor equations \eqref{eq:D4quantum}. Whenever an inverse trigonometric function appears in this list, we take the principal branch value. We have only included here the solutions for which $y_0>0$, as these are the only solutions that can possibly lie within the large complex structure region. However, even if a solution is included here, it may not lie near the large complex structure point for all Calabi-Yau manifolds.}	
	\end{center}
\end{table}
As we do not have a general formula for the radius of the large complex structure region, we will include all solutions for which a positive solution for $y_0$ exists.

In addition to the cases listed in \tref{tab:D4_Alignment_x00}, there are degenerate cases with  $\wt\beta=0$ or $\alpha^2=1$. In the case $\wt{\beta} = 0$, we have that $\alpha \to \infty$ in such a way that
\begin{align} \notag
\alpha \wt{\beta} \= -\frac{Y_{110}+\Upsilon}{Y_{111}}~, \qquad \text{and} \qquad \wt{\beta}\sqrt{\alpha^2 - 1} \, \to \, \left|\frac{Y_{110}+\Upsilon}{Y_{111}} \right|.
\end{align}
These expressions resolve the ambiguities in \tref{tab:D4_Alignment_x00}.

In the two cases where either $\alpha^2{\=}1$ or $\wt{\beta} {\=} \wt{\beta}\alpha {\=}0$, both with $\sigma<0$, the only possibly admissible solution is
\begin{align} \notag
x_0 \= 0~, \qquad y_0 \= |\sigma|^{1/3}.
\end{align}
The values of charges that are not included in the \tref{tab:D4_Alignment_x00}, or among the two degenerate cases, do not have a solution which lies near the large complex structure point. We provide some intuition for the situations where multiple $y_{0}$ are possible for a set of charges in \fref{fig:Cubic_Plot}.
\vskip5pt
\begin{figure}[H]
	\begin{center}
		\framebox[\textwidth]{
		\begin{minipage}[c]{0.98\textwidth}
				\centering 
				\vspace{30pt}
				\includegraphics[width=0.5\textwidth]{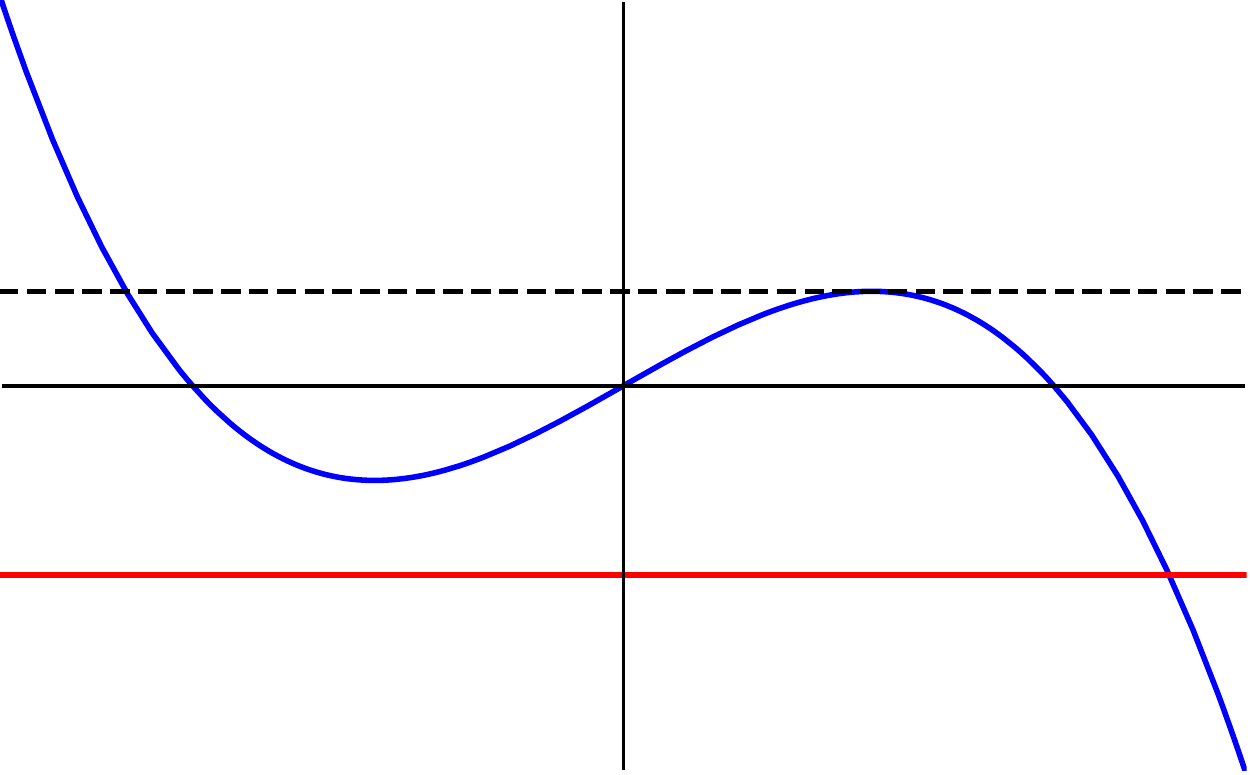}\vskip0pt
				\place{5}{1.5}{$1$}
				\place{5}{1.15}{$y$}
				\place{3}{2.3}{$z=f(y)$}
				\place{1.23}{2.3}{$-y^3+4y$}	
				\place{1.8}{0.55}{$\sigma\wt\beta^{-3}(\alpha^{2}-1)^{-3/2}$}
		\end{minipage}}
	\vskip10pt
	\capt{6in}{fig:Cubic_Plot}{In the case where $\beta\in\IR$ and $|\alpha|>1$ the number of positive solutions for $y_{0}$ varies, subject to the value of $\sigma$. The red line moves up and down as $\sigma$ is varied, with our solution being the value of $y$ such that the red line intersects the blue curve. As can be seen from the figure, when the red line lies in the region $0<z<1$ there are two positive solutions, but when $z>1$ there are no positive solutions. A similar plot can be borne in mind for other charge ranges.}
	\end{center}	
\end{figure}
\vskip-20pt
It is also useful to expand $y_0$ as a series in $\sigma$. This gives a perturbative expansion for $y_0$ which is convergent when $\sigma/(\wt{\beta}\sqrt{\alpha^2-1})$ is small. This ratio is suitably small in the large charge ratio limits, for example.

If $\sigma{\=}0$, then there are solutions to \eqref{eq:D4quantum} with positive $y_{0}$ only if $\wt\beta^{2}\left(\alpha^{2}-1\right)>0$. These solutions with $\sigma{\=}0$ are
\begin{equation*} \notag
x_{00}\=\wt\beta\alpha~, \qquad y_{00}\=\sqrt{3}\left\vert\wt\beta\sqrt{\alpha^{2}-1}\right\vert.
\end{equation*}
This $y_{00}$ is the zeroth-order term in the $\sigma$ expansion, with the full series given by
\begin{equation}\label{eq:d6zeroquantumcoefficients}
y_{0}\=y_{00}\left[1-\frac{1}{12\pi} \sum_{n=1}^{\infty} \frac{\Gamma(\frac{n}{2}-\frac{1}{6})\, \Gamma(\frac{n}{2}+\frac{1}{6})}{n!}\left(\frac{2 \, \sqrt{27} \, \sigma}{y_{\text{00}}^{3}}\right)^{n} \; \right].
\end{equation}
By separating the sum into odd and even terms, we find an expression in terms of hypergeometric functions
\begin{equation*}\notag
\frac{y_{0}}{y_{00}}\=\twoFone{\frac16}{-\frac{1}{6}}{\frac12}{\frac{27\sigma^{2}}{y_{00}^{6}}}-\frac{\sigma}{y_{00}^{3}}\,\twoFone{\frac13}{\frac23}{\frac32}{\frac{27\sigma^{2}}{y_{00}^{6}}}.
\end{equation*}

\subsubsection*{Instanton corrections} 
\vskip-10pt
Having written the alignment equations in terms of $x$ and $y$, it turns out that the most convenient way of writing the instanton corrections is as a series in $q_y$. In other aspects the procedure for solving the alignment equations is nearly identical to that used for the orthogonality equation. 

Using the $q_y$ expansions \eqref{eq:xy_q_j_expansion} for $x$ and $y$, the second instanton-corrected equation \eqref{eq:D4_Strominger_Equations} becomes
\begin{align} \notag
\sum_{k=1}^\infty q_y^k \left( x_k - \sum_{j=1}^k \eta_{k-j} \, \cS_2(j) \right) \= 0~.
\end{align} 
Demanding that the coefficient of $q_y^k$ vanishes gives that
\begin{align} \notag
x_k \= \sum_{j=1}^k \eta_{k-j} \, \cS_2(j)~.
\end{align}
This lends itself to an iterative solution as the left-hand side only involves $x_j$ and $y_j$ with~$j<k$.  

Next we use the $N$-expansions \eqref{eq:x_y_ansatse}, \eqref{eq:defn_N_p_expansions}, and \eqref{eq:defn_C_S} to give $x_j$, $\eta_j$, and $\cS_2(j)$ in terms of the rational functions $R^{x_{\,}}_\fp$ and $R^{y_{\,}}_\fp$:
\begin{align} \notag
\sum_{\fp \in \pt(k)} N_\fp R^{x_{\,}}_\fp \= \sum_{\fp \in \pt(k)} N_\fp (\bm{\eta} \sr \cS_2)_\fp~.
\end{align}
The convolution $(\bm{\eta}\sr \cS_2)_\fp$ involves $R^{x_{\,}}_\fq$ and $R^{y_{\,}}_\fq$ with $|\fq|<k$, so $R^x_\fp$ with $|\fp|=k$ appears only on the right-hand side. By construction, the $R^{x_{\,}}_\fp$ are themselves independent of the Gromov-Witten invariants, so the coefficients of different $N_\fp$ satisfy the equation independently. Thus we have
\begin{align}
R^{x_{\,}}_\fp &\= (\bm{\eta} \sr \cS_2)_\fp~, \label{eq:D4_Strominger_Equations_Rx_Recurrence}
\end{align}
which determines the $R^{x_{\,}}_\fp$ recursively.

The requirement that the coefficient of $q_y^k$ vanishes in the first instanton-corrected equation \eqref{eq:D4_Strominger_Equations} can be written as
\begin{align}
\begin{split}
0 &\=\frac{1}{6} \sum_{\fp \in \pt^0_2(k)} \hskip-8pt c_\fp (3x_\fp - y_\fp)-\frac{\sigma}{3} \eta_k- 2 \sum_{i=1}^k \eta_{k-i} \, \cC_3(i) - \cC_2(k) - \hskip-12pt \sum_{0 \leq j \leq i \leq k-1} \hskip-12pt x_{i-j}  \, \eta_{j} \, \cS_2(k-i)~.
\end{split} \nonumber
\end{align}
Here $y_k$ appears only in $\eta_k$ and in the first sum when $\fp$ takes the value $\fp = k+0$. Thus $y_k$ appears multiplied by the factor 
\begin{equation*} \nonumber
\frac{\sigma}{3 y_0^2} - \frac{1}{3} y_0 \= -\frac{1}{3} y_0 (1-\zeta)~.
\end{equation*}
We rearrange the equation to express $y_k$ in terms of $x_i$ with $i \leqslant k$ and $y_i$ with $k<i$.
\begin{equation*} \nonumber
y_k {=} \frac{3}{y_0 (1{-}\zeta)} \hskip-2pt \left( \hskip-2pt \frac{1}{6} \sum_{\fp \in \pt_{2}(k)} \hskip-8pt c_\fp (3 x_\fp {-} y_\fp ) {+} x_0 x_k {-} \frac{\sigma}{3} \hskip1pt \eta'_k - 2 \sum_{i=1}^k \eta_{k-i} \, \cC_3(i) - \cC_2(k) - \hskip-12pt \sum_{0 \leqslant j \leqslant i \leqslant k-1} \hskip-12pt x_{i-j}  \, \eta_{j} \, \cS_2(k{-}i) \hskip-2pt \right).
\end{equation*}
We have solved for $x_k$ in terms of $x_i,y_i$ with $i<k$ above. This is therefore a well-defined recurrence relation for $y_k$. To find the recurrence for $R^{y_{\,}}_\fp$, we use the expressions \eqref{eq:x_y_ansatse} to obtain 
\begin{align}
\begin{split}
R^{y_{\,}}_\fp(x_0,y_0) &{=} \frac{3}{y_0 (1{-}\zeta)} \left(  \frac{1}{2}r^{x_{\,}}_{(2,2),\fp} {-} \frac{1}{6}r^{y_{\,}}_{(2,2),\fp} {+} x_0 R^{x_{\,}}_\fp {-} \frac{y_0^3 \zeta}{3} \, \bm{\eta}'_\fp {-}2 \left(\cC_3 \sr \bm{\eta}\right)_\fp \right. \hskip-2pt - \cC_{2\fp} {-} \left(\cS_2 \sr R^{x_{\,}} \! \! \sr \bm{\eta}\right)_\fp \hskip-2pt \biggr). \label{eq:D4_Strominger_Equations_Ry_Recurrence}
\end{split}
\end{align}
The denominators are again simple, given essentially by the prefactor appearing in the expression above. We therefore extract the numerators to give us polynomials.
\begin{equation}
\begin{aligned}
P^{x_{\,}}_\fp(u,v) \; &\= \+ Y \pi v R^{x_{\,}}_\fp\left( \frac{u}{\pi}, \frac{v}{\pi} \right), \ \ &l(\fp) = 1~, \\[5pt]
P^{x_{\,}}_\fp(u,v) \; &\= -\frac{\pi \IDelta(v)^{\length(\fp)}}{v^{2}(1-\zeta)^{3}} \, R^{x_{\,}}_\fp\left( \frac{u}{\pi}, \frac{v}{\pi} \right), \ \ &l(\fp) > 1~,  \\[5pt]
P^{y_{\,}}_\fp(u,v) \; &\= -\frac{\pi \IDelta(v)^{\length(\fp)}}{v (1-\zeta)} \, R^{y_{\,}}_\fp \left( \frac{u}{\pi}, \frac{v}{\pi} \right),
\end{aligned} \label{eq:D4_Strominger_Equations_RP_relations}
\end{equation}
where, in this subsection,
\begin{align*} \notag
\IDelta(v) &\= Y v^3 (1-\zeta)^{2}.
\end{align*}
We record the first few polynomials
\begin{align*}
&\begin{split}
P_{\{\hskip2pt 1 \hskip2pt \}}^{x_{\,}}(u,v) &\;= \+ \; \frac{1}{4} \sin(2 u)~,\\[5pt]
P_{\{\hskip2pt 2 \hskip2pt \}}^{x_{\,}}(u,v) &\;= \+ \frac{1}{16} \sin(4 u)~,\\[5pt]
P_{\{1^2\}}^{x_{\,}}(u,v) &\;= -\frac{1}{32} \Big(3 + 9 v + 2 (4-\zeta) v^2 \Big) \sin(4 u)~,\\[5pt]
P_{\{\hskip2pt 1 \hskip2pt \}}^{y_{\,}}(u,v) &\;= \+ \; \frac{3}{4}(1+v)\cos(2u)~,\\[5pt]
P_{\{ \hskip2pt 2 \hskip2pt \}}^{y_{\,}}(u,v) &\;= \+ \frac{3}{32}(1+2v)\cos(4u)~,\\[5pt]
P_{\{1^2\}}^{y_{\,}}(u,v) &\;= \+ \frac{3}{64} \Big(9+6(4-\zeta)v + 3(8-4\zeta-\zeta^2)v^2 + 4(2-\zeta-\zeta^2)v^3 \\[5pt]
&\+ \+ \+ \hskip3pt +\left( 9 +6(4-\zeta)v + 3(10-8\zeta+\zeta^2) v^2 + 4(4-5\zeta+\zeta^2) v^3 \right) \cos(4u) \hskip-1pt \Big)~.
\end{split} \nonumber
\end{align*}
In summary, the full solution is
\begin{align*}
\begin{split}
x &\= x_0 - \sum_{k=0}^\infty \me^{-2\pi k y_0} \left( \sum_{\fp \in \pt_{>1}(k)}  \hskip-5pt N_\fp\frac{\pi  y_0^2 (1-\zeta)^3}{\IDelta(\pi y_0)^{\length(\fp)}}P^{x_{\,}}_\fp(\pi x_0, \pi y_0) - \frac{N_k P^{x_{\,}}_k(\pi x_0, \pi y_0)}{Y \pi^2 y_0}\right),\\[5pt]
y &\= y_0 - \sum_{k=0}^\infty \me^{-2\pi k y_0} \hskip-5pt \sum_{\fp \in \pt(k)}  \hskip-5pt N_\fp \frac{y_0 (1-\zeta)}{\IDelta(\pi y_0)^{\length(\fp)}}P^{y_{\,}}_\fp(\pi x_0, \pi y_0)~,
\end{split} \nonumber
\end{align*} 
where $x_0$ and $y_0$ are the perturbative solutions to \eqref{eq:D4quantum}. The polynomials $P^{x_{\,}}_\fp$ and $P^{y_{\,}}_\fp$ are given in terms of the rational functions $R^{x_{\,}}_\fp$ and $R^{y_{\,}}_\fp$ in \eqref{eq:D4_Strominger_Equations_RP_relations}. These, in turn, are given by the recurrence relations \eqref{eq:D4_Strominger_Equations_Rx_Recurrence} and  \eqref{eq:D4_Strominger_Equations_Ry_Recurrence}.

\newpage
\section{The D6-D2-D0 System}\label{sect:D6-system}
\vskip-10pt 
In this section, we consider a charge vector that corresponds to a black hole consisting of some D6 branes bound to a set of D2 and D0 branes. In other words, we will take the charge vector to be given by 
\begin{align} \notag
Q_{D6} \= \kappa\begin{pmatrix}
\Lambda\\
\Upsilon\\
1\\
0
\end{pmatrix}.
\end{align}
Although for generic charges of this form the alignment equations are the most complicated of the four cases studied here, these will simplify considerably when we specialise to the important special case $\Lambda=0$, corresponding to attractor points lying on the imaginary axis of the moduli space. This special case allows us to present a closed-form solution to the alignment equations, as we will discuss in detail in \sref{sect:Solutions}. The equations in this section are of interest for studying rank two attractor points as the explicit example found recently in \cite{Candelas:2019llw} has a charge vector of this form with $\Lambda=0$.
\subsection{The orthogonality equation}
\vskip-10pt 
The condition that the central charge vanishes is again best written as a single complex equation
\begin{align}
t^3 - 3 \beta^2 t - 2 \ii \sigma +2 \beta^3 \gamma - \frac{12}{Y} \, \cI + \frac{6}{Y} \, t \cI' \= 0~.
\label{eq:D6zerofull}
\end{align}
\subsubsection*{The perturbative equations}
\vskip-10pt
Neglecting the instanton contributions from this equation leaves a cubic equation, which is already of a form which, when $\beta\neq0$, we can solve using a trigonometric substitution as in \sref{sect:D4-system}. We write the equation as
\begin{align}\label{eq:D6zerotrig}
-4\left(\frac{t_0}{2\beta}\right)^3 +3  \frac{t_0}{2\beta} \=   \gamma -  \ii \frac{\sigma}{\beta^{3}}~,
\end{align}
and the solution is given by
\begin{equation*}\notag
t_0\=2\beta\sin\theta~, \qquad \text{where} \qquad \sin3\theta\=\gamma-\ii\frac{\sigma}{\beta^{3}}~.
\end{equation*}
Again, the three solutions correspond to $\theta$, $\theta + \frac{2\pi}{3}$ and $\theta + \frac{4\pi}{3}$. In addition, we shall only be concerned with the values of $\theta$ that give a $t_0$ potentially lying in the large complex structure region, that is with $\Im \, t_0>0$. Whether such a solution actually lies within the large complex structure region depends on the exact value of $t_0$ and the radius of the large complex structure region. It is useful to also write down the real and imaginary parts of the coordinate $t_0$ explicitly. In order to do this we will write 
\begin{equation*}\notag
\theta\=\theta_{1}+\ii\theta_{2}~,
\end{equation*}
with $\theta_{1}$ and $\theta_{2}$ real. The range of values $\theta_1$ and $\theta_2$ take depends on the solution that is chosen. The real coordinates $x_{0}$ and $y_{0}$ are then given by \tref{tab:D6_Orthogonality_x0}.

\begin{table}[H]
	\renewcommand{\arraystretch}{1.5}
	\begin{center}
		\begin{tabular}{ |c|c|c|}
			\hline
			\hfil \textbf{Phase of $\beta$} & \textbf{$x_{0}$} &\textbf{$y_{0}$}  \\[2pt]
			\hline\hline
			$~\beta\in\IR^{+}$   
			&$\phantom{-}2\beta\sin\theta_{1}\cosh\theta_{2}$
			&$2\beta\cos\theta_{1}\sinh\theta_{2}$\\[2pt]
			\hline
			$\beta\in\ii\IR^{+}$    
			&$-2|\beta|\cos\theta_{1}\sinh\theta_{1}$
			&$2|\beta|\sin\theta_{1}\cosh\theta_{2}$\\[2pt]
			\hline
		\end{tabular}
		\vskip10pt
		\capt{5in}{tab:D6_Orthogonality_x0}{The real and imaginary parts of $t_0$. We refrain in this case from giving as detailed a table as in the section on the D4 alignment equations. The expressions are conceptually straightforward but typographically lengthy.}
	\end{center}
\end{table}
\vskip-20pt
One should take a suitable branch of  $\arcsin\left(\gamma-\ii\frac{\sigma}{\beta^{3}}\right)$ such that $y_0$ is positive. For all possible $\gamma$, $\sigma$, and nonzero $\beta$ there is at least one suitable choice, and at most two. 

The analysis runs slightly differently when $\beta$ vanishes. When this occurs, $\gamma$ necessarily tends to infinity in such a way that $\beta^{3}\gamma \to -\frac{3\Lambda}{Y}$, which is a finite constant. In this case the perturbative equations reduce to
\begin{equation*}\notag
t_{0}^{3} \; \defineas \; r_0^3 \, \ee^{3 \ii \phi_0}=-2\beta^{3}\gamma+2\ii\sigma~, \qquad \text{where } \, r_0>0 \, \text{ and } \, 0\leqslant \phi_0<\frac{2}{3}\pi~.
\end{equation*}
The solutions with positive imaginary part are then obtained by choosing the correct branch of the cube root, and are given by the table.
\begin{table}[H]
	\renewcommand{\arraystretch}{1.5}
	\begin{center}
		\begin{tabular}{ |c|c|}
			\hline
			\textbf{Charge Data} & \textbf{$t_{0}$}   \\[1pt]
			\hline\hline
			$\beta^{3}\gamma<0$ and $\sigma=0$   
			&$r_0 \, \me^{\ii \left(\phi_0 + \frac{2}{3}\pi \right)}$\\[1pt]
			\hline
			$\sigma>0$   
			&$r_0 \, \me^{\ii \phi_0},\;\; r_0 \, \me^{\ii \left(\phi_0 + \frac{2}{3}\pi \right)}$\\[1pt]
			\hline
			otherwise    
			&$r_0 \, \me^{\ii \phi_0}$\\[1pt]
			\hline
		\end{tabular}
		\vskip10pt
		\capt{5in}{tab:D6_Orthogonality_zerobeta}{Solutions with positive imaginary part when $\beta$ vanishes.}
	\end{center}
\end{table}
\vskip-30pt
We turn now to the discussion of solving the equations \eqref{eq:D6zerofull} for $y_0$ as a power series in $\sigma$. There are a number of cases that require separate consideration. These are listed in \tref{tab:D6_Orthogonality_x00}.
\vskip5pt
\begin{table}[H]
	\renewcommand{\arraystretch}{1.4}
	\begin{center}
		\begin{tabular}{ |l|c|c|c|}
			\hline
			\hfil \textbf{Charge data} & \textbf{$x_{00}$} &\textbf{$y_{00}$} & \textbf{Definition of $\nu$} \\[2pt]
			\hline\hline
			$~\beta\in\IR^{+}$, $\gamma>1$   &$\phantom{-}\beta\cosh \nu $&$\sqrt{3}\beta\sinh \nu$ & $\hskip-5pt \nu = \frac{1}{3}\cosh^{-1} \gamma$\\[2pt]
			\hline
			$\beta\in\IR^{+}$, $\gamma<-1$    &$-\beta\cosh \nu $&$\sqrt{3}\beta\sinh \nu$  & $\+ \hskip-5pt \nu = \frac{1}{3} \cosh^{-1} -\gamma$\\[2pt]
			\hline
			$\beta=\ii \IR^+$ &$\phantom{-}|\beta|\sinh \nu$ &$\sqrt{3}|\beta|\cosh \nu$ & $\+ \nu = \frac{1}{3} \sinh^{-1} \Im \, \gamma$\\[2pt]
			\hline
			$\beta=0$    &$\frac{\nu}{2}\,|2\beta^{3}\gamma|^{1/3}$ &$\frac{\sqrt{3}}{2}\,|2\beta^{3}\gamma|^{1/3}$ & $ \hskip-17pt \nu = \text{sign} \, \beta^{3}\gamma$\\[2pt]
			\hline
		\end{tabular}
		\vskip10pt
		\capt{5.5in}{tab:D6_Orthogonality_x00}{Table of solutions for $x_{00}$ and $y_{00}$ with $y_{00}>0$. All the inverse trigonometric functions appearing in the table are understood to have their values in the principal branch. The case $|\gamma|<1$ does not appear in the table as there are then no solutions with $\Im \, t_0 >0$.}
	\end{center}
\end{table}
\vskip-30pt
We now give the $\sigma$-expansions around these zeroth-order solutions. In this context, let us write $P_{n}(z)$ for the Jacobi polynomial $P_{n}^{(\frac{1}{3},-\frac{1}{3})}(z)$. Additionally, we introduce the quantity
\begin{equation*}\notag
\text{T}(k) \= 2\frac{d^{k}}{dz^{k}}\sin(z)\bigg\vert_{z=\frac{2\pi}{3}} \= \begin{cases}
\phantom{-}\sqrt{3}~, & k\=0\text{ mod } 4\\
\hspace{9pt}-1~, & k\=1\text{ mod } 4\\
-\sqrt{3}~, & k\=2\text{ mod } 4\\
\hspace{10pt}\phantom{-}1~, & k\=3\text{ mod } 4.
\end{cases}
\end{equation*}
It is shown in appendix \ref{app:D6_Orth_Polynomials} that when $\beta$ is real we can write
\begin{equation}\label{eq:D6sqpert1}
\frac{y_{0}}{\beta}\=\sqrt{3}\sinh\nu+\sum_{n=1}^{\infty}\frac{\text{T}(n)}{6n}\bigg[\ee^{\nu}P_{n-1}\big({-}\coth 3\nu\big){-}\ee^{-\nu}P_{n-1}\big(\coth 3\nu\big)\bigg]\left(\frac{\sigma}{\beta^{3}\sinh 3\nu}\right)^{n}.
\end{equation}
In the case where $\beta \in \ii \IR^+$, we have instead that
\begin{equation}\label{eq:D6sqpert2}
\frac{y_{0}}{|\beta|}\=\sqrt{3}\cosh\nu+\sum_{n=1}^{\infty}\frac{\text{T}(n)}{6n}\bigg[\ee^{\nu}P_{n-1}\big({-}\tanh 3\nu\big){+}\ee^{-\nu}P_{n-1}\big(\tanh 3\nu\big)\bigg]\left(\frac{\sigma}{|\beta|^{3}\cosh 3\nu}\right)^{n}.
\end{equation}
In the $\beta$=0 case, we have
\begin{equation*}\notag
t_{00} \= \exp\left[\left(1-\frac{\nu}{3}\right)\frac{\pi\ii}{2}\right]\left(\frac{6|\Lambda|}{Y}\right)^{1/3},\; \; \; \; t_{0} \= t_{00}\left(1+\frac{\ii Y \sigma}{3\Lambda}\right)^{1/3}.
\end{equation*}
This can be expanded in $\sigma$. Picking out the imaginary part, we can write
\begin{equation*}\notag
y_{0} \= \left(\frac{3|\Lambda|}{4Y}\right)^{1/3}\sum_{n=0}^{\infty}\text{T}(n)\frac{\left(-\frac{1}{3}\right)_{n}}{n!}\left(-\frac{\sigma Y}{3|\Lambda|}\right)^{n}.
\end{equation*}
The polynomials $P_n$ satisfy useful identities that follow from the general properties of Jacobi polynomials, and which we briefly recall. 

The $P_n$ satisfy the differential equation:
\begin{align*} \notag
(1-z^2) \frac{d^2}{d z^2} P_n(z) - \left(2z + \frac{2}{3} \right) \frac{d}{d z}P_n(z) + n(n+1)P_n(z) \= 0~.
\end{align*}
In addition, these are easily generated by means of a recurrence relation
\begin{align*} \notag
9n^2(n-1)P_n(z)+9n(n-1)(1-2n)z P_{n-1}(z) + n (3n-2)(3n-4) P_{n-2}(z) \= 0~,
\end{align*}
with initial values
\begin{align*} \notag
P_{0}(z)\=1~, \qquad \text{and} \qquad P_{1}(z)\=\frac{1}{3}+z.
\end{align*}
After solving for $y_0$, the $\sigma$-expansions for $x_0$ can be easily obtained. After separating the equation \eqref{eq:D6zerotrig} into its real and imaginary parts, one of these will be a quadratic equation for $x_{0}$. Precisely which one is quadratic depends on the phase of $\beta$. Solving this gives 
\begin{align*} \notag
x_0 \= \pm \sqrt{\beta^2 + \frac{y_0^3 + 2\sigma}{3y_0}}~, 
\end{align*}
where the $\pm$ sign is chosen so that consistency with \tref{tab:D6_Orthogonality_x00} is maintained in the $\sigma\rightarrow 0$ limit.
\subsubsection*{Instanton Corrections}
\vskip-10pt
As before, we require that the coefficient of every power vanishes in the $q_t$ expansion and this leads to recurrence relations for $R^t_\fp$. Explicitly, the condition that the coefficient of $q_t^k$ vanishes can be written compactly as
\begin{align*}
\frac{1}{6} \sum_{\fp \in \pt_3^0(k)} \hskip-6pt c_\fp t_\fp - \frac{1}{2} \beta^2 t_k + \sum_{i=0}^k t_{k-i}\cE_2(i) - 2 \cE_3(k) \= 0~. \nonumber
\end{align*}
Here $t_k$ appears in the first term when $\fp {\=} 0\,{+}\,0\,{+}\,k$, with the coefficient $c_{0{+}0{+}k}=3$, and also in the second term, so overall it is multiplied by $\frac{1}{2} (t_0^2 - \beta^2)$. Using this, it is easy to write down the recurrence relation for $t_k$:
\begin{align*} \notag
t_k &\= \frac{2}{\beta^2 - t_0^2} \left( \frac{1}{6} \sum_{\fp \in \pt_3(k)} \hskip-6pt c_\fp t_\fp + \frac{1}{2} t_0 \hskip-5pt \sum_{\fp \in \pt_2(k)} \hskip-6pt t_\fp + \sum_{i=0}^k t_{k-i}\cE_2(i) - 2 \cE_3(k) \right). 
\end{align*} 
We can use this to derive the recurrence relation for the functions $R^t_\fp$,
\begin{align}
R^{t}_\fp &\= \frac{2}{\beta^2-t_0^2} \left( \frac{1}{6} r^t_{(2,3),\fp} + \left(\cE_2 \sr R^t\right)_\fp - 2 \cE_{3,\fp} \right)~. \label{eq:D6_RxRy_Zero_Recurrence}
\end{align} 
Once again we can define functions $P^{t}_\fp$, which are polynomials in~$t_0$, that correspond to the numerators of $R^t_\fp$. The denominators of $R^t_\fp$ with $t_0$ dependence come from the overall prefactor in \eqref{eq:D6_RxRy_Zero_Recurrence}, so we define
\begin{align*} \notag
P^{t}_\fp(w) &\= \frac{\ii \pi \, \IDelta(\ii w)^{\length(\fp)}}{\pi^2 \beta^2 + w^2} R^{t}_\fp \left( \ii \frac{w}{\pi} \right),
\end{align*}
where in this section
\begin{align*} \notag
\IDelta(w) &\= Y \left( \pi^2 \beta^2 - w^2 \right)^2.
\end{align*}
These polynomials can be then computed from the recurrence relation \eqref{eq:D6_RxRy_Zero_Recurrence}, together with the above relation between the rational functions $R^t_\fp$ and the polynomials $P^t_\fp$. We list here the first few polynomials:
\begin{align*}
\begin{split}
P^{t}_{\{1\}}(w) &\= \hskip4pt \frac{1}{2} \hskip3pt \left(1+ w\right), \\[5pt]
P^{t}_{\{2\}}(w) &\= \hskip2pt \frac{1}{16} \hskip1pt \left(1+ 2 w\right), \\[5pt]
P^{t}_{\{1^2\}}(w) &\= \hphantom{1} \frac{1}{4} \hskip3pt \left(1+ w\right) \Big(w+2w^2+2w^3+\pi^2 \beta^2 \left(1+2w\right) \Big), \\[5pt]
P^{t}_{\{1,2\}}(w) &\= \;\frac{1}{32} \left(2w+9w^2+18w^3+12w^4+\pi^2 \beta^2 \left(3+14w+12w^2\right) \vphantom{\left(w^2\right)^2} \hskip-3pt \right).
\end{split} \nonumber
\end{align*}
The location of the attractor point $t$ can be written in terms of these polynomials, giving us a relatively simple solution to the attractor equations.
\begin{align*} \notag
\begin{split}
t &\= t_0 - \ii \pi \left( \beta^2 - t_0^2 \right) \sum_{k=1}^\infty \me^{2\pi \ii k t_0} \sum_{\fp \in \pt(k)} \frac{N_\fp}{ \IDelta(\pi t_0)^{\length(\fp)}} P_\fp^{t}(-\ii \pi t_0)~.
\end{split}
\end{align*}

\subsection{The alignment equations} \label{sect:D6_Alignment_Equations}
\vskip-10pt 
As discussed in \sref{sect:Method_Overview} we use \eqref{eq:aandb} to reduce the alignment equations to a pair:
\begin{align}
\begin{split}\label{eq:D6_Strominger_Equations}
x(x^2+y^2) - \frac{\sigma x}{y}+\frac{3}{Y}\left( 2  \cI_1 - \frac{2x}{y} \cI_2 + \frac{x^2+y^2}{y}\cI_2'\right)-\beta^{3}\gamma &\= 0~,\\[10pt]\+ x^2+y^2  +\frac{2}{Y}\left(\frac{x}{y} \cI_2'- \cI_1' \right)-\beta^2\phantom{\gamma}&\= 0~.
\end{split}
\end{align}
\subsubsection*{The perturbative equations} 
\vskip-10pt 
In this case, solving the perturbative equations reduces to finding the roots of a quartic recorded in \eqref{eq:D6quantumrecast}, which allows for a closed-form solution in terms of radicals. However, the resulting expression is not particularly illuminating. We will first solve the equations in the special case $\sigma=0$ and use the solutions $x_{00}$, $y_{00}$ to give us the explicit solution in terms of radicals, as well as an expansion in $\sigma$. Each of these formulations of the solution is useful for slightly different purposes although ultimately both solutions are of course equivalent.

Neglecting instantons, the equations for $x_{0}$ and $y_{0}$ read
\begin{align} \notag
x_{0}\, y_{0}&\=\beta\gamma \, y_{0}+\frac{\sigma}{\beta^{2}}\, x_{0}~,&
x_{0}^2+y_{0}^2&\=\beta^2~.\label{eq:D6quantum}\\
\intertext{In the special case $\sigma=0$, these are solved by}
x_0 &\= x_{00} \=\beta\gamma~, & y_0 &\= y_{00}\=\beta\sqrt{1-\gamma^2}~.
\end{align}
We also note for later discussion that the second equation in \eqref{eq:D6quantum} tells us that an attractor point only exists for this brane configuration if $\beta$ is real, and therefore so is $\gamma$. The modulus \hbox{$|t_{0}|{\=}\beta$} is independent of $\sigma$. For $\sigma{\=}0$, we are also forced to have ${|\gamma|<1}$, and so we introduce a real angle $\phi_{00}\;{\in}\;(0,\pi)$ by
\begin{equation*} \notag
\gamma\=\cos\phi_{00}~, \qquad \text{ so that } \qquad x_{00}\= \beta\cos\phi_{00} \;\; \text{ and } \;\; y_{00}\=\beta\sin\phi_{00}~.
\end{equation*}
It is optimal to work with the rescaled parameter $\widehat{\sigma}=\frac{\sigma}{\beta^{3}}$. We define also the rescaled coordinates 
\begin{equation*}\notag
\widehat{x}\=\frac{x_{0}}{\beta}~,\hspace{30pt}\widehat{y}\=\frac{y_{0}}{\beta}~.
\end{equation*}
In the rescaled variables the alignment equations are
\begin{equation}\label{eq:d6_a_recaled}
\widehat{x}\widehat{y}-\gamma\widehat{y}-\widehat{\sigma}\widehat{x}\=0~,\hskip30pt \widehat{x}^{2}+\widehat{y}^{2}\=1~.
\end{equation}
The first equation above provides an expression for $\widehat{x}$ in terms of $\widehat{y}$:
\begin{equation}\label{eq:yq}
\widehat{x}\=\frac{\gamma\widehat{y}}{\widehat{y}-\widehat{\sigma}}~.
\end{equation}
On substituting \eqref{eq:yq} into the second of equations \eqref{eq:d6_a_recaled} and clearing the denominators, we obtain a quartic equation for $\widehat{y}$:
\begin{equation}\label{eq:D6quantumrecast}
(\widehat{y}^2-1)\left(\widehat{y}-\widehat{\sigma}\right)^2+\widehat{y}^{2}\gamma^{2}\=0~.
\end{equation}
Some general remarks about this equation are in order. The highest term is $\widehat{y}^{4}$, and the constant term is $-\widehat{\sigma}^{2}$, so the product of the four roots of this equation is $-\widehat{\sigma}^{2}\leqslant 0$.

We are interested in solutions of \eqref{eq:D6quantumrecast} with positive positive imaginary part, which is necessary for them to lie in the large complex structure region, and how the number of these solutions varies as a function of $\widehat{\sigma}$. To any quartic polynomial with real coefficients is associated a discriminant~$D_{4}$, which for our polynomial is
\begin{equation*}\nonumber
D_{4}\=16\,\gamma^{2}\widehat{\sigma}^{2}\left(\left(1-\gamma^{2}-\widehat{\sigma}^{2}\right)^{3}-27\gamma^{2}\widehat{\sigma}^{2}\right).
\end{equation*}
When $D_{4}$ is positive, the quartic has either four or no real roots, while when $D_{4}$ is negative the quartic has precisely two real solutions. 

When $|\gamma|\geqslant 1$, this discriminant is negative for all $\widehat{\sigma}$. So when $|\gamma|\geqslant 1$ there are two real solutions to \eqref{eq:D6quantumrecast}, and by our earlier observation that the product of the roots is $-\widehat{\sigma}^2$, precisely one of these is positive. So in the following we will restrict our attention to the case $|\gamma|=\cos\phi_{00}<1$.  When $\widehat{\sigma}=0$, the quartic has four real roots. These are $\pm\sin\phi_{00}$ and $0$, the latter with multiplicity two. When $\widehat{\sigma}$ is small but non-vanishing, the solutions $\pm \sin\phi_{00}$ are perturbed slightly and both of the previously vanishing solutions become nonzero. 

Since a root can only vanish when $\widehat{\sigma}=0$, these new nonzero roots cannot change sign as $|\widehat{\sigma}|$ increases. Furthermore, the four roots multiply together to give a negative number. Since there is always one positive and one negative root, this forces both newly nonzero roots to have the same sign. Solving to leading order in $\widehat{\sigma}$, these solutions are
\begin{equation*}\notag
\widehat{y}\=\frac{\widehat{\sigma}}{1-\gamma}+\mathcal{O}\left(\widehat{\sigma}^{2}\right)~,\ \ \ \ \ \widehat{y}\=\frac{\widehat{\sigma}}{1+\gamma}+\mathcal{O}\left(\widehat{\sigma}^{2}\right)~.
\end{equation*}
So, irrespective of the value of $\phi_{00}$, the newly nonzero solutions are both of the same sign as $\widehat{\sigma}$.

The sign of $D_{4}$ is determined by the cubic
\begin{equation*}\notag
d(\phi_{00},\xi) \; \defineas \; \left.\frac{D_{4}}{16\gamma^{2}\widehat{\sigma}^{2}}\right\vert_{\widehat{\sigma}^{2}=\xi}\=\left(\sin^{2}\phi_{00}-\xi\right)^{3}-27\cos^{2}\phi_{00}\,\xi~.
\end{equation*}
For all $\phi_{00}\,{\in}\,(0,\pi)$, this quantity $d(\phi_{00},\xi)$ is monotonically decreasing as a function of $\xi$. Such a monotonic cubic has one real root, which here is located at
\begin{equation*}\notag
\left(1-|\gamma|^{2/3}\right)^{3}\; \defineas\; \Xi(\gamma)^{2}.
\end{equation*}
The sign of $D_{4}$ changes when $\widehat{\sigma}^{2} = \Xi(\gamma)^2$. 

When $\widehat{\sigma}\,{=}\,\left(1-|\gamma|^{2/3}\right)^{3/2}$, the solution flowing from $\sin\phi_{00}$  meets the larger solution that vanishes at $\widehat{\sigma}=0$. For larger $\widehat{\sigma}$ they both become nonreal, and so lose relevance for us. The solution flowing from $\sin\phi_{00}$ remains a solution for every negative value of $\widehat{\sigma}$.

These observations are summarised in the  following \tref{tab:D6_A_numbers}, where we also include equation numbers for the closed-form solutions that we find later. This phenomenon of the splitting of attractors as $\sigma$ varies was first noted in \cite{Bellucci:2007eh}.
\begin{table}[H]
	\renewcommand{\arraystretch}{1.2}
	\begin{center}
		\begin{tabular}{ |c|l|c|}
			\hline
			\textbf{Case} & \hfil \textbf{~Topological/charge data~{}} & \hfil \textbf{~Solutions with $y_{0}>0$ ~{}} \\[2pt]
			\hline\hline
			i & \quad $|\gamma|\geqslant1$,   $\,\,\,\,\phantom{0<}\widehat{\sigma}= 0$   & none \\[1pt]
			\hline
			ii & \quad $|\gamma|\geqslant1$,   $\,\,\,\,\phantom{0<}\widehat{\sigma}\neq 0$   & $y_0^*$ \\[1pt]
			\hline
			iii & \quad $|\gamma|<1$,   $\,\,\,\,\phantom{0<}\widehat{\sigma}\leqslant0$    & $y_0$\\[1pt]
			\hline
			iv  & \quad $|\gamma|<1$,   $\,\,0<\widehat{\sigma}<\Xi(\gamma)$ & $y_0$, $y_0^{\pm}$ \\[1pt]
			\hline
			v  & \quad$|\gamma|<1$,   $\,\,\,\,\phantom{0<}\widehat{\sigma}=\Xi(\gamma)$ & $y_0^{\pm}$ \\[1pt]
			\hline
			vi  &\quad $|\gamma|<1$,   $\,\,\,\,\phantom{0<}\widehat{\sigma}>\Xi(\gamma)$ & $y_0^{+}$ or $y_0^{-}$ \\[1pt]  
			\hline
		\end{tabular}
		\vskip10pt
		\capt{5.5in}{tab:D6_A_numbers}{Solutions to the alignment equations in different cases. The quantities $y_0$, $y_0^{\pm}$ and $y_0^*$ are defined in \eqref{eq:D6_Alignment_Radical}, \eqref{eq:d6a_caseiii_sol12}, and \eqref{eq:d6a_caseii_sol}.}
	\end{center}
\end{table}
\vskip-30pt
We can illustrate the above cases with some plots that perhaps clarify the changes in number of solutions. 
The locus of points defined by the second of the equations \eqref{eq:d6_a_recaled} is a circle, while the locus corresponding to the first equation is a rectangular hyperbola with asymptotes $\widehat{x}=\widehat{\sigma}$ and $\widehat{y}=\gamma$. The following \fref{fig:d6a_conics} illustrates the behaviour described in \tref{tab:D6_A_numbers}. Note the progression iv$\rightarrow$v$\rightarrow$vi for increasing positive $\widehat{\sigma}$.
\begin{figure}[H]
\centering
\captionsetup[subfigure]{labelformat=empty}
\subfloat[i]{\includegraphics[width = 1.3in]{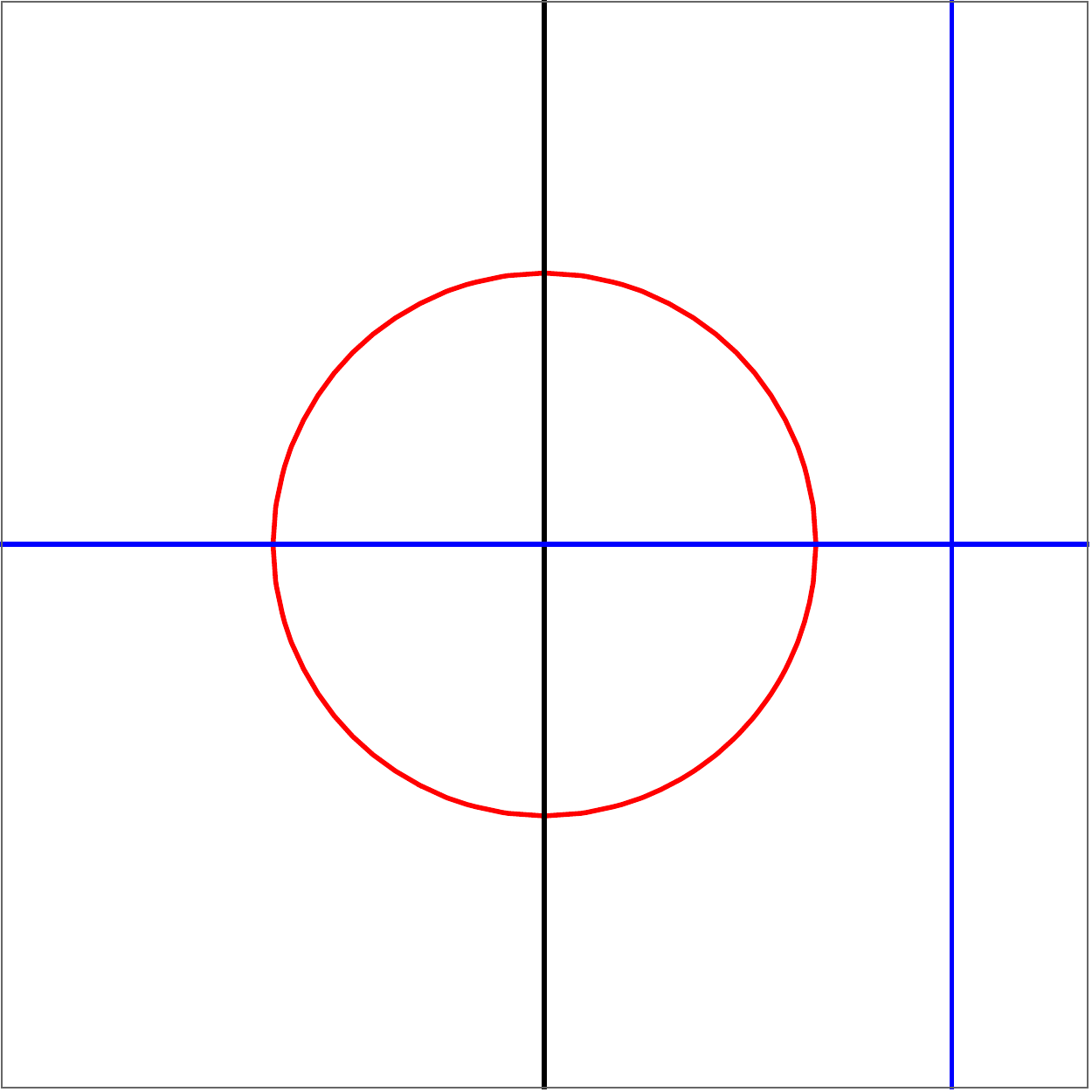}}\hskip30pt 
\subfloat[ii]{\includegraphics[width = 1.3in]{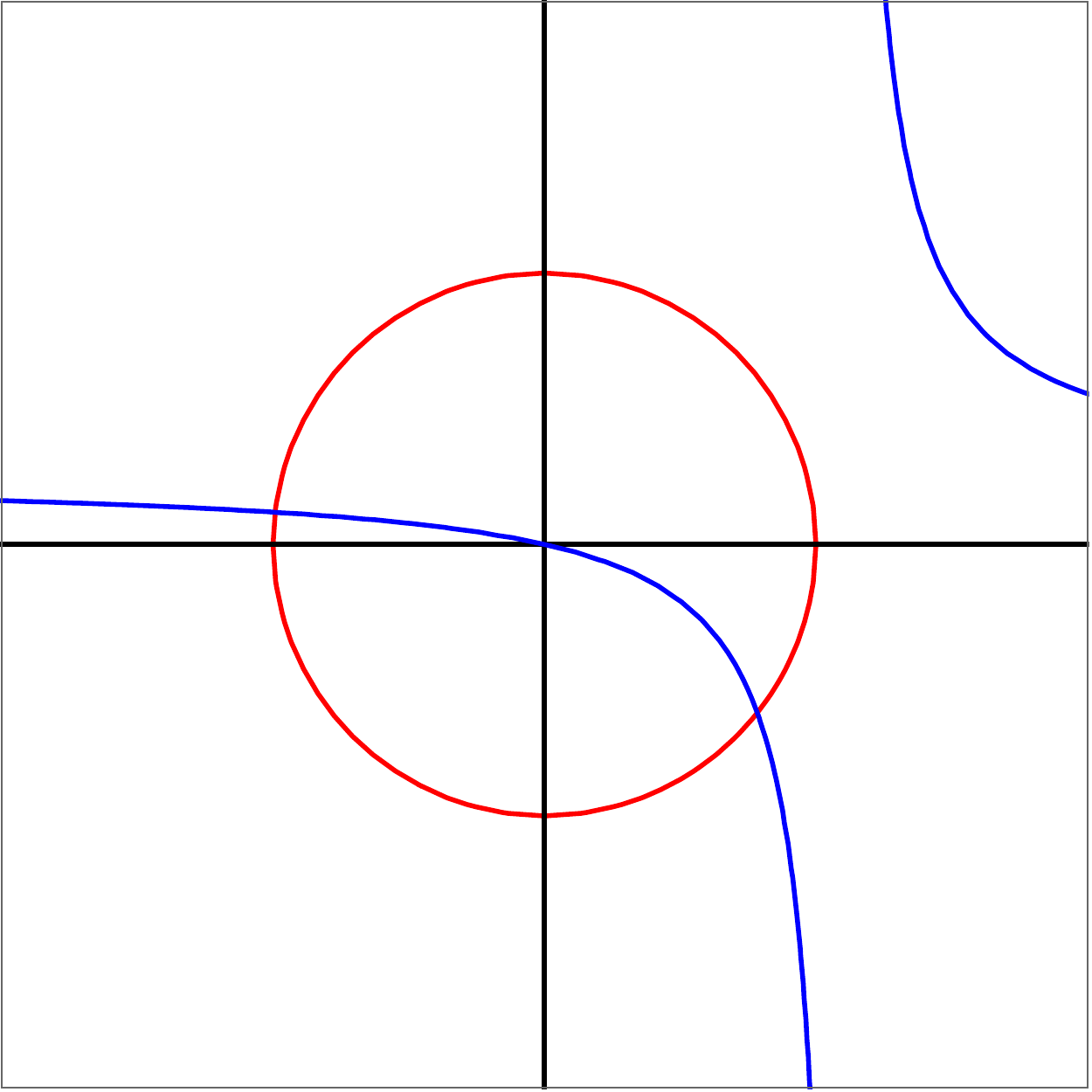}}\hskip30pt
\subfloat[iii]{\includegraphics[width = 1.3in]{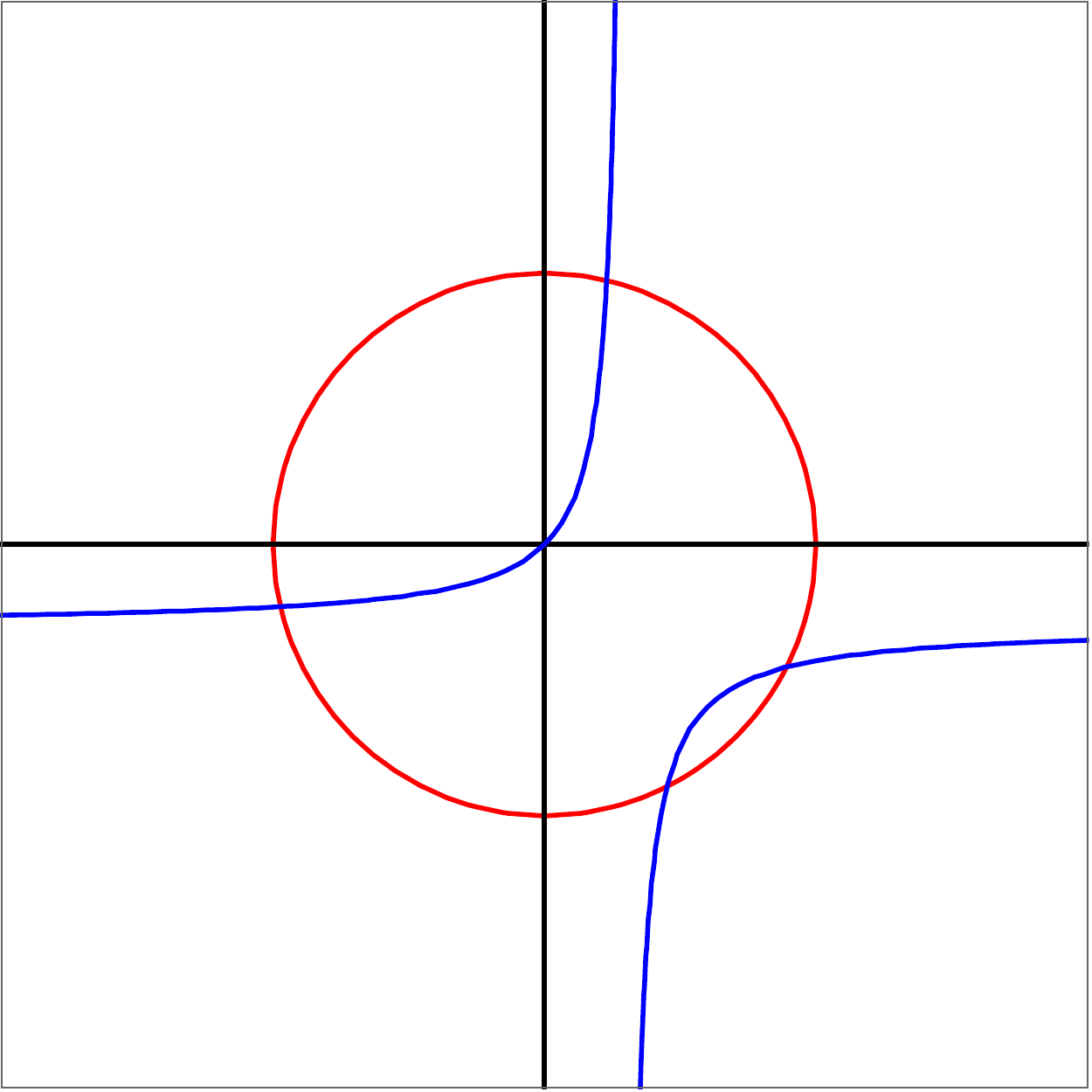}}\\
\subfloat[iv]{\includegraphics[width = 1.3in]{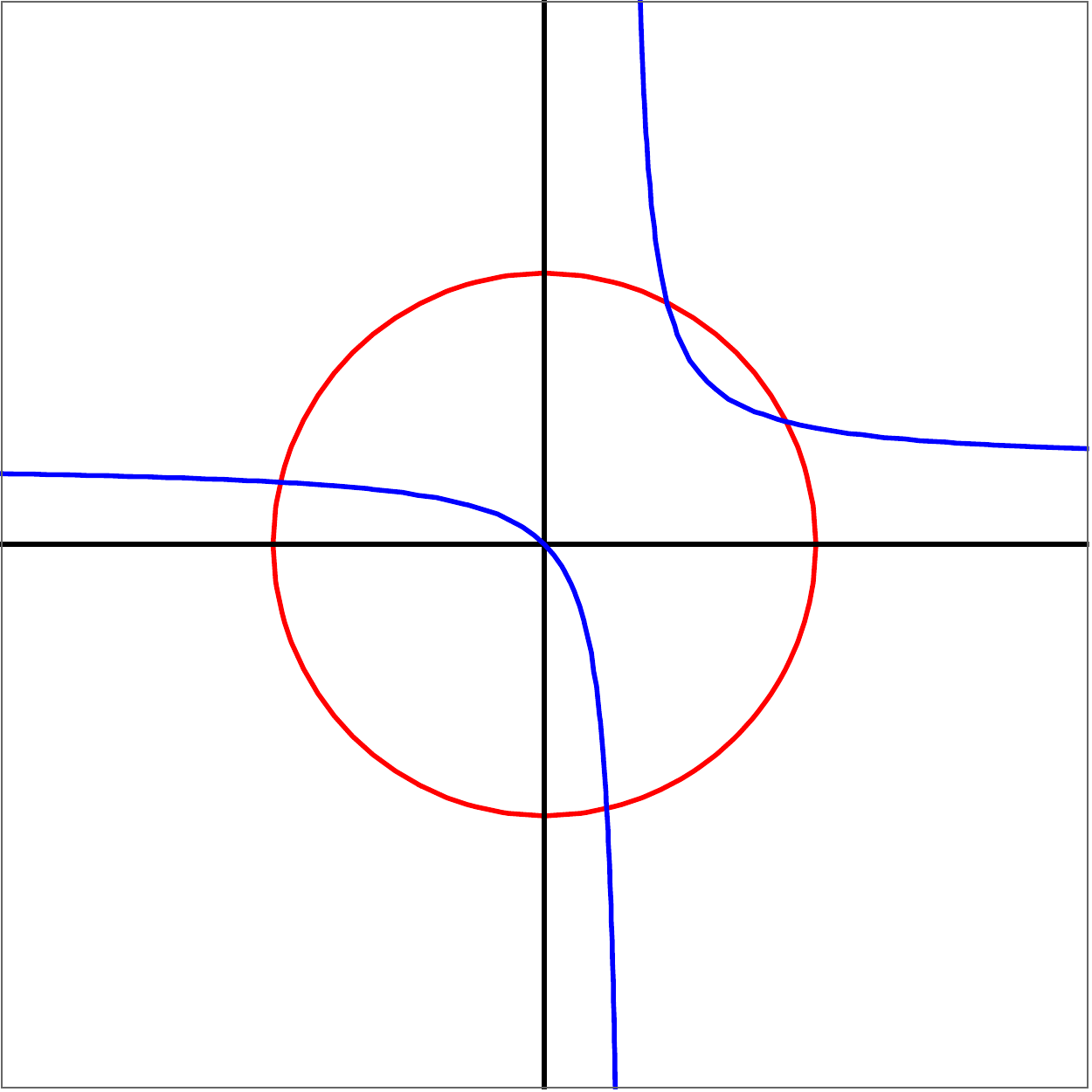}}\hskip30pt
\subfloat[v]{\includegraphics[width = 1.3in]{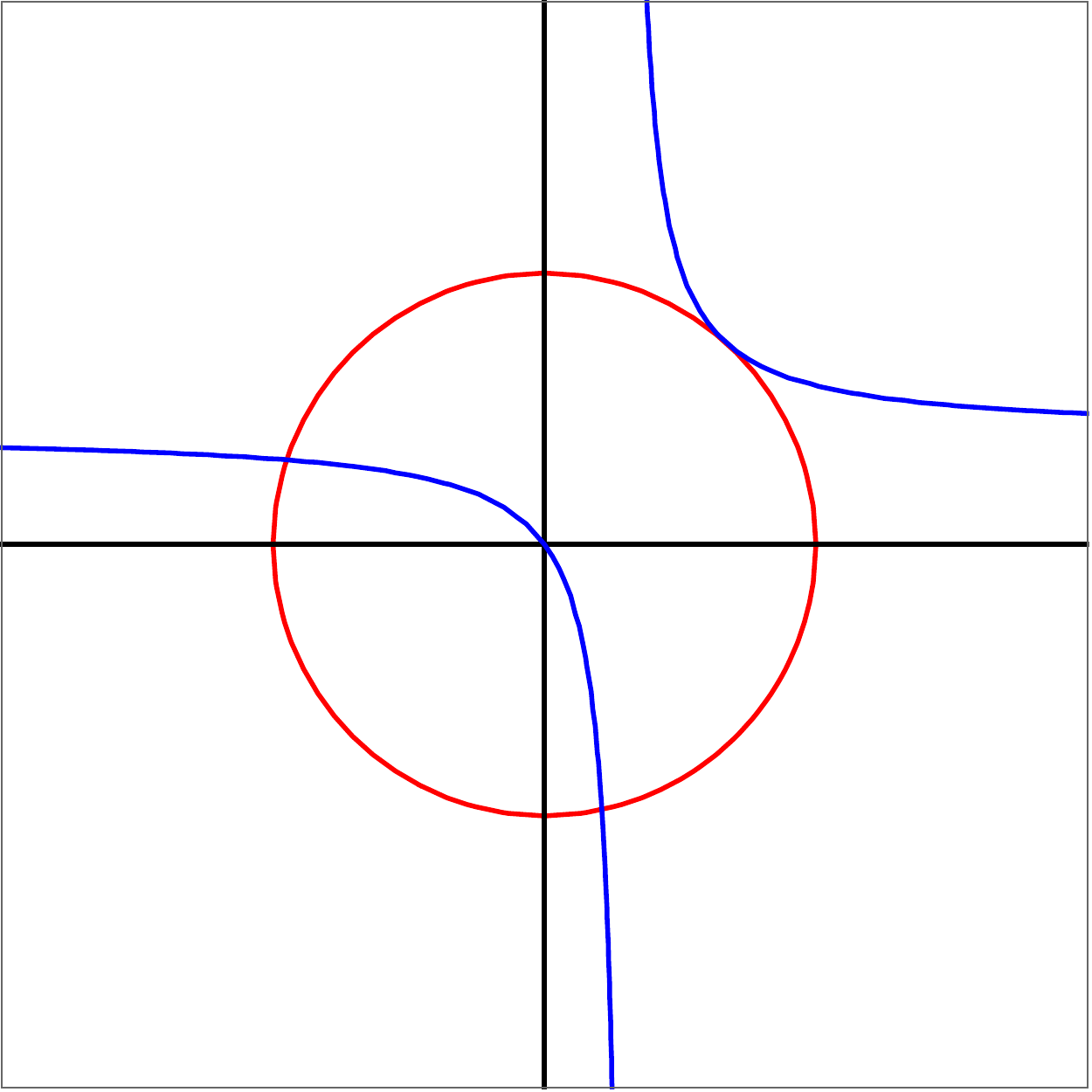}}\hskip30pt
\subfloat[vi]{\includegraphics[width = 1.3in]{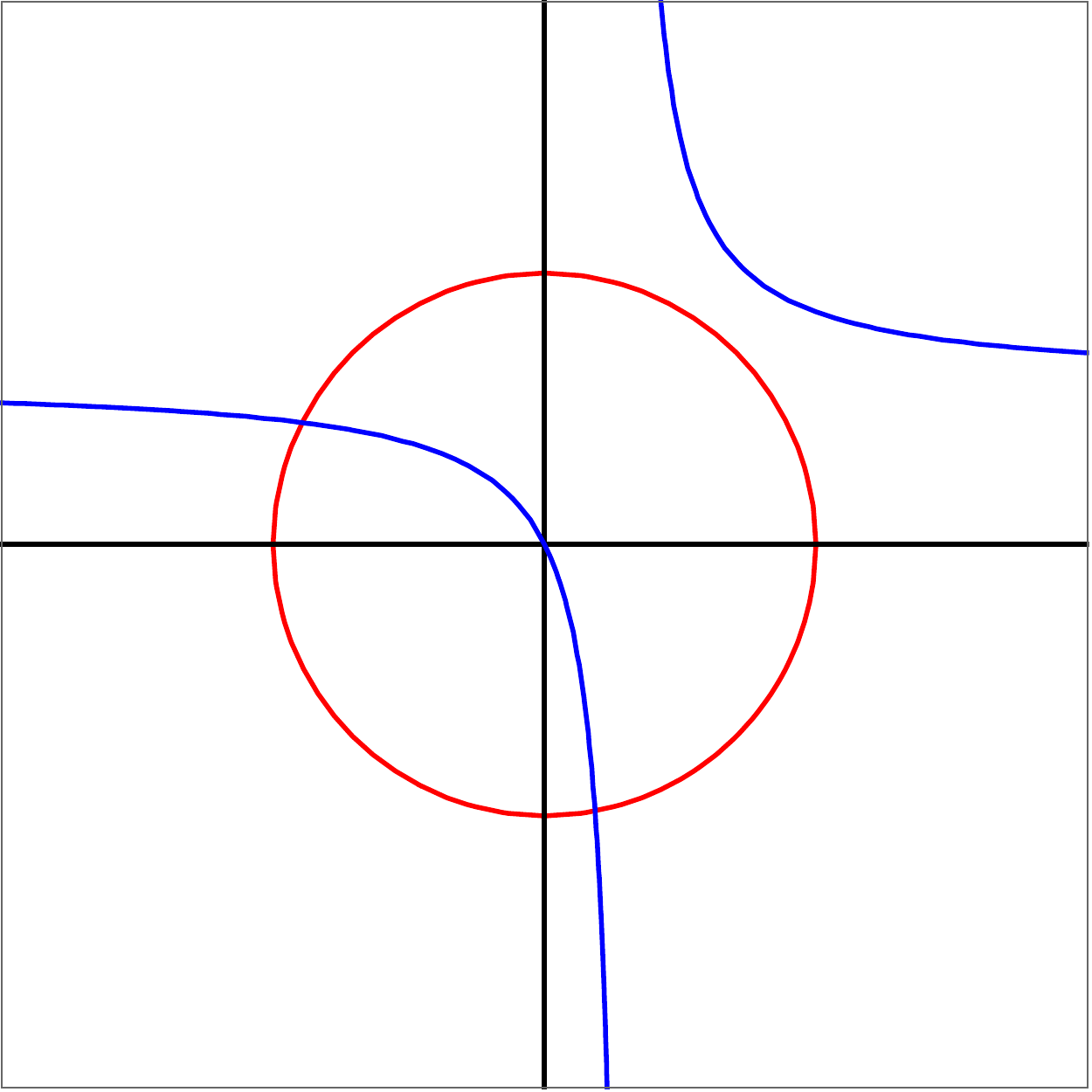}}\\
\vskip5pt
\capt{5.92in}{fig:d6a_conics}{Illustrations of the conic intersections. The black lines are the coordinate axes $\widehat{x}=0$ and $\widehat{y}=0$. Our interest is in the solutions with positive $\widehat{y}$, that is the intersections above the $\widehat{x}$-axis.}
\end{figure}
When $\widehat{\sigma}=0$ and $|\gamma|<1$, the hyperbola is degenerate and intersects the circle once in the upper half plane. Turning $\widehat{\sigma}$ on will perturb this solution. For a small positive $\widehat{\sigma}$ this perturbed solution is the uppermost point of intersection in the Figure iv above. We are able to give explicit expressions for $y_0$ as a power series in $\widehat{\sigma}$ about $y_{00}$, in terms of hypergeometric functions of argument~$\gamma^{2}$. The details of this can be found in Appendix \ref{app:D6_Polynomials}.
\begin{equation} \label{eq:D6_Alignment_Hypergeometric}
\begin{split} 
y_{0}\=y_{00}-\frac{\beta\gamma^{2}}{2}\,\sum_{k=1}^{\infty}(k+1)\,\threeFtwo{\frac{k+1}{2}}{\frac{k+2}{2}}{\frac{k+3}{2}}{\frac32}{2}{\gamma^{2}}\widehat{\sigma}^{k}~.
\end{split}
\raisetag{100pt}
\end{equation}
It is also possible to solve the equations \eqref{eq:D6quantum} in closed form in terms of radicals. The simplest expression for the root $y_{0}$ that becomes $y_{00}$ when $\sigma=0$ is
\begin{align}\label{eq:D6_Alignment_Radical}
\frac{2y_{0}}{\beta}&\=\widehat{\sigma}+\sqrt{Q_{2}+\widehat{\sigma}^{2}}+\sqrt{2\sin^{2}\phi_{00}-Q_{2}-\frac{2\widehat{\sigma}\left(2-\sin^{2}\phi_{00}\right)}{\sqrt{Q_{2}+\widehat{\sigma}^{2}}}}~,
\end{align}
where we have introduced auxiliary quantities $Q_1$ and $Q_2$ with
\begin{align*} \notag
\begin{split}
Q_{1}&\=\frac{3}{\sin^{2}\phi_{00}-\widehat{\sigma}^{2}}\left(\,\widehat{\sigma}\cos\phi_{00}+ \ii \sqrt{\left(\frac{\sin^{2}\phi_{00}-\widehat{\sigma}^{2}}{3}\right)^{3}-\widehat{\sigma}^{2}\cos^{2}\phi_{00}}\;\right)^{2/3},\\[10pt]
Q_{2}&\=\frac{\sin^{2}\phi_{00}-\widehat{\sigma}^{2}}{3}\left(Q_{1}+2+\frac{1}{Q_{1}}\right).
\end{split}
\end{align*}
We are using the branch of the function $z^{1/3}$ that is $|z|^{1/3} \ee^{\ii \arg(z)/3}$, with $-\pi < \arg(z) \leqslant \pi$.

The two additional solutions in case iv are given by 
\begin{equation}\label{eq:d6a_caseiii_sol12}
\frac{2y_{0}^\pm}{\beta}\=\widehat{\sigma}-\sqrt{Q_{2}^{\pm}+\widehat{\sigma}^{2}}+\sqrt{2\sin^{2}\phi_{00}-Q_{2}^{\pm}+\frac{2\widehat{\sigma}\left(2-\sin^{2}\phi_{00}\right)}{\sqrt{Q_{2}^{\pm}+\widehat{\sigma}^{2}}}}~,
\end{equation}
with
\begin{align*} \notag
\begin{split}
Q_{1}^{\pm}&\=\frac{3}{\sin^{2}\phi_{00}-\widehat{\sigma}^{2}}\left(\,\pm\widehat{\sigma}\cos\phi_{00}+ \ii \sqrt{\left\vert\frac{\sin^{2}\phi_{00}-\widehat{\sigma}^{2}}{3}\right\vert^{3}-\widehat{\sigma}^{2}\cos^{2}\phi_{00}}\;\right)^{2/3},\\[10pt]
Q_{2}^{\pm}&\=\frac{\sin^{2}\phi_{00}-\widehat{\sigma}^{2}}{3}\left(Q_{1}^{\pm}+2+\frac{1}{Q_{1}^{\pm}}\right).
\end{split}
\end{align*}
At the critical value $\widehat{\sigma}=\Xi(\gamma)$ the larger\footnote{$y_{0}^{-}>y_{0}^{+}$ if and only if $\cos\phi_{00}>0$.} of these becomes equal to solution \eqref{eq:D6_Alignment_Radical}, which is the case~v in the \tref{tab:D6_A_numbers}. For $\widehat{\sigma}>\Xi(\gamma)$ one of these is no longer real and we enter the case vi with the real member of this pair giving the only solution.

In the case ii where $|\gamma|{\,\geqslant\,}1$ and $\widehat{\sigma}{\,\neq\,}0$ there is always a single relevant solution. This solution is 
\begin{equation}\label{eq:d6a_caseii_sol}
\frac{2y_{0}^*}{\beta}\=\widehat{\sigma}-\text{sign}(\sigma)\sqrt{Q^{*}_{2}+\widehat{\sigma}^{2}}+\sqrt{\frac{2|\widehat{\sigma}|(2+\sinh^{2}\varphi)}{\sqrt{Q_{2}^{*}+\widehat{\sigma}^{2}}}-2\sinh^{2}\varphi-Q_{2}^{*}}~,
\end{equation}
where 
\begin{align*} 
\begin{split}
\gamma&\=\pm\cosh\varphi,\\[10pt]
Q_{1}^{*}&\=\frac{3}{\sinh^{2}\varphi+\widehat{\sigma}^{2}}\left(-\widehat{\sigma}\cosh\varphi + \sqrt{\left(\frac{\sinh^{2}\varphi+\widehat{\sigma}^{2}}{3}\right)^{3}+\widehat{\sigma}^{2}\cosh^{2}\varphi} \;\right)^{2/3},\\[10pt]
Q_{2}^{*}&\=\frac{\sinh^{2}\varphi+\widehat{\sigma}^{2}}{3}\left(Q_{1}^{*}-2+\frac{1}{Q_{1}^{*}}\right).
\end{split}
\end{align*}
In any nondegenerate case with $\widehat{\sigma}\neq0$, a value for $y_{0}$ singles out one point on the relevant hyperbola in \fref{fig:d6a_conics}, which uniquely specifies $x_{0}$:
\begin{equation}\label{d6_a_x0}
x_{0}\=\frac{\beta\gamma y_{0}}{y_{0}-\beta\widehat{\sigma}}~.
\end{equation}

\subsubsection*{Instanton corrections} 
\vskip-10pt 
The instanton corrections to the alignment equations take the form of a series in $q_y$. To solve these equations perturbatively, we again require that the coefficient of every power of $q_y$ vanishes. This leads to two conditions
\begin{align*} \notag
\begin{split}
0&\=-\frac{1}{3} \sum_{0 \leqslant j \leqslant i \leqslant k} \hskip-5pt (x_{k-i} x_{i-j} x_j + x_{k-i} y_{i-j} y_j) + \frac{\sigma}{3} \sum_{i=0}^k \eta_{k-i} x_i + 2 \cS_3(k)\\
&\hskip2cm \+ + 2\hskip-9pt \sum_{0 \leqslant j \leqslant i \leqslant k} \hskip-9pt \cC_{3}(k-i) \eta_{i-j} x_j + \hskip-15pt \sum_{0 \leqslant l \leqslant j \leqslant i \leqslant k} \hskip-13pt x_{k-i} x_{i-j} \eta_{j-l} \cS_2(l) + \sum_{i=0}^k y_{k-i}\cS_2(i)~,\\[10pt]
0&\= \frac{1}{2} \sum_{i=0}^k (x_{k-i} x_{i} + y_{k-i} y_{i}) + \cC_2(k) - \hskip-10pt \sum_{0 \leqslant j \leqslant i \leqslant k} \hskip-5pt x_{k-i} \eta_{i-j} \cS_2(j)~.
\end{split}
\end{align*}
In the first equation, $x_k$ and $y_k$ appear in the first three terms, while other terms contain only $x_i$ and $y_i$ with $i<k$. The prefactors of $x_k$ and $y_k$ are
\begin{align*} \notag
\frac{1}{3} \Big( (1-\zeta) y_0^2 - 3 x_0^2 \Big), \qquad \text{and} \qquad -\frac{1}{3} (2+\zeta) x_0 y_0~.
\end{align*}
In the second equation the coefficients are $x_0$ and $y_0$, respectively. We can then solve the two equations above for $x_k$ and $y_k$. In this way, we get a recurrence relation for $x_k$ and $y_k$ in terms of $y_i$ and $x_i$ with $i<k$, which is of the form
\begin{align*} \notag
\begin{split}
x_k \= \frac{A(k) + (2 + \zeta) B(k) x_0}{(1-\zeta)\beta^2}~, \quad \qquad y_k \= -\frac{A(k) x_0 + B(k) \big(3 x_0^2 + (1-\zeta) y_0^2 \big)}{y_0 (1-\zeta)\beta^2}~,
\end{split}
\end{align*}
where $A(k)$ and $B(k)$ can be written as $N$-expansions
\begin{align*} \notag
A(k) &\= \sum_{\fp \in \pt(k)} N_\fp A_\fp~, \qquad B(k) \= \sum_{\fp \in \pt(k)} N_\fp B_\fp~,
\end{align*}
with the coefficients $A_\fp$ and $B_\fp$ given by
\begin{align} \label{eq:D6_AB}
\begin{split}
A_\fp &\= - r^{x_{\,}}_{(2,3),\fp} - \big(r^{y_{\,}}_{(1,2)} \hs R^{x_{\,}} \! \big)_\fp - r^{y_{\,}}_{(2,2),\fp} x_0 + \zeta y_0^3 \left( (\bm{\eta} \hs R^{x_{\,}} \! )_\fp + \bm{\eta}'_\fp x_0 \right) \\
&\hskip2cm \+ + 6 \, \cS_{3,\fp} + 6 \, (\cC_3 \sr \bm{\eta} \sr R^{x_{\,}} \! )_\fp + 3 \, (\cS_2 \sr \bm{\eta} \sr R^{x_{\,}} \! \! \sr R^{x_{\,}} \!)_\fp + 3 \, (\cS_2 \sr R^{y_{\,}} \! )_\fp~,\\[10pt]
B_\fp &\= \frac{1}{2} \left( R^{y_{\,}} \!  \hs R^{y_{\,}} + R^{x_{\,}} \! \hs R^{x_{\,}} \! \right)_\fp + \cC_{2,\fp} - (\cS_2 \sr \bm{\eta} \sr R^{x_{\,}} \! )_\fp~.
\end{split}
\end{align} 
The quantities $A(k)$ and $B(k)$ can be seen to only depend on $R^{x_{\,}}_\fq$ and $R^{y_{\,}}_\fq$ with $|\fq|{\,<\,}k$. This leads immediately to the following recurrences:
\begin{align}
\begin{split}
R^{x_{\,}}_\fp &\= \frac{ A_\fp + (\zeta+2) B_\fp x_0}{(1-\zeta)\beta^2}~, \qquad \qquad R^{y_{\,}}_\fp \= -\frac{ A_\fp x_0 + B_\fp (3 x_0^2 + (1-\zeta) y_0^2)}{y_0 (1-\zeta)\beta^2}~.
\end{split} \label{eq:D6_Rxy_Strominger_Recurrence}
\end{align}
It is clear that in general $R^{x_{\,}}_\fp$ and $R^{y_{\,}}_\fp$ are rational functions. Furthermore, it can be seen that their denominators take on a simple form. This allows us, by a suitable scaling, to find functions $P^{x_{\,}}_\fp$ and $P^{y_{\,}}_\fp$ which are polynomials in the quantities $x_0$, $y_0$, $\sin(2\pi x_0)$, and~$\cos(2\pi y_0)$,
\begin{align} \label{eq:D6_PR_Relation}
\begin{split}
P^{x_{\,}}_\fp (u,v) &\= -\frac{\pi \IDelta(v)^{\length(\fp)}}{v^2(1-\zeta)} R_\fp^{x_{\,}}\left(\frac{u}{\pi},\frac{v}{\pi} \right), \qquad \qquad P^{y_{\,}}_\fp (u,v) \= -\frac{\pi \IDelta(v)^{\length(\fp)}}{v (1-\zeta)} R_\fp^{y_{\,}}\left(\frac{u}{\pi},\frac{v}{\pi} \right),
\end{split}
\end{align}
where, in this section, $\IDelta$ denotes the quantity
\begin{align*} \notag
\IDelta(v) \= Y (1-\zeta)^2 v^3 \beta^2~.
\end{align*}
These polynomials get rather complicated quite quickly. We list here the first few polynomials:
\begin{align*}
\begin{split}
P^{x_{\,}}_1(u,v) &\= \hskip2pt -\frac{1}{4} \hskip3pt \Big( \hskip-3pt \left((1-\zeta) u^2+3 v (v+1)\right)\sin(2u)+u\left((2+\zeta) \vphantom{{}^2} v+3\right)\cos(2u) \Big)\, ,\\[5pt]
P^{x_{\,}}_2(u,v) &\= -\frac{1}{32} \Big( \hskip-3pt \left(2(1-\zeta)u^2+3v(2 v+1)\right)\sin(4u)+u(2 (2+\zeta) v+3)\cos(4u)\Big)\, ,\\[5pt]
P^{y_{\,}}_1(u,v) &\=\+ \hskip2pt \frac{1}{4} \hskip3pt \Big(\hskip-3pt \left(3 u^2 (v+1)+(1-\zeta) v^3\right)\cos(2u)+uv((2+\zeta) v+3)\sin(2u)  \Big)\, ,\\[5pt]
P^{y_{\,}}_2(u,v) &\= \+ \frac{1}{32} \Big( \hskip-3pt\left(u^2 (6 v+3)+2 (1-\zeta) v^3\right)\cos(4u)+uv (2(2+\zeta)v+3)\sin(4u) \Big)\, .
\end{split}\nonumber
\end{align*}
We rewrite \eqref{eq:x_y_ansatse} in terms of these polynomials, which gives the following $N$-expansions for $x$ and~$y$
\begin{align*} \notag
x &\= x_0 - \pi y_0^2 (1-\zeta)\sum_{k=1}^\infty \me^{-2\pi k y_0} \sum_{\fp \in \pt(k)} \frac{N_\fp}{\IDelta(\pi y_0)^{\length(\fp)}} P_\fp^{x_{\,}}(\pi x_0,\pi y_0)~,\\[5pt]
y &\= y_0 - \phantom{\pi} y_0^{\phantom{2}} (1-\zeta) \sum_{k=1}^\infty \me^{-2\pi k y_0} \sum_{\fp \in \pt(k)} \frac{N_\fp}{\IDelta(\pi y_0)^{\length(\fp)}}P_\fp^{y_{\,}}(\pi x_0,\pi y_0)~. \notag
\end{align*}  
These polynomials are in general quite complicated. However, in important special cases there are closed-form expressions for them.
\newpage

\section{Solutions} \label{sect:Solutions}
\vskip-10pt
In the previous sections, we have introduced the general formalism for solving attractor equations, which we will use in this section to study some important special cases for which this formalism undergoes considerable simplification. In particular, we find simple solutions to the orthogonality equation in the case of the D4-D2-D0 system, and for a special case of the alignment equations for the D6-D2-D0 system with $x=0$.

We first consider the D6-D2-D0 alignment equations in the case $x=0$. We find that the perturbative equations \eqref{eq:D6quantum} do not, in fact, involve $\sigma$, so  $y_0=\beta$, which is a great simplification compared to the analysis of \sref{sect:D6_Alignment_Equations}. The recurrence relations for $R^y_\fp$ have also a relatively simple form.

The slightly more involved D4-D2-D0 orthogonality equation is also studied. We can solve the attractor equations in this case without any additional assumptions. These solutions have a similar form to those of the D6-D2-D0 case, which can be understood by examining intersections of linear subspaces with the charge lattice.

One of the most interesting consequences of these solutions is that we are able to give a systematic expansion, including the instanton corrections, for the central charge and thus for the entropy of the black holes corresponding to the aforementioned systems.

Although our results are not limited to rank two attractor points, an important class of concrete examples are provided by the rank two attractor points found in \cite{Candelas:2019llw}. We briefly review some results on these points before verifying that our method reproduces these from a different point of view. The intriguing number theoretic properties of the rank two attractor points provide us with an interesting identity \eqref{eq:L-k_indentity} connecting Bessel functions and Gromov-Witten invariants to L-functions.
\subsection{The D6-D2-D0 alignment equations} \label{sect:Solutions_D6-D2-D0}
\vskip-10pt
The alignment equations \eqref{eq:D6_Strominger_Equations} simplify considerably when we restrict to the special case where $\Lambda = 0$. By looking at the equations and seeing that $\gamma$ also vanishes, we see that the perturbative solution for $x$ is given by $x_0=0$. It is also easy to verify that $x=0$ is a solution to the instanton-corrected equations as well. Thus we can take $x_i = 0$ for all $i$ and $R^{x_{\,}}_\fp = 0$ for all $\fp$. 

As $\sigma$ only appears multiplied by $x$, it drops out of the equations. The perturbative coordinate $y_0$ is then simply
\vskip-25pt
\begin{align*} \notag
y_0  \= \beta \=  \sqrt{\frac{Y_{100} + 2 \Upsilon}{Y}}~. 
\end{align*}
More significantly, upon setting $x=0$ the recurrence relation for the instanton corrections, given by \eqref{eq:D6_Rxy_Strominger_Recurrence} together with \eqref{eq:D6_AB}, simplifies considerably. One is left with the relation
\begin{align}
R^y_\fp(y_0) &\= -\frac{1}{2y_0} (R^y {\hs} R^y)_\fp - \frac{1}{y_0} \cC_{2,\fp}~. \label{eq:D6_Special_Case_Simplified_Recurrence}
\end{align}
To proceed, we write out $\cC_{2,\fp}$ in terms of the functions $R^y_\fp$. Again, for $x=0$, the coefficients $\cC_{2,\fp}$ simplify and can be written as
\begin{align} \label{eq:D6_Special_Case_C_2}
\cC_{2,\fp} \= \frac{1}{Y} \sum_{m \in \fp} \sum_{j = 0}^{l(\fp)-1} \frac{(-2\pi m)^{j-2}}{j!} r^{y_{\,}}_{(j,j),\fp - m} \= \frac{1}{Y} \sum_{m \in \fp} \sum_{j = 1}^{l(\fp)-1} \frac{(-2\pi m)^{j-2}}{j!} \overbrace{{(R^y {\hs} R^y {\hs} ...  {\hs} R^y)}}^{j \text{ times}}\hskip-2pt{}_{\fp-m}~.
\end{align}
Thus the recurrence relation is brought to the form
\begin{align*} \notag
R^y_\fp(y_0) \= -\frac{1}{2y_0} (R^y {\hs} R^y)_\fp - \frac{1}{y_0} \sum_{m \in \fp} \sum_{j = 1}^{l(\fp)-1} \frac{(-2\pi m)^{j-2}}{j!} \overbrace{{(R^y {\hs} R^y {\hs} ...  {\hs} R^y)}}^{j \text{ times}}\hskip-2pt{}_{\fp-m}~.
\end{align*}
As previously, it is most convenient to define a set of polynomials $P^y_\fp(v)$ by separating the denominators of $R^y_\fp$. We could in principle use the definition \eqref{eq:D6_PR_Relation}, but here it is useful to give a slightly different definition adapted to this special case:
\begin{align*} \notag
\widetilde{P}^y_\fp(v) &\; \defineas \; -\frac{(4 \pi^2 v^2 Y)^{l(\fp)}}{a_\fp v}  R^y_\fp(v)~,
\end{align*}
where $a_\fp$ are partition-dependent rational constants that we will later fix to get a convenient form for the solution. With this definition, it can be seen that $\widetilde{P}^y_\fp(v)$ are polynomials with coefficients in~$\IQ(\pi)$. In particular, for partitions of length 1, $\fp = \{p_1\}$, we find that
\begin{align*} \notag
\widetilde{P}^y_{\{p_1\}}(v) &\= 4 \pi^2 v Y R^y_{\{p_1\}}(v) \= \frac{1}{a_{\{p_1\}} p_1^2}~.
\end{align*}
An important observation is that the polynomials $\widetilde{P}^y_\fp(v)$ depend on the partition $\fp$ only via its length $l(\fp)$ and modulus $|\fp|$. This motivates us to define a second set of polynomials
\begin{align*} \notag
\widetilde{P}^y_{\fp}\left(\frac{v}{2\pi |\fp|} \right) \; \defineas \; f_{\length(\fp)}(v)~.
\end{align*}
The polynomials $f_s(v)$ only depend on the length $\length(\fp)=s$ of the partition~$\fp$. In addition, $f_s$ turn out to have rational coefficients that depend on the $a_\fp$. By a suitable choice of these constants the polynomials $f_s(v)$  have integral coefficients.

By numerical experimentation, we find that the $a_\fp$ fit a combinatorial pattern. Any coefficient associated to the partition \mbox{$\sum_{j=1}^\infty \mu_j j = \left\{ 1^{\mu_1},2^{\mu_2},...\right\}$} factorises into coefficients associated to the partitions $\left\{1^{\mu_1} \right\}, \left\{2^{\mu_2} \right\},...$, and moreover
\begin{align}
a_{\left\{ 1^{\mu_1},2^{\mu_2},...\right\}} \= \prod_{j=1}^{\infty} a_{\left\{j^{\mu_j}\right\}} \= \prod_{j=1}^{\infty} \frac{1}{j^{2\mu_j} (\mu_j!)}~. \label{eq:defn_a_fp}
\end{align}
The first few functions $f_s(z)$ are
\begin{align*}
\begin{split}
f_1(z) &\= z+1~,\\
f_2(z) &\= z^2+3 z+3~,\\
f_3(z) &\= z^3+6 z^2+15 z+15~,\\
f_4(z) &\= z^4+10 z^3+45 z^2+105 z+105~,\\
f_5(z) &\= z^5+15 z^4+105 z^3+420 z^2+945 z+945~,\\
f_6(z) &\= z^6+21 z^5+210 z^4+1260 z^3+4725 z^2+10395 z+10395~,\\
f_7(z) &\= z^7+28 z^6+378 z^5+3150 z^4+17325 z^3+62370 z^2+135135 z+135135~.
\end{split} \nonumber
\end{align*}
These are seen to obey a second-order differential equation of confluent hypergeometric type
\begin{align*} \notag
z \frac{d^2}{d z^2} f_s - 2(s+z) \frac{d}{dz} f_s + 2 s f_s \= 0~,
\end{align*}
which implies that they have the expansion:
\begin{align*} \notag
f_s(z) \= \sum_{j=0}^s \frac{(2s-j)!}{2^{s-j}(s-j)!j!} \, z^j~.
\end{align*}
The fact that the differential equation is of confluent hypergeometric type suggests a connection with Bessel functions. Indeed, the polynomials $f_s$ can be expressed in terms of the modified spherical Bessel functions of the third kind, $\bm{k}_n(z)$:
\vskip-30pt
\begin{align}
f_s(y) &\= \me^{y} y^{s+1} \bm{k}_{s}(y)~, \quad \text{where} \quad \bm{k}_n(z) \= \sqrt{\frac{2}{\pi z}} K_{n+\frac{1}{2}}(z)~, \label{eq:D6-D0_P_s_M}
\end{align}
with $K_\nu(z)$ the modified Bessel functions of the second kind \cite{Erdelyi2}. These functions have useful properties. Among others, they satisfy the following recurrence relation and differential identity:
\begin{align*} \notag
\begin{split}
\bm{k}_n(z) &\= \bm{k}_{n-2}(z) + \frac{2n-1}{z} \bm{k}_{n-1}(z)~, \qquad \bm{k}_{n+1}(z) \;= - z^n \frac{d}{dz}\left(z^{-n}\bm{k}_n(z)\right)~.
\end{split}
\end{align*}
As a consequence of the first identity, there is a recurrence relation for the polynomials~$f_s$,
\begin{align*} \notag
f_s(z) &\= z^2 f_{s-2}(z) + (2s-1)f_{s-1}(z)~.
\end{align*}
Incorporating these results, we find the following expression for $t$:
\begin{align} \label{eq:t_solution_D6_special_Equations_f_s}
t \= \ii \b \left(1- \ \sum_{j=1}^\infty \me^{-2\pi j \beta} \sum_{\fp \in \pt(j)} \frac{a_\fp N_\fp}{(4 \pi^2 \b^2 Y)^{\length(\fp)}} f_{\length(\fp)-1}(2 \pi j \b)\right),
\end{align} 
where the polynomials are given by \eqref{eq:D6-D0_P_s_M}, and we have recalled that $y_0 {\,=\,} \b$. Expressing this in terms of the Bessel functions $\bm{k}_n(y)$ gives an equivalent formula, in which the relevance of the Bessel functions is explicit.
\begin{align}
\begin{split}
t &\= \ii y~, \qquad \text{where} \qquad \frac{y}{\b} \= 
1 - \sum_{j=1}^\infty \sum_{\fp \in \pt(j)} a_\fp N_\fp \left( \frac{j}{2 \pi \b Y} \right)^{\length(\fp)} \bm{k}_{\length(\fp)-1}(2 \pi j \b)~.
\end{split} \label{eq:t_solution_D6_special_Equations}
\end{align}
Summarising, we have found the fully instanton-corrected solutions for the attractor point coordinate $t$ when the charge vector $Q$ takes the form
\begin{align} \label{eq:Q_D6_Special_Equations}
Q \= \kappa\begin{pmatrix}
0\\
\Upsilon\\
1\\
0
\end{pmatrix}.
\end{align}
Strictly speaking, as we had to resort to numerical experimentation to arrive at this result, the result remains a conjecture rather than a rigorously proven result. We have, however, verified the validity of \eqref{eq:t_solution_D6_special_Equations} formula in a number of cases, involving both rank one and rank two attractor points, to a numerical accuracy of at least 50 digits, with sums containing over $8.1 {\,\times\,} 10^7$ partitions. For the rank two attractor point of \cite{Candelas:2019llw}, using 11 iterations of a Shanks transformation, we find agreement to 80 digits. Rank one attractor points can be found arbitrarily close to the large complex structure point, which makes it easy to find agreement to hundreds of digits in many cases.

These formulae can be applied for any one-parameter family of Calabi-Yau manifolds by substitution of the Gromov-Witten invariants and the correct $Y_{ijk}$. However, care must be taken as not all charge vectors correspond to $t_0$ that lies near the large complex structure point. Below, in appendix \ref{sect:Asymptotics_Convergence}, we discuss the conditions for the series \eqref{eq:t_solution_D6_special_Equations_f_s} and \eqref{eq:t_solution_D6_special_Equations} to converge, and thus lie in the large complex structure region.
\subsection{The D4-D2-D0 orthogonality equation}
\vskip-10pt
In addition to the alignment equations, we can also find similarly simple solutions to the D4-D2-D0 orthogonality equation. In this case we do not have to make the assumption $x=0$ to simplify the equations, since we can formulate them in terms of a single coordinate $t$. This allows us to use the same methods to reduce the functions $R^t_\fp$ to the polynomials $f_s$.

We found in \sref{sect:D4-system} that the recurrence relation for the rational functions $R^t_\fp$ is given by \eqref{eq:D4_R^t_Recurrence},
\begin{align*} \notag
R^{t}_\fp(t_0) \= \frac{1}{\alpha \wt{\beta} - t_0} \left( \frac{1}{2} (R^t \hskip-1pt \hs R^t)_\fp + \cE_{2,\fp} \right),
\end{align*}
which is reminiscent of \eqref{eq:D6_Special_Case_Simplified_Recurrence}, especially when we recall that $\alpha \wt{\beta} - t_0 = -y_0$. Writing out $\cE_{2,\fp}$ gives
\begin{align*} \notag
\cE_{2,\fp} \= \frac{1}{Y} \sum_{m \in \fp} \sum_{j = 1}^{l(\fp)-1} \frac{(2\pi \ii m)^{j-2}}{j!} \overbrace{{(R^t \hskip-1pt {\hs} \hskip-1pt R^t {\hs} ...  {\hs} R^t)}}^{j \text{ times}}\hskip-2pt{}_{\fp-m}~,
\end{align*}
which differs from the relation \eqref{eq:D6_Special_Case_C_2} only by factors of $\ii$ and the replacement of $R^y_\fp$ by $R^t_\fp$. It follows that the solution to the recurrence relation for $R^t_\fp$ is also determined by the polynomials $f_s$. In~fact,
\begin{align*} \notag
\ii \frac{2 \pi |\fp|}{a_\fp v} \left(\frac{Y v^2}{|\fp|^2}\right)^{\length(\fp)} R^{t_{\,}}_\fp \left(\alpha \wt{\beta} +\ii \frac{v}{2\pi |\fp|} \right) \= f_{\length(\fp)-1}(v)~.
\end{align*}
Similarly to the previous section, the left-hand side of this relation depends on the partitions $\fp$ only through their lengths. 

Using the relation \eqref{eq:D6-D0_P_s_M} between the polynomials $f_s(v)$ and the Bessel functions $\bm{k}_s(v)$, we can give $N$-expansions for $t$:
\begin{align*} \notag
\begin{split}
t &\= t_0 - \ii y_0 \sum_{j=1}^\infty \me^{2\pi \ii j t_0} \!\! \sum_{\fp \in \pt(j)} \frac{a_\fp N_\fp}{(4 \pi^2 y_0^2 Y)^{\length(\fp)}} f_{\length(\fp)-1}(2 \pi j y_0)~,\\[2pt]
&\=t_0 - \ii y_{0}\sum_{j=1}^\infty \me^{2\pi \ii j x_0} \!\! \sum_{\fp \in \pt(j)} a_\fp N_\fp \left( \frac{j}{2 \pi y_0 Y} \right)^{\length(\fp)} \! \bm{k}_{\length(\fp)-1}(2 \pi j y_0)~.
\end{split}
\end{align*}
Recall that the perturbative solution $t_0$ is given by
\begin{align*} \notag
t_0 \= \wt{\beta}\left(\alpha+\sqrt{\alpha^2-1} \right).
\end{align*}
This is very similar to the solution \eqref{eq:t_solution_D6_special_Equations} which we found to the D6-D2-D0 alignment equations in the case $x{\=}0$. The only differences are due to the different zeroth-order solutions and that this expansion is in terms of $q_t$ instead of $q_y$. Generically, of course, the coordinate $t$ has both real and imaginary parts non-zero. However, these are closely related as we see from equation \eqref{eq:D4_orthogonality_equations_solution_x_y}, for example. Finally, note that when $x_0 = 0$ but $y_0 \neq 0$, that is when $\alpha = 0$, this actually reduces to solution \eqref{eq:t_solution_D6_special_Equations}. 
\subsubsection*{Relation to the D6-D2-D0 alignment equations}
\vskip-10pt
We can explain the striking similarity of the situation we have just discussed to the alignment equations in the D6-D2-D0 case by inspecting the period vector in the special case $t=\ii y$. We have already seen in \sref{sect:Introduction} that, at rank two attractor points, the alignment and orthogonality equations are related. Even in the case of rank one attractor points, a very similar relation holds in certain special cases, such as when $x=0$. In this, the period vector is
\begin{align*} \notag
\Pi \= \begin{pmatrix}
0\\
a(y)\\
1 \\
0
\end{pmatrix} + \ii y \begin{pmatrix}
~b(y) \\
-Y_{110}\\
0 \\
1
\end{pmatrix}, 
\end{align*}
where $a$ and $b$ are real functions of $y$. The existence of an attractor point of the type discussed in \sref{sect:Solutions_D6-D2-D0} is equivalent to $a(y)$ being a rational number, so that, up to an overall scale, $\Re[\Pi]$ is a vector with integral components. Such an attractor point is of rank two if and only if $b(y)$ is also a rational number, as this makes the imaginary part of $\Pi$ also projectively integral. 

Since we have discussed the case of rank two attractor points in \sref{sect:Preamble}, we will concentrate on the case where $b(y)$ is irrational. In this case, there is, up to an overall scale, exactly one vector with integral components that is orthogonal to $\Pi$:
\begin{align*} \notag
Q \= \kappa \begin{pmatrix}
~~a(y) \\
-Y_{110} \\
0 \\
1
\end{pmatrix},
\end{align*}
where $\kappa$ is chosen so as to make $Q$ integral. 

The converse is also true: if there is a solution to the orthogonality equation with the charge vector of this form, $a(y)$ must be rational, which in turn implies that there is a solution to the alignment equations with charge vector
\begin{align*} \notag
Q' \= \kappa' \begin{pmatrix}
0 \\
a(y) \\
1 \\
0
\end{pmatrix},
\end{align*}
which, for suitable $\kappa'$, is integral.

\subsection{Entropy at rank two attractor points}
\vskip-10pt
A particularly interesting consequence of the formula \eqref{eq:t_solution_D6_special_Equations} is that we can compute the area of black holes corresponding to rank two attractor points. At rank two attractor points, the plane~$V$ spanned by $\Re \Pi$ and $\Im \Pi$ intersects the charge lattice in a lattice plane. As a consequence, there is a simple relation between the central charge and $t$, which, combined with earlier formulae, allows us to write a relatively simple expression for the black hole area.

To find the relation, we consider rank two attractor points with a coordinate $t$, corresponding to a charge lattice basis given by
\begin{align*} \notag
Q_1 \=
\kappa \begin{pmatrix}
0\\
\Upsilon\\
1 \\
0
\end{pmatrix}, \qquad Q_2 \= \lambda \begin{pmatrix}
\Lambda\\
0\\
0 \\
1
\end{pmatrix}.
\end{align*}
These two charge vectors satisfying the alignment equations \eqref{eq:Strominger_Equations_Original} at $t$ are charge vectors respectively corresponding to \hbox{D6-D2-D0} and D4-D2-D0 brane configurations\footnote{See \sref{sect:Particular_Brane_Systems} and appendix \ref{app:D-Brane_Interpretation_of_Q} for the relation of charge vectors and branes.}. It is not currently known whether there are more generic brane configurations corresponding to rank two attractor points in the large complex structure region, but the currently known examples \cite{Candelas:2019llw} are of this form. It would nonetheless be easy to repeat the discussion with more general pairs of charge vectors.

For attractor points on the imaginary axis, these charges are proportional to the real and imaginary parts of the period vector, respectively. Therefore we are able to write the charge vectors and so the central charge directly in terms of the period vector. We have
\begin{align*} \notag
Q_1 \= \kappa \Re \, \Pi~, \qquad Q_2 \= \frac{\lambda}{y} \Im \, \Pi~, \qquad \frac{1}{z_0}\Pi \= \frac{1}{\kappa} Q_1 + \ii \, \frac{y}{\lambda} Q_2~.
\end{align*}
Now, consider a black hole with a charge vector $Q_1$. It follows from the expressions above, together with the definition \eqref{eq:Central_Charge_Definition}, that the central charge  $\cZ_1 \defineas \cZ(Q_1)$ satisfies the relations
\begin{align}
|\cZ_1|^2 \= \frac{y}{2}\left| \frac{\kappa}{\lambda} \, Q_1^T \Sigma Q_2\right| \=  \frac{y \kappa^2}{2}|\Lambda-\Upsilon|~, \label{eq:rank_two_central_charge_1}
\end{align}
where $y$ is given by \eqref{eq:t_solution_D6_special_Equations}. In a similar way, a black hole with charge vector $Q_2$ has central charge $\cZ_2 \defineas \cZ(Q_2)$, which satisfies the relation
\begin{align}
|\cZ_2|^2 \= \frac{1}{2 y} \left| \frac{\lambda}{\kappa} Q_1^T \Sigma Q_2\right| \= \frac{\lambda^2}{2y}|\Lambda-\Upsilon|~. \label{eq:rank_two_central_charge_2}
\end{align}
We make a few observations on $\cZ_1$: First, the central charge scales as $\kappa^2$ which means in particular that in the large charge limit $\kappa \to \infty$ the central charge will be proportional to the Wald entropy of the black hole with the charge vector $Q_1$. We will review this in more detail in \sref{sect:Microstate_Counting}. The second important point is that this formula depends only on the charge ratios $\Upsilon$ and $\Lambda$, and the topological data of the Calabi-Yau manifold. Given a pair of charge vectors corresponding to a rank two attractor point, we can immediately find the central charge associated to one of the charges. In practice this formula also gives numerical results to good accuracy, since near the large complex structure point the coefficients $a_\fp$ decrease rapidly and ensure the rapid convergence of the sum.

As discussed in \sref{sect:Preamble}, the central charge is directly related to the Bekenstein-Hawking entropy in the supergravity theory with at most two-derivative corrections as
\begin{align*} \notag
S_{\text{BH}}(Q_1) \= \pi |\cZ_1|^2~.
\end{align*}
We discuss this further in \sref{sect:Applications}.

\subsection{An example: a rank two attractor point on AESZ34}
\vskip-10pt
It was found in \cite{Candelas:2019llw} that there is a rank two attractor point in the large complex structure region, for a one-parameter family of Calabi-Yau manifolds which is the number 34 on the AESZ list \cite{Almkvist:2010}. This gives an explicit example of a very interesting attractor point which we can study using the methods developed in the preceding sections. By studying the zeta function and its factorisation, the authors of \cite{Candelas:2019llw} found that there is a rank two attractor point located at
\begin{align*} \notag
\varphi \= 33 - 8 \sqrt{17} \= 0.0151...
\end{align*}
which we should compare with the $\varphi$ value at the nearest singularity to the origin, $\varphi {\,=\,} \frac{1}{25} {\,=\,} 0.04$. Therefore this lies in the large complex structure region in the sense discussed in \sref{sect:Preamble}. It will also be argued in appendix \ref{sect:Asymptotics_Convergence} that the series \eqref{eq:t_solution_D6_special_Equations} converges for the corresponding value of $y_0$.

Using the mirror map, the $t$ coordinate corresponding to the attractor point was found to be
\begin{align*} \notag
t \= \ii \frac{5}{16 \cdot 17} (9 + \sqrt{17}) \frac{\pi \lambda_4(1)}{\lambda_4(2)}~,
\end{align*}
where $\lambda_4$ is the real part of the L-function associated to the modular form with the LMFDB \cite{LMFDB} designation \textbf{34.4.b.a}. 

The charges associated to this attractor point are 
\begin{align*} \notag
Q_1 \= \kappa \begin{pmatrix}
0\\
3 c \\
1\\
0
\end{pmatrix}, \qquad Q_2 \= \kappa \begin{pmatrix}
-2 c \\
0 \\
0\\
5
\end{pmatrix},
\end{align*}
where $c=1$ or $c=2$ correspond to taking different quotients in the construction of the manifold (see \cite{Candelas:2019llw}). Without any essential loss of generality, we take $c=1$ for the following discussion. We note that $Q_1$ has exactly the form \eqref{eq:Q_D6_Special_Equations}, which makes it possible to apply formula \eqref{eq:t_solution_D6_special_Equations} to find both the attractor point location and the value of the central charge at this attractor point. To specialise to AESZ34 we need only to substitute in these charges, the Yukawa couplings, and the Gromov-Witten invariants. 

In this case, the quantities appearing in the perturbative part of the prepotential are given by
\begin{align*} \notag
Y \= 12~, \ \ Y_{110} \= 0~, \ \ Y_{100} \= -1~, \ \ \sigma \= - \frac{2\zeta(3)}{(2\pi )^3}~,
\end{align*}
To $Q_1$ correspond a set of values for our shorthand Greek symbols.
\begin{equation*} \notag
\begin{aligned}
\alpha_{1} &\= \frac{\sqrt{3}}{2} \ii~, & \wt\beta_{1} &\= \frac{\sqrt{3}}{6}\ii~, & \beta_{1} &\= \frac{\sqrt{15}}{6}~, & \gamma_{1} &\= 0~.\\[5pt]
%\alpha_{2} &\= 0~, & \wt\beta_{2} &\= \frac{\sqrt{15}}{10}\ii~, & \beta_{2} &\= \frac{\sqrt{3}}{6}\ii~, & \gamma_{2} &\= \frac{12\sqrt{3}}{6}\ii~.
\end{aligned}
\end{equation*}
The first few instanton numbers $n_k$ and scaled Gromov-Witten invariants are listed in \tref{tab:GW-invariants}.
\vskip3pt
\begin{table}[H]
	\begin{center}
		\begin{tabular}{|l|l|l|}
			\hline
			\vrule height14pt depth7pt width0pt \hfil \textbf{$k$} & \hfil \textbf{$n_k$} & \hfil \textbf{$N_k$}          \\ \hline \hline
			\vrule height12pt width0pt 1 & 12 & 12           \\
			2 & 24 & 204          \\
			3 & 112 & 3036         \\
			4 & 624 & 40140        \\
			5 & 4200 & 525012       \\
			6 & 31408 & 6787356      \\
			7 & 258168 & 88551636     \\
			8 & 2269848 & 1162202316   \\
			9 & 21011260 & 15317211576  \\
		   10 & 202527600 & 202528125204 \\[3pt] \hline
		\end{tabular} \vskip5pt
		\capt{3.5in}{tab:GW-invariants}{The first 10 instanton numbers $n_k$ and scaled Gromov-Witten invariants $N_k$ for the manifold AESZ34.}
	\end{center}
\end{table}
The perturbative solution is given by 
\vskip-25pt
\begin{align*} \notag
y_0 \= \beta_{1} \= \frac{\sqrt{15}}{6} ~.
\end{align*}
Our formulae imply the following relation between the L-function values and the Gromov-Witten invariants:
\begin{align}
\frac{1}{17}(9+\sqrt{17})\frac{\pi \lambda_4(1)}{\lambda_4(2)} \= \frac{8}{\sqrt{15}} \left\{1 {-} \sum_{j=1}^\infty \sum_{\fp \in \pt(j)}a_\fp N_\fp \left(\frac{j}{4 \pi \sqrt{15}} \right)^{\length(\fp)} \bm{k}_{\length(\fp)-1}\left(\frac{\sqrt{15}}{3} \pi j \right) \right\}. \label{eq:L-k_indentity}
\end{align}
This remarkable identity works also as a check of our results. We can compute the sum on the right-hand side as far as $j=75$, which includes over $8.1 {\,\times\,} 10^7$ partitions and is enough to give us accuracy of over 55 digits. We see that there is indeed an equality, to this accuracy. After applying eleven Shanks transformations to our sequence of partial sums we get agreement to 80 digits. The above identity is also an interesting companion to the those relating to the rank two attractor points on AESZ34 found in \cite{Candelas:2019llw}. 

A curious aspect of this relation is that the quantities on the left-hand side, the coordinate values and L-function values, correspond to the number field $\IQ(\sqrt{17})$ while the right-hand side involves $y_0$, which is an element of of $\IQ(\sqrt{15})$.
\vskip5pt
\begin{figure}[H]
	\centering
		\begin{tikzpicture}
			\usetikzlibrary{arrows.meta}

			\draw[{Latex[length=3mm, width=2mm]}-{Latex[length=3mm, width=2mm]}] (2,3.7)--(4,3.7);	
			\node[] at (5.2,3.9) {$\zeta$-function};		
			\node[] at (5.2,3.5) {factorisation};	

			\node[] at (0,4) {attractor point at};			
			\node[] at (0,3.5) {$\varphi = 33 - 8 \sqrt{17}$};
			
			\node[] at (1.2,2) {mirror map};			
			\draw[{Latex[length=3mm, width=2mm]}-{Latex[length=3mm, width=2mm]}] (0,1)--(0,3);
			\node[] at (0,0.5) {attractor point at};			
			\node[] at (0,-0.1) {$t = \ii \frac{5}{16 \cdot 17} \scriptstyle{(9 + \sqrt{17})} \frac{\pi \lambda_4(1)}{\lambda_4(2)}$};	
				
			\node[] at (3,0.7) {\eqref{eq:t_solution_D6_special_Equations}};
			\draw[{Latex[length=3mm, width=2mm]}-{Latex[length=3mm, width=2mm]}] (2,0.3)--(4,0.3);	
			\node[] at (5.3,0.3) {$Q = \kappa \begin{pmatrix}
				0\\
				3 c \\
				1\\
				0
				\end{pmatrix}$};		
		\end{tikzpicture}
	\vskip10pt
	\capt{6in}{fig:Modularity_Relations}{Summary of the relations between different quantities associated to the rank two attractor point at $\varphi = 33 - 8 \sqrt{17}$. Roughly speaking, the method used in \cite{Candelas:2019llw} to study the attractor point moves anti-clockwise in the figure whereas our method moves clockwise.}	
\end{figure}
\newpage

\section{Applications}\label{sect:Applications}
\vskip-10pt
To conclude, we give here two simple applications of the formulae we have found and take the opportunity to explore some of their qualitative properties. The first application is to classify the attractor points into those with vanishing and non-vanishing central charges, based on the values of the black hole charge ratios $\Upsilon$ and $\Lambda$. These two types of attractor points have distinct physical interpretations: It is conjectured that only the attractor points with $|\cZ| \neq 0$ have corresponding BPS states in the full quantum theory of gravity \cite{Denef:2007vg,Denef:2000nb,Denef:2001xn,Moore:1998pn,Moore:1998zu}, while the attractor points with $|\cZ|=0$ are related to supersymmetric flux compactifications \cite{Kachru:2020abh,Kachru:2020sio}. In addition, knowing which charges have an attractor point with vanishing central charge can be useful for analysing split attractor flows \cite{Denef:2007vg,Collinucci:2008ht}.

Another, more speculative, application arises from comparing the entropy formulae derived in the previous sections to the counting of microstates. We argue that the correct interpretation is to consider them the leading order contributions in the large charge expansion. The large complex structure limit is interesting for analysing the microscopic description of the black holes, as in this region the black holes can be thought of as a bound state of D-branes \cite{Maldacena:1996ky}. These branes, in turn, can be realised as divisors, curves or points on the Calabi-Yau manifold. The microstate indices can then be given by the Witten indices of moduli spaces of these curves, which appear as coefficients of a modified elliptic genus \cite{Gaiotto:2006wm,Gaiotto:2007cd}. 

Here we show that our formulae, which converge rapidly in the appropriate large charge ratio limits, also reproduce the correct leading order expressions in the large charge limit, so reproducing the well-known results of \cite{Maldacena:1997de,Shmakova:1996nz} in the regime where these results are simultaneously valid. More interesting, however, is the fact that our expressions provide a systematic expansion that includes the perturbative corrections in $\sigma$ and the exponentially small instanton contributions. It is intriguing to ponder on how this expansion relates to the celebrated Rademacher expansion \cite{Rademacher78}, which provides expressions for the coefficients of modular forms from which asymptotics can be read off. We hope to return elsewhere to the relation between these expansions, which we do not attempt to match here. 

\subsection{Classification of attractor points}
\vskip-10pt \label{sect:Classification_of_Attractor_Points}
We have been considering the $N$-expansions relating to the D6-D2-D0 and D4-D2-D0 systems, and have assumed that the value of $y_0$ is large enough for the $N$-expansions to converge. Recall that each such a solution devolves from either a solution of the alignment or orthogonality equations. Moreover, every solution to the perturbative equations, with large enough $y_0$, leads to an associated solution to the fully instanton-corrected equations, having $t_0$ as a zeroth-order term. From these considerations, we can draw a conclusion: the central charge vanishes if and only if $t_0$ is a perturbative solution to the orthogonality equation.

The classification of attractor points into those with vanishing and non-vanishing central charges is of central importance if the BPS state existence conjecture holds \cite{Denef:2007vg,Denef:2000nb,Denef:2001xn,Moore:1998pn,Moore:1998zu}. As discussed in greater detail in \sref{sect:Preamble}, the conjecture states that the existence of BPS states in the full quantum theory of gravity can be deduced from the existence of attractor flows.

We concentrate on the case where at least one of the charge ratios $\Lambda$ or $\Upsilon$ is large, and further $\big\vert|\Upsilon|-|\Lambda|\big\vert$ is also large. In this regime we inevitably approach the large complex structure limit, which allows us to make statements that are true for any one-parameter family. We are then able to derive simple conditions for existence of attractor points with large complex structure. However, the fact that we are limited to the large complex structure region has the consequence that we cannot make meaningful statements about flows that terminate elsewhere.

We should point out an important complication due to split attractor flows. These are flows which split into multiple branches \cite{Denef:2000ar,Denef:2000nb,Denef:2001xn,Denef:2002ru,Denef:2007vg}, due to the existence of multi-centred black holes. Owing to both the existence of split flows and the possible existence of different basins of attraction for single attractor flows, the results here do not completely determine the BPS spectrum of one-parameter Calabi-Yau manifolds, even in the large charge ratio limits. For example, even if an attractor point with a non-vanishing central charge exists near the large complex structure point, the basin of attraction for this point is not necessarily the whole large complex structure region; there could be another basin of attraction that flows to an attractor point outside the large complex structure region. Such a point does not enter our classification here. Even within the basin of attraction of an attractor point included in the classification below, there can exist split flows with at least one branch ending at an attractor point that does not lie within the large complex structure region.

\begin{figure}[H]
	\centering
	\framebox[.55\textwidth]{
		\includegraphics[width=0.50\textwidth]{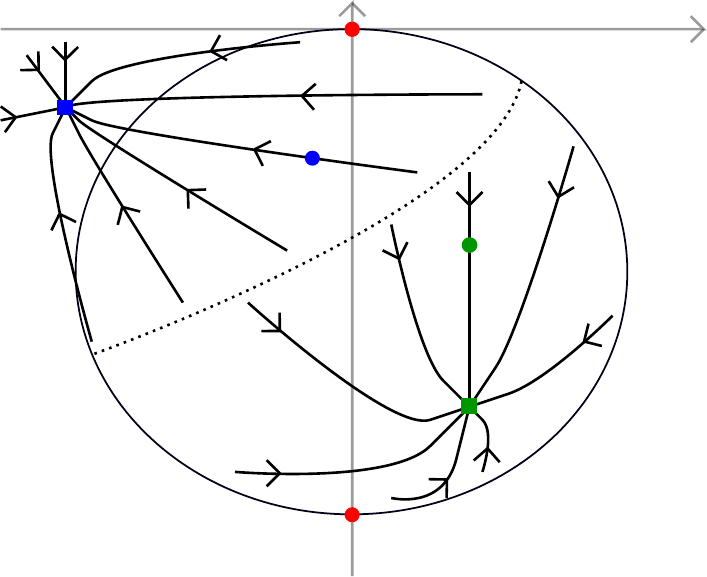}}
	\vskip5pt	
	\capt{6.1in}{fig:basin_boundary_sketch}{A schematic picture of a basin boundary inside the large complex structure region. The large complex structure region is represented by a circle on which the large complex structure singularity and the nearest other singularity lie. It is possible that there are two attractor points corresponding to the same charge, one inside the large complex structure region, and one outside it. Further, there can exist a basin boundary inside the large complex structure region, represented here by a dotted line. On one side of this boundary the attractor flows end at an attractor point inside the large complex structure region, represented by a green square. On the other side, the flows end at an attractor point outside the large complex structure region, corresponding to a blue square.}  	
\end{figure}
\subsubsection*{The D4-D2-D0 system}
\vskip-10pt
The perturbative solution to the D4-D2-D0 alignment equations is given by \eqref{eq:D4quantum}, \eqref{eq:D4_y_0_Solution}, and \eqref{eq:D4_theta_defn}:
\begin{align*} \notag
\begin{split}
x_0 \= \alpha \wt{\beta}~, \qquad \text{and} \qquad y_0 \= 2 \wt{\beta} \sqrt{\alpha^2 - 1} \sin \theta~, \quad \text{with} \quad \sin 3 \theta \= \frac{\sigma}{\wt{\beta}^3 (\alpha^2-1)^{3/2}}~.
\end{split}
\end{align*}
In the large charge ratio limits where $|\wt{\beta}|$ or $|\alpha|$ are large, $\sin 3 \theta$ is small. It was shown in \sref{sect:D4_Strominger_Equations} that the solution, with $y_0$ large, in this limit is given by $\th = \frac{\pi}{3}$, not by $\theta = 0$. So we have 
\begin{align*} \notag
x_0 \= \alpha \wt{\beta}~, \qquad y_0 \= \wt{\beta} \sqrt{3(\alpha^2-1)} + \cO\left(\frac{\sigma}{\wt{\beta}^3 (\alpha^2-1)^{3/2}} \right),
\end{align*} 
where the full expansion of $y_0$ as a series in $\sigma$ is given by \eqref{eq:d6zeroquantumcoefficients}.

In order for this to be a valid solution that lies near the large complex structure limit, both $x_0$ and $y_0$ need to be real, and in addition $y_0$ should be large. Writing out the shorthand in terms of the manifestly real $Y$, $Y_{100}$, $Y_{110}$, and the charge ratios, we see that $x_0$ is real for any charge ratios $\Lambda,\Upsilon$ and any Yukawa couplings. The condition for $y_0$ to be large and real becomes
\begin{align}
(Y_{110}+\Upsilon)^2 \; \gg \; Y(Y_{100} +2 \Lambda)~. \label{eq:D4_alignment_condition}
\end{align}
Here and in the following we use $A \gg B$ to indicate that for every one-parameter Calabi-Yau manifold, there exists a manifold-specific constant $C$ so that when $A-B > C$, then an attractor point exists in the large complex structure region.

We can treat the orthogonality equation similarly. In this case, $\sigma$ does not enter explicitly, and the perturbative solution to the orthogonality equation is simply given by \eqref{eq:D4zero}. We require again that the imaginary part of $t_0$ is positive and large. As noted previously, $\alpha \wt{\beta}$ is manifestly real, so the condition for the imaginary part to be large is
\begin{align*} \notag
\wt{\beta}^2(\alpha^2-1) \; \ll \; 0~.
\end{align*}
Expanding the shorthand, we find the condition that is the opposite to \eqref{eq:D4_alignment_condition}:
\begin{align*} \notag
Y(Y_{100} + 2 \Lambda) \; \gg \; (Y_{110} + \Upsilon)^2~.
\end{align*}
Thus we find the following conditions for the existence of attractor points in the large charge ratio limits we are considering.
\begin{table}[H]
	\begin{center}
		\renewcommand{\arraystretch}{1.3}
		\begin{tabular}{|c|c|c|}
			\hline
			Limit & $|\cZ|=0$ & $|\cZ| \neq 0$ \\ \hline \hline
			$\Upsilon^2 \gg 2Y\Lambda$ &  \xmark   &  \cmark    \\ \hline
			$\Upsilon^2 \ll 2Y\Lambda$ &  \cmark   &  \xmark    \\ \hline
			
		\end{tabular}\\
		\vskip10pt
		\capt{5in}{tab:D4_BPS_Existence}{Large charge ratio limits for which an attractor point near the large complex structure point is guaranteed to exists in the D4-D2-D0 case.}
	\end{center}
\end{table}
\vskip-20pt
Significantly, in large charge ratio limits where $\big\vert|\Upsilon|-|\Lambda|\big\vert$ is sufficiently large, one of the conditions in the table is always satisfied, and the solutions to the orthogonality and alignment equations are mutually exclusive. If a solution with vanishing central charge exists then there is no solution with non-vanishing central charge and vice versa. While one may expect a result like this based on the BPS state existence conjecture, a different result would not necessarily contradict the conjecture as there can in principle be basin boundaries in the moduli space. Crossing these walls can change the BPS spectrum discontinuously. Thus having two attractor points for the same charge near the large complex structure point with both vanishing and non-vanishing central charge would not violate the conjecture, but rather imply that there is a boundary between the basins of attraction of the two distinct attractor points. However, our analysis implies that sufficiently near the large complex structure we have only one attractor point, which can be of either kind, and no such wall exists.

We can illustrate the conditions for the existence of solutions to the alignment equations in the case of AESZ34 by using the recurrence relations uncovered in \sref{sect:D4_Strominger_Equations} to solve the attractor equations to order 10 in $q_y$. We can then plot families of solutions where we fix one of $\Lambda$ or $\Upsilon$ and vary the other. We will also plot the locations of some attractor points. The exact form these plots take depend on the choice of the family of Calabi-Yau manifolds, but most of the qualitative properties are similar for any family of manifolds. 

In \fref{fig:D4_Strominger_Equation_Solutions_Fixed_Upsilon} (i), we plot $1/t$, the reciprocal of the attractor point coordinate, keeping $\Lambda$ fixed and varying $\Upsilon$ from $-1000$ to $1000$. This illustrates several qualitative aspects of the solutions. In particular, as $\Lambda$ approaches $\infty$ or $-\infty$ or $\Upsilon \to \infty$, the attractor points approach the large complex structure point at $t = \infty$. For $\Lambda <0$, the curves traced out as $\Upsilon$ varies are closed, while for $\Lambda >0$ the curves end on the boundary of the large complex structure region, beyond which a solution no longer exists. These endpoints are in accord with \tref{tab:D4_BPS_Existence}, from which we see that if $\Lambda$ too greatly exceeds $|\Upsilon|$, then there are no solutions to the alignment equations. Finally, note that the plot is symmetric about the vertical axis following from the fact that when $Y_{110}=0$, $x_0$ is proportional to $\Upsilon$ and $y_0$ depends on $\Upsilon$ only through $\Upsilon^2$.

To give another perspective on the D4-D2-D0 alignment equations, we can draw a similar figure but now keep $\Upsilon$ fixed and vary $\Lambda$. In \fref{fig:D4_Strominger_Equation_Solutions_Fixed_Upsilon} (ii), we let $\Upsilon$ to vary from $-10 000$ to $10 000$ and plot the solutions where they exist. We notice again some of the properties listed in \tref{tab:D4_BPS_Existence} reflected in the plot. For example, we see that solutions do not exist when $\Lambda$ greatly exceeds $\Upsilon$ and $\Lambda > 0$, while large negative $\Lambda$, they exist for all values of $\Upsilon$. As expected, we also see the attractor points tending towards the origin for $\Upsilon \to - \infty$ and $|\Lambda| \to \infty$. We also note the symmetry under $\Upsilon \to -\Upsilon$, which this time maps one family of solutions to another.
\begin{figure}[!htb]
	\centering
	\begin{minipage}{.47\textwidth}
		\vskip10pt
		\centering
		\includegraphics[width=0.97\linewidth, height=0.3\textheight]{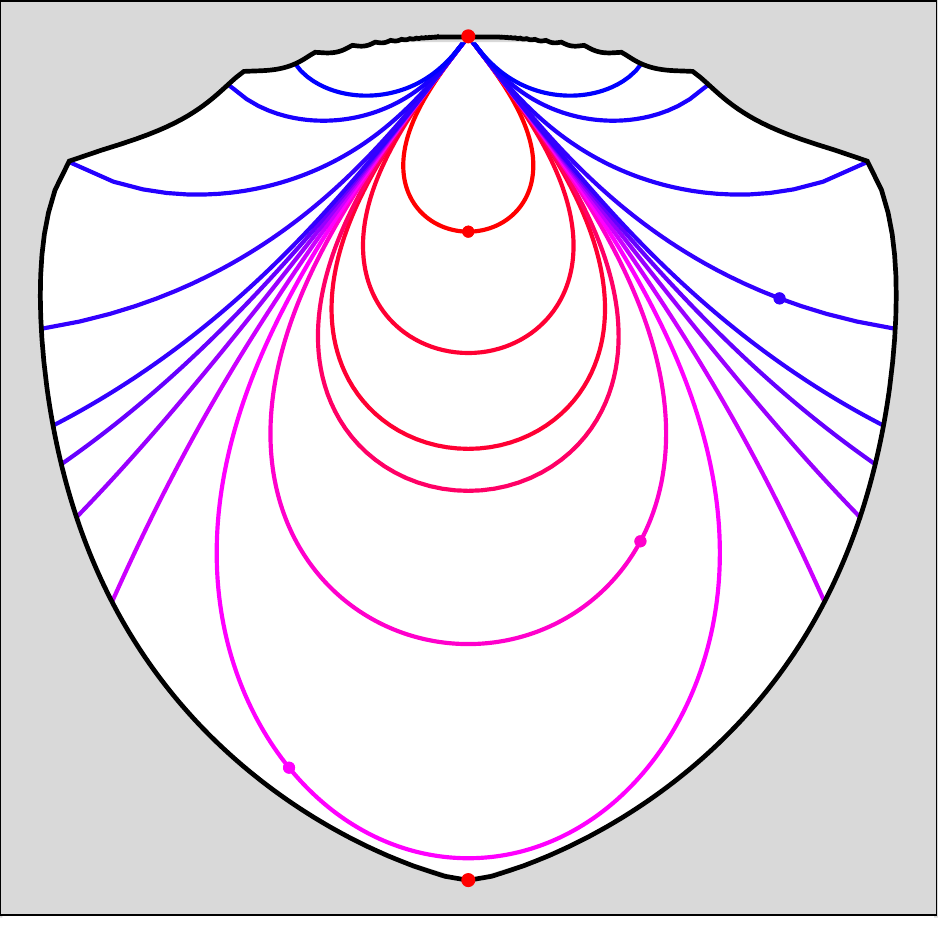}
		\caption*{}
	\end{minipage}
	\begin{minipage}{.47\textwidth}
		\vskip10pt		
		\centering
		\includegraphics[width=0.97\linewidth, height=0.3\textheight]{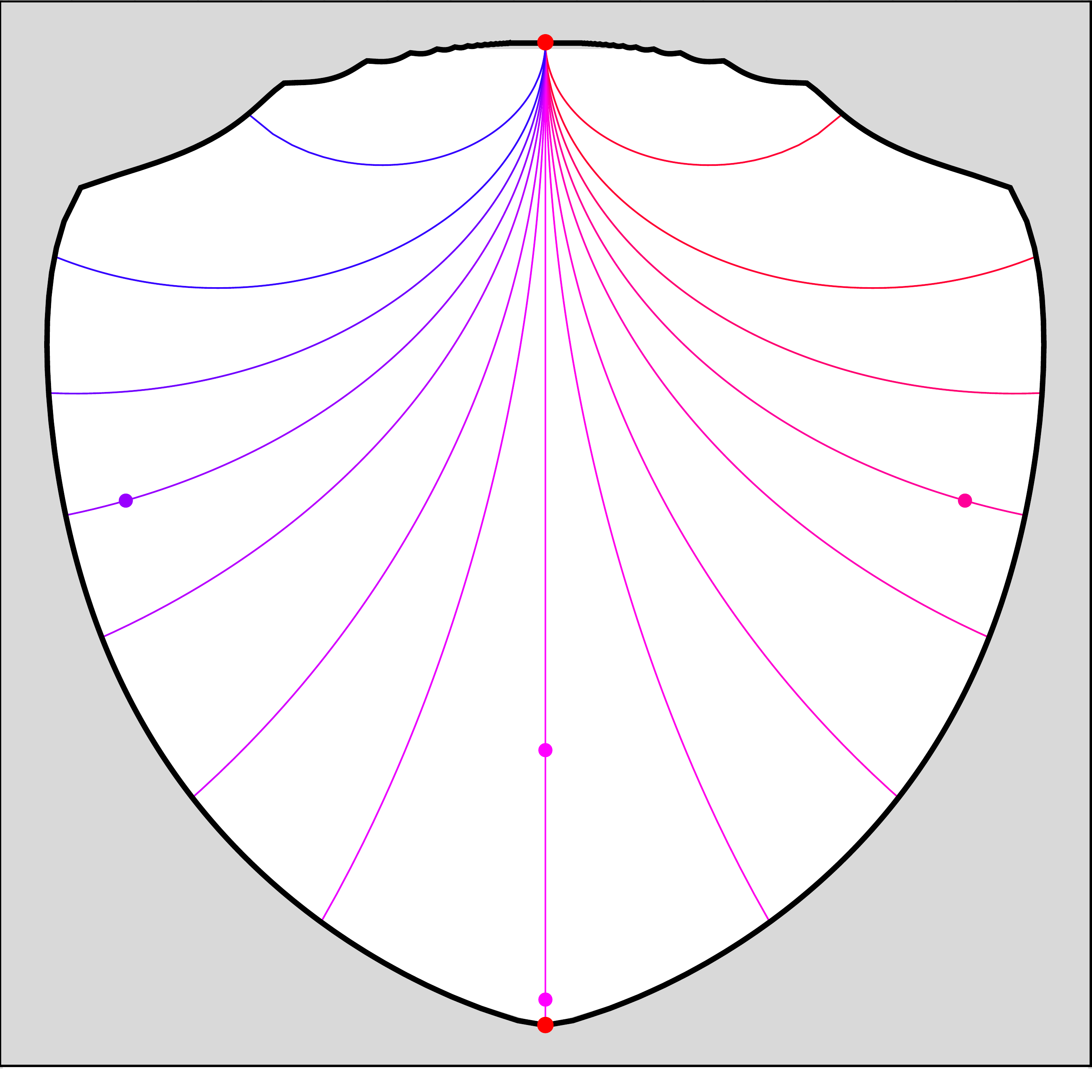}
		\caption*{}		
	\end{minipage}\vskip-20pt
	\place{1.48}{2.02}{\small{$(-6,0)$}}
	\place{1.6}{1.25}{\small{$(-1/4,2)$}}
	\place{1.15}{0.67}{\small{$(0,1)$}}
	\place{2.57}{2.07}{\small{$(2,-8)$}}	
	
	\place{5.55}{1.73}{\small{$(3/4,-5)$}}
	\place{4.83}{0.42}{\small{$(0,0)$}}
	\place{4.83}{1.05}{\small{$(-1/4,0)$}}	
	\place{3.5}{1.73}{\small{$(3/4,5)$}}			
	
	\place{0.3}{0.35}{i}
	\place{3.4}{0.35}{ii}						
	\capt{6in}{fig:D4_Strominger_Equation_Solutions_Fixed_Upsilon}{i) Families of solutions to the D4-D2-D0 alignment equations on the $1/t$ plane. The origin is the large complex structure point, and the large complex structure region (see \sref{sect:Preamble}) has the white background. On each curve, $\Lambda$ is fixed with $\Upsilon$ ranging from $-1000$ to $1000$. For large positive values of $\Lambda$, solutions do not exist when $|\Upsilon|$ is small. $\Lambda$ takes values $\Lambda \in \{-6,-2,-1,-3/4,-1/4,0,1/4,1/2,3/4,1,2,6,20,40 \}$. We have also highlighted some points along the curves. The labels are the values of the charge ratios in the form $(\Lambda,\Upsilon)$. The two unlabelled red points denote the singularities at $t = \infty$ and $t = \ii \, y_{\text{\emph{sing}}}\approx 2.4 \ii$.\\[5pt]
		ii) Here $\Upsilon$ is kept fixed on each curve, and $\Lambda$ varies from -10 000 to 10 000. In addition, we have highlighted some attractor points on the curves. These are labelled by the charge ratios $(\Upsilon,\Lambda)$. Corresponding to different curves, $\Upsilon$ takes the values $\Upsilon \in \pm \{0,1,2,7/2,5,7,10,20\}$.}
	\vskip-10pt
\end{figure}
\subsubsection*{D6-D2-D0 system}
\vskip-10pt
The perturbative solution to the alignment equations \eqref{eq:D6quantum} is given by one of the somewhat complicated expressions \eqref{eq:D6_Alignment_Radical}, \eqref{eq:d6a_caseiii_sol12}, and \eqref{eq:d6a_caseii_sol}. While it is possible to use these expressions to study the existence of solutions with large imaginary part, it is more convenient to directly work on the equations \eqref{eq:D6quantum} in different limits, 
\begin{align*} \notag
x_{0}y_{0}&\=\beta\gamma y_{0}+\frac{\sigma}{\beta^{2}}x_{0}~,&
x_{0}^2+y_{0}^2&\=\beta^{2}~.
\end{align*}
The different cases are $(|\Lambda|,|\Upsilon|) {\; \to \; } (\infty,c),\,$ $(|\Lambda|,|\Upsilon|)  {\; \to \; }  (\infty,\infty)$ and $(|\Lambda|,|\Upsilon|)  {\; \to \; }  (c,\infty)$, with $c$ a finite, possibly zero, constant. It is straightforward to check that in all these cases, we can neglect $\sigma$. Thus we have, in all cases that
\begin{align*} \notag
x_{0}&\=\beta\gamma~, \qquad y_{0}\=\beta\sqrt{1-\gamma^2}~,
\end{align*}
where $\gamma$ can be zero.

The combination $\beta \gamma$ is manifestly real, so to have a valid solution near the large complex structure point, we need to demand that $y_{0}$ is real, large, and positive. This requires that
\begin{align} \label{eq:D6_alignment_large_charge_ratio_condition}
(2 \Upsilon + Y_{100})^3 \; \gg \; 9 Y \Lambda^2~.   
\end{align} \vskip-10pt
We can analyse the orthogonality equation in a similar manner. With instantons neglected, the complex coordinate $t_0$ satisfies the equation \eqref{eq:D6zerotrig}. We can again study the three same large charge ratio limits as earlier, and easily see we can again neglect $\sigma$ in this context. Thus we can use the results collected in \tref{tab:D6_Orthogonality_x00}; When $\beta \in \IR^+$, the conditions for $y_{00}$ to be large and positive are either $|\gamma|$ large, or $\beta$ large with $|\gamma|>1$, whereas when $\beta \in \ii \IR^+$, we just need that either $|\beta|$ or $|\gamma|$ is large. In the limits we consider, these four conditions can be equivalently expressed as a single condition \vskip-22pt
\begin{align*} \notag
9Y\Lambda^2 \; \gg \; \left(Y_{100}+2\Upsilon \right)^{3}.
\end{align*}\vskip-10pt
Together with \eqref{eq:D6_alignment_large_charge_ratio_condition}, we find again a simple set of conditions collected in \tref{tab:D6_BPS_Existence}.
\vskip-5pt
\begin{table}[H]
	\begin{center}
		\renewcommand{\arraystretch}{1.3}
		\begin{tabular}{|c|c|c|}
			\hline
			Limit & $|\cZ|=0$ & $|\cZ| \neq 0$ \\ \hline \hline
			$9Y\Lambda^2 \ll \Upsilon^{3}$ &  \xmark   &  \cmark    \\ \hline
			$9Y\Lambda^2 \gg \Upsilon^{3}$ &  \cmark   &  \xmark    \\ \hline	
		\end{tabular}\\
		\vskip10pt
		\capt{5in}{tab:D6_BPS_Existence}{Large charge ratio limits for which an attractor point near the large complex structure point is guaranteed to exists in the D4-D2-D0 case, together with the type of the corresponding attractor point.}
	\end{center}
\end{table}
We see that the solutions with $|\cZ|{\,=\,}0$ and $|\cZ| {\,\neq\,} 0$ are again mutually exlusive in the same manner as in the D4-D2-D0 case. This implies that at least in the cases we consider here, for every charge vector, an attractor point is uniquely of either type, and the type can be read from the charge~vector. 
\subsection{Comparison to microscopic entropy} \label{sect:Microstate_Counting}
\vskip-10pt
The attractor points we have considered are related to solutions to the two-derivative $\cN{\=}2$, $d{\=}4$ supergravity action, with the prepotential appearing in the supergravity theory identified with the prepotential of the Calabi-Yau manifold. However, in general, the effective actions of string- and M-theories have an infinite series of higher-derivative terms. A subset of these corrections \cite{Antoniadis:1993ze} are determined by a holomorphic function $\cF(z,W^2)$, called the \emph{all-genus prepotential}, which depends on not only on the vector multiplet scalars $z$, but also on the Weyl multiplet, which is a chiral multiplet $W$ (for details, see for example~\cite{Behrndt:1998eq}).

The function $\ccF(z,W^2)$ can be expanded as a power series in $W^2$:
\begin{align}
\ccF(z,W^2) &\= \sum_{g=0}^\infty \cF^{(g)}(z) W^{2g}~, \label{eq:All-Genus_Prepotential_Expansion}
\end{align}
with $\cF^{(g)}(z)$ identified as the \textit{genus $g$ prepotential}. In particular, the genus 0 term $\cF^{(0)}(z)$ is the prepotential $\cF(z)$ that we have been using thus far. 

An important point is that the all-genus prepotential has the homogeneity property
\begin{align}\label{eq:All-Genus_Prepotential_Homogeneity}
\ccF(\lambda z, \lambda^2 W^2) \= \lambda^2 \ccF(z,W^2)~.
\end{align}
From the homogeneity property \eqref{eq:All-Genus_Prepotential_Homogeneity} of the all-genus prepotential and the expansion \eqref{eq:All-Genus_Prepotential_Expansion}, it follows~that 
\begin{align*} \notag
\cF^{(g)}(\lambda z) \= \lambda^{2-2g} \cF^{(g)}(z)~.
\end{align*} 
In \cite{Behrndt:1998eq}, it is argued that to take the higher-genus corrections into account, the alignment equations in \eqref{eq:Strominger_Equations} have to be modified to take the form
\begin{align} \label{eq:stabilisation_equations}
Q \= \left(\begin{array}{c}
q_0\\q_{\phantom{0}}\\p^0\\p^{\phantom{0}}
\end{array}\right) \= \Im \left[ C \begin{pmatrix}
\partial_0 \ccF(z,W^2)\\
\partial_1 \ccF(z,W^2)\\
z^0\\
z^1
\end{pmatrix} \right],
\end{align}
where 
\begin{align*} \notag
C \= 2\me^{\ccK/2} \, \overline{\cZ}~,
\end{align*} 
with $\ccK$ an all-genus K\"ahler potential, which generalises the genus zero K\"ahler potential of special geometry:
\begin{align*} \notag
\me^{-\ccK} \; \defineas \;  \ii \left( \overline{z}^a \partial_a \ccF(z,W^2) - z^a \overline{\partial_a \ccF(z,W^2)} \right).
\end{align*}
The authors of \cite{LopesCardoso:1999cv} show that by defining the rescaled variables
\begin{align*}
\wt{Z}^a \= C z^a, \qquad \wt{W}^2 \=  C^2 W^2, \nonumber
\end{align*}
the equations \eqref{eq:stabilisation_equations} become simply 
\begin{align*} \notag
Q \= 2 \, \Im \begin{pmatrix}
\partial_0 \ccF(\wt{Z},\wt{W}^2)\\[3pt]
\partial_1 \ccF(\wt{Z},\wt{W}^2)\\[3pt]
\wt{Z}^0\\[3pt]
\wt{Z}^1
\end{pmatrix}, \qquad \wt{W}^2 \= -64~.
\end{align*}
From this it follows that if we make replacement $Q \to \lambda Q$ with $\lambda$ large, then we need to scale $\Im \, \wt{Z}^a \to \lambda \, \Im \,\wt{Z}^a$ to preserve equality. The real parts must be chosen to satisfy the two remaining equations. By using the explicit expressions for the polynomial part of $\cF$, it can be shown that for large real $\lambda$, the real part of $\wt{Z}^a$ scales also as $\Re \, \wt{Z}^a \to \lambda \, \Re \, \wt{Z}^a$. Thus the all-genus expansion behaves, for large $\lambda$, as
\begin{align*} \notag
\ccF(\lambda \wt{Z},\wt{W}^2) \= \sum_{g=0}^\infty \lambda^{2-2g} \cF(\wt{Z}) \wt{W}^{2g}~,
\end{align*}
which allows us to take $\ccF(\lambda \wt{Z},\wt{W}^2) \sim \lambda^2 \cF^{(0)}(\wt{Z}) = \lambda^2 \cF(\wt{Z})$ in the large charge limit $\lambda \to \infty$.

Using this and going back to the original unscaled variables, the stabilisation equations can be written in the large charge limit as
\begin{align*} \notag
Q \= \Im \left[C \begin{pmatrix}
\frac{\partial}{\partial z^0} \cF^{(0)}\\[5pt]
\frac{\partial}{\partial z^1} \cF^{(0)}\\[3pt]
z^0\\
z^1
\end{pmatrix} \right].
\end{align*}
This is of course nothing but the alignment equations \eqref{eq:Strominger_Equations_Original}. The upshot is that the genus-corrected equations \eqref{eq:stabilisation_equations} reduce to the alignment equations in the large charge limit. 

This observation is important not only for the consistency, but also a key to relating our results for the central charge to the microscopic entropy of a black hole. In \cite{LopesCardoso:1999cv}, it was shown that the Wald entropy for a black hole in four-dimensional $\cN=2$ supergravity takes the form
\begin{align*} \notag
S_{\text{Wald}} \= \pi |\cZ|^2 - 256 \pi \, \Im\left[ \frac{\partial \ccF(Z,W^2)}{\partial W^2} \right], \qquad \text{with} \qquad W^2 = -64 \overline{\cZ}^{-2} \me^{-\ccK}~.
\end{align*}
(For an alternative approach that arrives at the same formula for the entropy, see for example \cite{Sen:2007qy,Sen:2005wa}.) In the large charge limit only the first term, corresponding to the Bekenstein-Hawking entropy, is of the leading order, and thus, in the large charge approximation, the Wald entropy is given by
\begin{align*} \notag
S_{\text{Wald}} \; \sim \; \pi |\cZ|^2~.
\end{align*}
If we assume that the microscopic entropy should be compared to the Wald entropy, as computed in supergravity, an assumption supported by \cite{Maldacena:1997de}, then this justifies comparing the microscopic entropy, in the large charge limit, to the expressions we have derived for the central charge for large complex structure. The approach of including the $\sigma$ and instanton corrections potentially allows us to extend the results of \cite{Maldacena:1997de,LopesCardoso:1998tkj} to the large charge limit.

\subsubsection*{Entropy from conformal field theory}
\vskip-10pt
As an elementary check of our results, we reproduce the results in \cite{Maldacena:1997de,LopesCardoso:1998tkj,Manschot:2008ria,Manschot:2007ha}  by specialising them to the large charge limit so as to approach the Wald entropy of the black hole. This allows matching the entropy, computed from the elliptic genus, to the leading-order behaviour of $|\cZ|^2$.

It is believed, see \cite{Maldacena:1997de}, that the microscopic theory describing the D4-D2-D0 brane system is a two-dimensional $\mathcal{N}=(4,0)$ superconformal field theory. In \cite{Gaiotto:2006wm,Gaiotto:2007cd,Maldacena:1997de}, it is proposed that the microstate numbers for a system of $\wt{p}$ D4-branes bound to $\wt q$ D2 and $\wt{q_0}$ D0 branes can be read from the coefficients $c_\mu(n) = c_\mu \left(\wt{q}_0 + \wt{q}^2/(2\wt{p}Y) \right)$ of a modified elliptic genus. Appendix \ref{app:D-Brane_Interpretation_of_Q} contains the relations between the microscopic charges $\wt q$ and $\wt{q_{0}}$ and the macroscopic charges $q$ and $q_{0}$.

The famous Cardy formula \cite{Cardy:1986ie} gives the asymptotic growth of the modular form coefficients $c_\mu(n)$ for large $n$ in terms of the central charge and the charges $\wt{q}$ and $\wt{q}_0$
\begin{align*} \notag
c \left(n \right) \; \sim \; \exp \left[ 4\pi \sqrt{\frac{c_R}{24} \left(-\wt{q}_0 +\frac{\wt{q}^2}{2\wt{p}Y}\right)}\; \right],
\end{align*}
which is valid in the limit where the quantity $-\wt{q}_0 + \wt{q}^2/(2\wt{p}Y)$ is much larger than the conformal field theory central chare $c_R$. Here we also ignored the index $\mu$ as the different $c_{\mu}(n)$ have the same leading term in the large $n$ limit. In \cite{Maldacena:1997de}, it was shown, under some relatively mild assumptions, that the central charge $c_R$ can be expressed in terms of the charges as
\begin{align*} \notag
c_R \= Y\left(\wt{p}^3 + c_2\, \wt{p}\right)~.
\end{align*}
To compare this to our results, we recall the form \eqref{eq:chargevectors} that our charge vector takes. From this we identify, using the transformations in appendix \ref{app:D-Brane_Interpretation_of_Q}, $\wt{p}=\kappa$, $\wt{q}=-\kappa \left(\Upsilon+Y_{110}\right)$ and $\wt{q}_0=\frac{\kappa}{2} \left(2\Lambda+Y_{100}\right)$. As in \cite{Manschot:2008ria}, we concentrate on the special case where $\wt{q}=0$ for clarity. From the Cardy formula, we find the asymptotic expression
\begin{align*} \notag
c \left(n \right) \; \sim \; \exp \left[ \pi \sqrt{-\frac{Y}{3}\left(\kappa^{4}+c_{2}\kappa^{2}\right)\left( 2\Lambda+Y_{100} \right)} \, \right].
\end{align*}
We are interested here in the large charge limit $\kappa \to \infty$, since this allows us to make contact with the Wald entropy. In this limit, the expression simplifies to give the asymptotic formula for the microscopic entropy as
\begin{align} \label{eq:Modular_Form_Leading_Asymptotics}
S &\= \log c(\wt{q}_0) \; \sim \; \pi \kappa^2 \sqrt{-\frac{Y}{3}  \left(2\Lambda+Y_{100}\right)}~.
\end{align}
The relative sign between the D4 charge $\kappa$ and the D0 charge $\frac{\kappa}{2}(2\Lambda+Y_{100})$ is determined by $(2\Lambda+Y_{100})$. The above formula is only sensible when this factor is negative, which conforms with the expectation that D4-branes bind to anti-D0-branes, or anti-D4-branes bind to D0-branes. This is the case for which we will derive the macroscopic formula below. 
\subsubsection*{Central charge from supergravity}
\vskip-10pt
To compare this leading-order result to the expression for the central charge, it is in principle enough to solve the attractor equations in the approximation where we disregard the instanton corrections and set $\sigma{\,=\,}0$. This is essentially equivalent to the approach taken in \cite{Maldacena:1997de,Shmakova:1996nz}, except that we here ignore the higher-genus corrections, as we are interested in the large charge limit. It is interesting, however, to see explicitly how the result, in this limit is, just the leading term in the $N$-expansion for the central charge. This general formula will also contain the subleading corrections to the entropy and, in this sense, extend the results of \cite{Maldacena:1997de,Shmakova:1996nz}. 

As with the expressions for the recurrence relations \eqref{eq:D4_Strominger_Equations_Ry_Recurrence} for the rational functions $R_x$ and $R_y$, we expect the general formula for $|\cZ|^2$ to be rather complicated. To make the essential structure of the expansion and comparison clear, we will concentrate on the special case where $\Upsilon {\,=\,} 0$, $Y_{110} {\,=\,} 0$ and $2\Lambda+Y_{100}<0$. This situation is somewhat analogous to that which allowed us to find the explicit solutions in \sref{sect:Solutions}. Once again, the rank two attractor points discovered in \cite{Candelas:2019llw} satisfy this condition. The generalisation to the other cases is conceptually straightforward.

The conditions $\Upsilon {\,=\,} 0$ and $Y_{110} {\,=\,} 0$ imply that $\alpha {\,=\,} 0$, and so $x_0 {\,=\,} 0$. Looking at the alignment equations \eqref{eq:D4_Strominger_Equations}, we see that this implies $x {\,=\,} 0$, even when the instanton corrections are taken into account. The central charge is given by \eqref{eq:Central_Charge_Definition}, which simplifies when $x{\,=\,}0$. The central charge can be written in the form
\begin{align*} \notag
\begin{split}
\cZ(Q_{D4}) &\= -\kappa \, \frac{-\frac{1}{2}Y y^2 + \frac{1}{2} Y\wt{\beta}^2 + \cI'_1}{\sqrt{\frac{4}{3}Y y^3 + \frac{2}{3}Y \sigma + 4 \cI_2 - 4 y \cI_1'}} \, \defineas \, -\kappa L \sqrt{\frac{Y}{M}}~,
\end{split}
\end{align*}
where 
\begin{align*} \notag
L \= -\frac{1}{2} y^2 + \frac{1}{2} \wt{\beta}^2 + \frac{1}{Y} \cI'_1~, \quad \text{and} \quad M \= \frac{4}{3} y^3 + \frac{2}{3} \sigma + \frac{4}{Y}\cI_2 - \frac{4y}{Y}\cI_1'~.
\end{align*}
Using the expansions in \sref{sect:Partitions}, we can write the quantities $K$ and $M$ appearing here as $N$-expansions:
\begin{align*} \notag
\begin{split}
L &\= -\frac{1}{2} y_0^2 + \frac{1}{2} \wt{\beta}^2 + \sum_{k=1}^\infty q_y^k \sum_{\fp \in \pt(k)} N_\fp \bm{L}_\fp~,\\[5pt]
\frac{1}{\sqrt{M}} &\= \frac{1}{\sqrt{ \frac{2}{3}\left( \sigma + 2 y_0^3 \right)}}  + \sum_{k=1}^\infty q_y^k \sum_{\fp \in \pt(k)} N_\fp \bm{M}_\fp~.
\end{split}
\end{align*}
We have written the $q_y$-independent part explicitly, as this will be important for comparing the leading-order result in this expansion with the modular form asymptotics and results of \cite{Maldacena:1997de,Shmakova:1996nz}. The coefficients of the invariants $N_\fp$ can be expressed as
\begin{align*} \notag
\begin{split}
\bm{L}_\fp &\= -\frac{1}{2} \left( R^{y_{\,}} \hskip-2pt \sr R^{y_{\,}} \hskip-1pt \right)_\fp - \cC_{2,\fp}~, \qquad \bm{L}_\emptyset \= \frac{1}{2}\left(\wt\b^2-y_0^2 \right)\\[10pt]
\bm{M}_\fp &\= \sum_{i=0}^\infty  \frac{(-1)^{i}\left( \frac{1}{2} \right)_i }{Y^i \left( \frac{2}{3} \sigma + \frac{4}{3} y_0^3 \right)^{i+1/2} i!} \sum_{\substack{\sum_{j=1}^i \fp_j = \mathfrak{p} \\ \fp_j \neq \emptyset}} c\big( \hskip-2pt \left \{\mathfrak{p}_j \right\}_{j=1}^i \big) \prod_{j=1}^i \bm{m}_{\mathfrak{p}_j}~,
\end{split}
\end{align*}
where the sum is defined as in \eqref{eq:defn_r^(x)} and the quantities $\bm{m}_\fp$ are given by
\begin{align*} \notag
\bm{m}_\fp &\= \frac{4}{3} Y r^{y_{\,}}_{(1,3),\fp} + 4 Y \cC_{3,\fp} + 4 Y (\cC_2 \sr R^{y_{\,}} \hskip-1pt)_\fp~.
\end{align*}
In terms of these quantities, the central charge becomes
\begin{align} \label{eq:D4_Central_Charge_Expansion}
\cZ(Q_{D4}) \=  \kappa \frac{\sqrt{3Y} \wt{\beta}^2  \left(1 +4 \sin^2\left( \th+\frac{4\pi}{3}\right) \right)}{\sqrt{8 \sigma + 128 \ii \wt{\beta}^3 \sin^3 \left(\th+\frac{4\pi}{3}\right)}} - \kappa \sqrt{Y} \sum_{k=1}^\infty \me^{-2\pi k y_0} \hskip-5pt \sum_{\fp \in \pt(k)} \hskip-5pt N_\fp \, (\bm{M} \sr \bm{L})_\fp~,
\end{align}
where we have written out the $\sigma$-corrected coordinates $x_0$ and $y_0$ in the first term explicitly in terms of $\wt{\beta}{\=}\sqrt{(2\Lambda+Y_{100})/Y}$, the Yukawa coupling $Y$, and the angle $\theta$, defined in \eqref{eq:D4_theta_defn}. It is $\theta$ that contains the contribution of~$\sigma$. It should be borne in mind that we defined $\theta$ as a particular value of an arcsine, and have incorporated a shift by $\frac{4\pi}{3}$ in \eqref{eq:D4_Central_Charge_Expansion}, as per the first row of \tref{tab:D4_Alignment_x00}. We have picked this solution from the pair in this row with an eye to a limit where $\sigma$ is neglected in the case where $2\Lambda+Y_{100}<0$.

Since we are only interested in the leading order here, we can disregard instanton corrections, and also set $\sigma{\;=\;}0$ (and thus $\theta=0$). The expression for the central charge then becomes
\begin{align*} \notag
\cZ(Q_{D4})  \; \sim \; \ii \frac{\kappa}{3^{1/4}} \sqrt{Y|\wt{\beta}|} \= \ii \kappa \left(\frac{Y}{3} \right)^{1/4} |Y_{100} + 2 \Lambda|^{1/4}. 
\end{align*}
Finally, we compute the Wald entropy in the large charge limit
\begin{align*} \notag
S_{\text{Wald}} \; \sim \; \pi |\cZ(Q_{\text{D4}})|^2 \; \sim \; \pi \kappa^2 \sqrt{-\frac{Y}{3}(2\Lambda+Y_{100})}~,
\end{align*}
which agrees precisely with the modular form asymptotics \eqref{eq:Modular_Form_Leading_Asymptotics}, as expected. Of course this result just verifies that the formulae here agree with the well-known results of \cite{Shmakova:1996nz,Maldacena:1997de} in the large charge limit. However, our approach to deriving the entropy from supergravity allows us to give systematic expansions, similar to \eqref{eq:D4_Central_Charge_Expansion}, for the subleading terms that were not included in \cite{Shmakova:1996nz,Maldacena:1997de}. It is intriguing that this gives especially simple expressions \eqref{eq:rank_two_central_charge_1} and \eqref{eq:rank_two_central_charge_2} for the charges associated to rank two attractor points. This also shows a striking similarity to results relating to the asymptotics of modular forms, see for example \cite{Manschot:2007ha,Rademacher78}.
\vskip1in
\section*{Acknowledgements}
\vskip-10pt
We are grateful for interesting conversations and comments to Christopher Beem, Xenia de la Ossa, Mohamed Elmi, Carmen Jorge-Diaz, and Lionel Mason. PK thanks the Osk.~Huttusen Säätiö for support. JM is supported by EPSRC.
\newpage

\newpage
\appendix
\addcontentsline{toc}{section}{Appendices}
\section{Symplectic Transformations} \label{app:Symplectic_Transformations}
\vskip-10pt
In this appendix we discuss the effect of a subgroup of the symplectic transformations that act on the period and charge vectors following \cite{Braun2015TwoOS}, but in the notation of \cite{Candelas:2019llw}. Distinguished among the set of possible bases for the period vector of a one-parameter Calabi-Yau manifold is the basis in which the period vector $\widehat{\varpi}$ is built out of a canonical (Frobenius) basis of solutions to the Picard-Fuchs equation. Near the large complex structure point this vector has the asymptotics 
\begin{equation*}\notag
\widehat{\varpi} \; \sim \; \left(\begin{matrix}1\\t\\t^2\\t^3\end{matrix}\right).
\end{equation*}
The change-of-basis matrix $\widehat{\rho}$ relates this to the vector  $\Pi = \widehat{\rho} \widehat{\varpi}$ in the integral basis: 
\begin{equation*}\notag
\widehat{\rho} \=-\left(\begin{matrix}\frac{1}{3}Y_{000} & \frac{1}{2}Y_{100} & 0 &-\frac{1}{6}Y_{111}\\[2pt]
\frac{1}{2}Y_{100} & Y_{110} & \frac{1}{2}Y_{111} & 0\\
-1\+&0&0&0\\0&-1\+&0&0\end{matrix}\right).
\end{equation*}
In the alignment equations \eqref{eq:Strominger_Equations_Original} the imaginary part of the period vector is proportional to the charge vector and in the orthogonality equation \eqref{eq:Orthogonality_Equations_Original} a skew-symmetric bilinear form vanishes. Both of these types of equation are still satisfied if one acts on both the period and charge vectors by the same symplectic matrix\footnote{The former equations are invariant under a wider set of linear transformations, but we only consider the symplectic ones as these are the transformations of the charges under electromagnetic duality in $d=4$, $\mathcal{N}=2$ supergravity.}.

Let us focus, for arbitrary integers $A,\,B$, and $C$, on the subgroup of symplectic matrices that are of the form
\begin{equation*}\notag
S \=\left(\begin{matrix}1&0&C&B\\0&1&B&A\\0&0&1&0\\0&0&0&1\end{matrix}\right).
\end{equation*}
This subgroup is isomorphic to $\IZ^{3}$ and furnishes a symmetry of the attractor equations. Noting that 
\begin{equation*}\notag
S\widehat{\rho} \=\left(\begin{matrix}C-\frac{1}{3}Y_{000} & B-\frac{1}{2}Y_{100} & 0 &\frac{1}{6}Y_{111}\\[2pt]
B-\frac{1}{2}Y_{100} & A-Y_{110} & -\frac{1}{2}Y_{111} & 0\\1&0&0&0\\0&1&0&0\end{matrix}\right),
\end{equation*}
one can view $S$ as acting on the quantities $Y_{ijk}$ as
\begin{equation*}\notag
\begin{aligned}
 Y_{111} & \;\rightarrow \; Y_{111}~,\\
 Y_{110}& \; \rightarrow \; Y_{110}-A~,\\
 Y_{100}& \; \rightarrow \; Y_{100}-2B~,\\ 
 Y_{000}& \; \rightarrow \; Y_{000}-3C~.
\end{aligned}
\end{equation*}
Throughout this paper, we have chosen a symplectic basis in which the $Y_{ijk}$ take the minimal form that we have assumed in the body of this paper. If one is working with a different set of values for the $Y_{ijk}$ then a transformation of the above form will relate the two bases. 

After acting on our $Y_{ijk}$ with such a transformation, the monodromy matrix will still be integral, but the $Y_{ijk}$ no longer have the minimal form. $S$ acts on a charge vector by
\begin{equation*}\notag
S \, \left(\begin{matrix}q_0 \!\!\\~q~\\p^0 \!\! \! \\p\end{matrix}\right) \= \left(\begin{matrix}q_0+Bp+Cp^0 \\q_{\phantom{0}}+Ap+Bp^0\\p^0 \!\!\\p\end{matrix}\right).
\end{equation*}
We see that $p^0$ and $p$ are fixed and identify the transformations for $q_0$ and $q$. 

In this paper we have worked with two cases: the D6 case where $p^0=1$ and $p=0$, and the D4 case where $p^0=0$ and $p=1$. For the D6 case the combinations 
\begin{align*} \notag
q+\frac{1}{2}Y_{100} &\qquad \text{and} \qquad q_0+\frac{1}{3}\text{Re}[Y_{000}]\\
\intertext{are invariant, while for the D4 case the invariant combinations are}
q+Y_{110} \hphantom{\frac{1}{2}} &\qquad \text{and} \qquad q_0+\frac{1}{2}Y_{100}~. \notag
\end{align*} 
This explains the combinations in \tref{tab:shorthand} that arise in the equations 
\eqref{eq:D4_Zero_Equations}, \eqref{eq:D4_Strominger_Equations}, \eqref{eq:D6zerofull}, and \eqref{eq:D6_Strominger_Equations}.
\newpage

\section{Monodromy} \label{app:Monodromy}
\vskip-10pt
Upon circling the large complex structure point $\varphi=0$, the $t$-coordinate changes by $t\rightarrow t+1$. This can be thought of as a consequence of the transformation $\Pi \to \text{M} \Pi$, of the period vector, with
\begin{equation*}\notag
\text{M} \= \left(\begin{matrix}1&-1&\+\frac{1}{6}Y_{111}-Y_{100}&\frac{1}{2}Y_{111}-Y_{110}\\[2pt]0&\+1&-\frac{1}{2}Y_{111}-Y_{110}&-Y_{111}\\0&\+0&1&0\\0&\+0&1&1\end{matrix}\right) \= \exp \text{L}~,
\end{equation*}
where
\begin{equation*} \notag
\text{L} \= \left(\begin{matrix}0&-1&-Y_{100}&-Y_{110}\\[2pt]0&\+0&-Y_{110}&-Y_{111}\\0&\+0&0&0\\0&\+0&1&0\end{matrix}\right).
\end{equation*}
Note that $L^4 = 0$ so the exponential is in fact a short finite sum. In these equations, we take the $Y_{ijk}$ to be the minimal $Y_{ijk}$ used previously, and M is understood not to act on these. On the charge vector, the action of the monodromy is
\begin{equation*}
Q^{(n)} \defineas \text{M}^n\left(\begin{matrix}q_0 \!\!\\~q~\\p^0 \!\! \! \\p\end{matrix}\right) = \exp (n \text{L}) \left(\begin{matrix}q_0 \!\!\\~q~\\p^0 \!\! \! \\p\end{matrix}\right) = \left(
\begin{array}{l}
\+\left(\frac{n^3}{6}Y- n Y_{100}\right) p^0+\left(\frac{n^2}{2}Y-nY_{110}\right)p-n q+q_0 \\[5pt]
-\left(\frac{n^2}{2}Y+nY_{110}\right)
p^0-n Yp+q \\[5pt]
\+\;p^0 \\[3pt]
\+\;np^0+p \\
\end{array}
\right). \nonumber
\end{equation*}
The monodromy matrix M is symplectic and the central charge depends on the period and charge vectors only through the symplectic invariant inner product, so the central charge is invariant under the monodromy. It follows that if $t$ is an attractor point corresponding to a charge vector $Q^{(0)}$, then $t+n$ is the coordinate of an attractor point corresponding to the charge vector $Q^{(n)}$. In virtue of this, the entire discussion of the D6~system in \sref{sect:D6-system} can be extended to include a restricted set of D4~charges.
\subsection{D4-D2-D0 system}
\vskip-10pt
The monodromy preserves $p^0$, and so in particular the condition $p^0 = 0$. In this case the monodromy can be thought of as acting on the charge ratios as
\begin{align*} \notag
\Lambda^{(n)} \= \Lambda^{(0)} +  \frac{n^2}{2} Y - n \left(Y_{110} + \Upsilon^{(0)} \right), \qquad \Upsilon^{(n)} \= \Upsilon^{(0)} - nY~.
\end{align*}
Since we have in \sref{sect:D4_Strominger_Equations} found an expression for the coordinates $x,y$ of the attractor point in terms of these charges, it is interesting to follow how this transformation gives rise to the coordinate transformation $x \to x + n$, $y \to y$. First note that it follows from the transformation above that the quantities $\alpha$ and $\wt{\beta}$ transform so that
\begin{align*} \notag
\wt{\beta} \; \to \;  \sqrt{\wt{\beta}^2 + n^2 + 2 n \alpha \wt{\beta}}~, \qquad \alpha \wt{\beta} \; \to \;  \alpha \wt{\beta} + n~. 
\end{align*}
It follows that $\wt{\beta} \sqrt{\alpha^2-1}$ is invariant. Thus the perturbative coordinates transform as
\begin{align*} \notag
x_{0} \= \alpha \wt{\beta} \; \to \; \alpha \wt{\beta} +n \= x_{0} + n~, \qquad y_0 \= 2 \wt{\beta} \sqrt{\alpha^2 - 1} \sin \theta \; \to \;  y_{0}~.
\end{align*}
To see that $\theta$ is invariant under the monodromy, note that this is required by the definition \eqref{eq:D4_theta_defn}, together with the above transformation laws.

The instanton corrections are not expected to modify these transformations and this is indeed the case as $R_\fp^{x_{\,}}$ and $R_\fp^{y_{\,}}$ depend on $x_0$ only via $\cos(2 m \pi x_0)$ or $\sin(2 m \pi x_0)$ with $m \in \IZ$.

\subsection{D6-D2-D0 system}
\vskip-10pt
A charge vector with $p\=0$ maps in general to a vector $p \; \neq \; 0 $. The charge ratios transform as
\begin{align*} \nn
\Lambda^{(n)} \= \Lambda^{(0)} + \frac{n^3}{6}Y - n (\Upsilon^{(0)} - Y_{100})~, \qquad \Upsilon^{(n)} \= \Upsilon^{(0)} - \frac{n^2}{2}Y - n Y_{110}~.
\end{align*}
Thus, by choosing $\Lambda^{(0)}$ and $\Upsilon^{(0)}$ suitably, we can always have arbitrary values for the first two components. The third component is kept fixed, and the last component is an integer multiple of $p^0$, the value of the third component. Conversely, given any charge of this form, we can always find a monodromy that transforms it into the form where $p=0$. So the solutions in \sref{sect:D6-system} relate to charge vectors of the form
\begin{align*} \notag
Q \= \left(\begin{matrix}
q_0\\
q\\
p^0\\
p
\end{matrix}\right), \qquad \text{with} \hskip5pt p^0|p \hskip3pt \text{ or } p^0 \= 0~.
\end{align*}

\newpage
 
\section{Microscopic Charge Vector} \label{app:D-Brane_Interpretation_of_Q}
\vskip-10pt
The discussion above has been cast in terms of the charge vector $Q$ (defined by \eqref{eq:defn_Q}), which is a convenient quantity when considering black holes from the macroscopic, or supergravity, perspective. For the purposes of discussing the microstructure of the black holes, it is more convenient to define a quantity
\begin{align*} \notag
\wt{Q} &\=\begin{pmatrix}
 \+\wt{q}_0\\
-\wt{q}_{\phantom{0}}\\
 \+\wt{p}^0\\
-\wt{p}^{\phantom{0}}
\end{pmatrix}.
\end{align*}
The charges that appear here correspond to the couplings to the RR-potentials in the D-brane world-volume action. Consider first the D6 brane with a world-volume flux $F$, the 2-form field $B$ and RR-potentials $C^{2n+1}$, $n=0,1,2,3$. Its world-volume action has the Wess-Zumino part
\begin{align*} \notag
S_{\text{WZ}} &\= \int_{X} C \wedge \me^{F-B} \wedge \sqrt{\widehat{A}(\wt{X})}~,
\end{align*}
where $C$ is a multi-form built out of the RR-potentials, $C = C^{(7)} + C^{(5)} + C^{(3)} + C^{(1)}$. We combine the world-volume flux and B-field into a single tensor $\cF = F - B$. The charges coupling to the RR-potentials can be computed from this action either directly, or alternatively by writing the charges as a multiform.
\begin{align*} \notag
\frac{\delta S_{\text{WZ}}}{\delta C^{2n+1}} &\= \left[\; \me^{\cF} \wedge \sqrt{\widehat{A}(\wt{X})} \; \right]_{6-2n}
\hskip-5pt = \hskip10pt \left[ \left( 1 + \cF + \frac{1}{2} \cF^2 + \frac{1}{6} \cF^3 \right) \wedge \left( 1 + \frac{c_2(\wt{X})}{24} \right) \right]_{6-2n},
\end{align*}
where $[*]_{6-2n}$ denotes the $(6-2n)$-form part of the multiform. We can combine all these parts into a microscopic charge multiform $\wt{\Gamma}$.
\begin{align*} \notag
\wt{\Gamma} &\= \sum_{n=0}^3 \frac{\delta S_{\text{WZ}}}{\delta C^{2n+1}} \= \me^{\cF} \wedge \sqrt{\widehat{A}(\wt{X})}~.
\end{align*}
To write this in a convenient form, we set $\cF = f e_1$, where $f$ is a scalar, and $e_1$ generates $H^2(X,\IZ)$. Thus we can write the charge form as
\begin{align*} \notag
\wt{\Gamma} &\= 1+ f e_1 + \left( \frac{1}{2}f^2 + \frac{c_2}{24}  \right) e_1^2 + \left( \frac{1}{6} f^3 + \frac{c_2}{24} f \right) e_1^3~.
\end{align*}
The form $\mho$, defining the microscopic period vector $\wt{\Pi}$, is given by the mirror of the holomorphic three form $\Omega$ on $X$:
\begin{align*} \notag
\mho \= \me^{B + \ii J} \= \me^{t e_1} \= 1 + t \, e_1 + \frac{1}{2} t^2 \, e_1^2 + \frac{1}{6} t^3 \, e_1^3~. 
\end{align*} 
The appropriate \cite{Denef:2007vg} inner product to use is the Dirac-Schwinger-Zwanziger product, defined by
\begin{align*} \notag
\wt{\Gamma} \cdot \mho \= \int_{\wt{X}} \wt{\Gamma} \wedge \mho^*~,
\end{align*}
where, on the right, $\mho^*$ is the form obtained by inverting the signs of the 2- and 6-form components of $\mho$, so that $\mho^* \= \me^{-t e_1}$. Note that this is not the complex conjugate. 

Alternatively, we can identify the components of the vectors $\wt{Q}$ and $\wt{\Pi}$ corresponding to $\wt{\Gamma}$ and $\mho$. 
\begin{align*} \notag
\wt{Q} \= \begin{pmatrix}
\+ \,\,\,\, \frac{1}{6} Y f^3 - \frac{Y_{100}}{2} f \,\\[3pt]
-\frac{1}{2}Y f^2 + \frac{Y_{100}}{2}\\
\+1\\
-f
\end{pmatrix}, \qquad \wt{\Pi} \= \begin{pmatrix}
\frac{1}{6}Y t^3\\[3pt]
\frac{1}{2}Y t^2\\[3pt]
1\\
t
\end{pmatrix},
\end{align*}
where $c_2$ has been expressed in terms of $Y_{100}$. With this definition, the Dirac-Schwinger-Zwanziger product becomes the familiar symplectic inner product.
\begin{align*} \notag
\wt{\Gamma} \cdot \mho \= \wt{Q}^T \Sigma \wt{\Pi}~.
\end{align*}
As a simple consistency check, we note that this is invariant under the monodromy $t \to t+1$ and $B \to B + 1$ as then $f \to f+1$. Note that this is the action of the monodromy despite the fact that $\cF \= F - B$, owing to the sign change in the vector $\wt{Q}$.

To find the relation between the microscopic charge vector $\wt{Q}$ and the macroscopic one $Q$, we note the relation between the corresponding period vectors
\begin{align*} \notag
\Pi_0 \big|_{\sigma=0} \= \text{T}_{\Pi} \wt{\Pi}~, \qquad \text{with} \qquad \text{T}_{\Pi} \= \begin{pmatrix}
1 & \+0 & 0 & -\frac{1}{2} Y_{100}\\[3pt]
0 & -1 & -\frac{1}{2} Y_{100} & - Y_{110}\\[3pt]
0 & \+0 &1 & 0\\[3pt]
0 & \+0 & 0 & 1
\end{pmatrix}.
\end{align*}
The central charge should be basis independent, requiring that $\wt{Q}^T \Sigma \wt{\Pi} {\=} Q^T \Sigma \Pi_0 \big|_{\sigma=0}$, which allows the identification
\begin{align*} \notag
\wt{Q} \= \text{T}^{-1}_{Q} Q~, \qquad \text{with} \qquad \text{T}_{Q} \= \left(\Sigma^{-1} \text{T}_{\Pi}^{-1} \Sigma \right)^T \=  \begin{pmatrix}
1 & \+0 & 0 & \frac{1}{2} Y_{100}\\[3pt]
0 & \+1 & -\frac{1}{2} Y_{100} & Y_{110}\\[3pt]
0 & \+0 &1 & 0\\[3pt]
0 & \+0 & 0 & -1\+
\end{pmatrix}.
\end{align*}

\newpage

\section{The \texorpdfstring{$\sigma$}{sigma}-Expansion for the D6 Orthogonality Equation} \label{app:D6_Orth_Polynomials}
\vskip-10pt
In this appendix we give the details of derivation of the series \eqref{eq:D6sqpert1} and \eqref{eq:D6sqpert2} for the $\sigma$-corrections in the D6-D2-D0 system.

The real and imaginary parts of the D6 orthogonality equation \eqref{eq:D6zerotrig} in the case $\beta\in\IR$ are such that, after defining $\widehat{y}=\frac{y_{0}}{\beta}$, the parameter $\beta$ disappears from the equation, leaving a pair of equations for $x_{0}$ and $\widehat{y}$. Eliminating $x_0$, one arrives at the single equation:
\begin{equation*}\notag
\left(3\widehat{y}+4\widehat{y}^{3}-\widehat{\sigma}\right)^{2}\left(3\widehat{y}+\widehat{y}^{3}+2\widehat{\sigma}\right)-27\, \widehat{y}^{3}\cosh^{2}3\nu\=0~,
\end{equation*}
where we have written $\gamma=\pm\cosh3\nu$. Informed by a certain amount of experimentation, we make the following ansatz for $\widehat{y}\,$: 
\begin{equation*}\notag
\widehat{y} \= -\sinh3\nu\,\ee^{-2\nu}\sum_{m=0}^{\infty}\text{T}(m)\Psi_{m}(\ee^{-2\nu})\left(-\frac{\widehat{\sigma}\,\ee^{3\nu}}{2\sinh^{2}3\nu}\right)^{m}.
\end{equation*}
The function $\Psi_{0}(z)$ is special, and is found to be $-(1+z+z^2)^{-1}$. For $m\geqslant1$, the $\Psi_{m}(z)$ are polynomials of degree $3m-2$. The first few are
\begin{equation*}\notag
\begin{aligned}
\Psi_{0}(z)&\=-\frac{1}{1+z+z^{2}}~,\\[10pt]
\Psi_{1}(z)&\=\smallfrac{1}{3} \hskip5pt \left(1-z\right)~,\\[5pt]
\Psi_{2}(z)&\=\smallfrac{1}{3^2} \hskip3pt \left(1+2z+2z^3+z^4\right)~,\\[5pt]
\Psi_{3}(z)&\=\smallfrac{1}{3^4} \hskip3pt \left(5-14z+35z^3-35z^4+14z^6-5z^7\right)~,\\[5pt]
\Psi_{4}(z)&\=\smallfrac{1}{3^5} \hskip3pt \left(10+35z+150z^3+210z^4+210z^6+150z^7+35z^9+10z^{10}\right)~,\\[5pt]
\Psi_{5}(z)&\=\smallfrac{1}{3^6} \hskip3pt \left(22-91z+572z^3-1001z^4+1716z^6-1716z^7+1001z^9-572z^{10}+91z^{12}-22z^{13}\right)~,\\[5pt]
\Psi_{6}(z)&\=\smallfrac{1}{3^8} \hskip3pt\left(154+728z+6160z^3+12740z^4+32032z^6+40040z^7+40040z^9+32032z^{10}\right.\\[-2pt]&\hskip56pt\left.+12740z^{12}+6160z^{13}+728z^{15}+154z^{16}\right)~,\\[5pt]
\Psi_{7}(z)&\=\smallfrac{1}{3^9} \hskip3pt \left(374-1976z+21318z^3-50388z^4+170544z^6-251940z^7+369512z^9-369512z^{10}\right.\\[-2pt]&\hskip56pt+251940z^{12}-170544z^{13}+50388z^{15}-21318z^{16}+1976z^{18}-374z^{19}\left.\right)~,\\[5pt]
\Psi_{8}(z)&\=\smallfrac{1}{3^{10}}\left(935+5434z+71995z^3+190190z^4+820743z^6+1385670z^7+2735810z^9\right.\\[-2pt]&\hskip56pt\left.+3233230z^{10}+3233230z^{12}+2735810z^{13}+1385670z^{15}+820743z^{16}\right.\\[1pt]&\hskip56pt\left.+190190z^{18}+71995z^{19}+5434z^{21}+935z^{22}\right)~.\\[5pt]
\end{aligned}
\end{equation*}
Notice that the coefficients are symmetrically distributed and that for odd $m$ the polynomial $\Psi_{m}(z)$ has a factor of $(1-z)^{2m-1}$. Most importantly for us, the polynomials only contain powers of $z$ that are 0 or 1 mod 3. The $\Psi_m$ are thus naturally divided into two polynomials in $z^{3}$. Recognising a pattern in the coefficients, we see that
\begin{equation*}\notag
\begin{split}
\Psi_{m}(z)&=\frac{1}{m!}\left[-\left(-\smallfrac{1}{3}\right)_{m}\twoFone{\frac{2}{3}{-}m}{1{-}m}{\frac{2}{3}}{z^{3\!}}+(-1)^{m}\left(\smallfrac{1}{3}\right)_{m}z \, \twoFone{\frac{4}{3}{-}m}{1{-}m}{\frac{4}{3}}{z^{3}\!}\right]\\[15pt]
&=\frac{(1{-}z^3)^{m{-}1}}{3\,m!}
\left[\left(\smallfrac{2}{3}\right)_{\hskip-2ptm{-}1}\hskip-3pt \twoFone{m}{1{-}m}{\frac{2}{3}}{\hskip-3pt{-}\frac{z^{3}}{1{-}z^{3}}}\!+(-1)^{m} \! \left(\smallfrac{4}{3}\right)_{\hskip-2ptm{-}1} \hskip-3pt z \, \twoFone{m}{1{-}m}{\frac{4}{3}}{\hskip-3pt{-}\frac{z^{3}}{1{-}z^{3}}}\right].
\end{split}
\end{equation*}
Owing to the fact that the $b$-parameter in the hypergeometric functions is a negative integer, these functions are polynomials. In passing to the second line, we rewrite this expression in a form that recognises that these are in fact related to Jacobi polynomials $P_{m}^{(\frac{1}{3},-\frac{1}{3})}$, by the definition \cite{Erdelyi3}:
\begin{equation*}\notag
P_{n}^{(\alpha,\beta)}(z) \= \frac{(\alpha+1)_{n}}{n!}\twoFone{1+\alpha+\beta+n\,}{-n}{\alpha+1}{\hskip-5pt\frac{1-z}{2}}.
\end{equation*}
Writing $z = \me^{-2\nu}$, we arrive at \eqref{eq:D6sqpert1}. The other series \eqref{eq:D6sqpert2} can be obtained from this.

\newpage
\section{The \texorpdfstring{$\sigma$}{sigma}-Expansion for the D6 Alignment Equations} \label{app:D6_Polynomials}
\vskip-10pt
For the D6-D2-D0 alignment equations we have given in \sref{sect:D6_Alignment_Equations} a perturbative solution in $\sigma$. The coefficients of this series are given in terms of a hypergeometric function of argument $\gamma^{2}$. Here we derive this expression.

We start with \eqref{eq:D6quantumrecast},
\begin{equation*}\nonumber
(\widehat{y}^{2}-1)(\widehat{y}-\widehat{\sigma})^{2}+\gamma^{2}\widehat{y}^{2}  \= 0~.
\end{equation*}
While we ultimately want to solve this as a power series in $\widehat{\sigma}$, the intermediate calculations are simpler if we begin by writing $\widehat{y}$ as a power series in $\gamma$. This approach requires us to work in the region of parameter space where
\begin{equation*}\notag
|\widehat{\sigma}|<\left(1-\gamma^{2/3}\right)^{3/2}.
\end{equation*}
When $|\gamma|<1$, we can expect to find a series about $\widehat{\sigma}=0$. If $\gamma=0$, however, the solution of interest is just $\widehat{y}=1$, rather than $\widehat{y} = \widehat{\sigma}$.

Our ansatz, in this case, is
\begin{equation}\label{eq:d6_alignment_app_series}
\widehat{y} \= 1-\sum_{n=1}^{\infty}\frac{\Phi_{n}(\widehat{\sigma})}{(1-\widehat{\sigma})^{3n-1}}\,\gamma^{2n}~.
\end{equation}
The $\Phi_{n}$ turn out to be polynomials. The first few of these are
\begin{equation*} \notag
 \begin{aligned}
 \Phi_{1}(z)&\=\frac{1}{2}~,\\[3pt]
 \Phi_{2}(z)&\=\smallfrac{1}{2^3}\hskip4pt\left(1+3z\right),\\[3pt]
 \Phi_{3}(z)&\=\smallfrac{1}{2^4}\hskip4pt\left(1+8z+5z^{2}\right),\\[3pt]
 \Phi_{4}(z)&\=\smallfrac{1}{2^7}\hskip4pt\left(5+73z+127z^{2}+35z^{3}\right),\\[3pt]
 \Phi_{5}(z)&\=\smallfrac{1}{2^{8}} \hskip4pt\left(7+158z+518z^{2}+398z^{3}+63z^{4}\right),\\[3pt]
 \Phi_{6}(z)&\=\smallfrac{1}{2^{10}}\left(21+667z+3466z^{2}+5046z^{3}+2217z^{4}+231z^{5}\right),\\[3pt]
 \Phi_{7}(z)&\=\smallfrac{1}{2^{11}}\left(33+1388z+10355z^{2}+24040z^{3}+20015z^{4}+5756z^{5}+429z^{6}\right),\\[3pt]
 \Phi_{8}(z)&\=\smallfrac{1}{2^{15}}\left(429+22901z+230033z^2+771209z^3+1025575z^4+557983z^5+114139z^6+6435z^7\right).
 \end{aligned}
\end{equation*}
One can proceed either by finding the recurrence relations for the coefficients in these polynomials, or alternatively by seeking a family of differential equations that the $\Phi_{n}$ satisfy. In this way we find
\begin{equation}\label{eq:new_phi_diffeq}
\begin{aligned}
&\hskip2.7cm-z\left(z-1\right)\left(z^{2}-1\right)\,\Phi_{n}'''(z)\\[5pt]
&\hskip20pt+\left(z-1\right)\Big(1+3(3n-1)z+3(n-2)z^{2}\Big)\,\Phi_{n}''(z)\\[5pt]
&\hskip-0.5cm \hskip40pt+\Big(2-6n-15n(n-1)z-3(n-1)(n-2)z^{2}\Big)\,\Phi_{n}'(z)\\[5pt]
&\hskip105pt \hskip1cm+n(n-1)\Big(8n-1+(n-2)z\Big)\,\Phi_{n}(z)\=0~.
\end{aligned}
\end{equation}
The initial conditions are
\begin{gather*} \notag
\Phi_{n}(0)\=-\frac{\left(-\frac{1}{2}\right)_{n}}{n!}~,\hskip10pt \qquad \Phi_{n}'(0)\=1+\frac{(3n-1)\left(-\frac{1}{2}\right)_{n}}{n!}~, \notag\\[8pt] \Phi_{n}''(0)\=2-6n-\frac{\left(2-11n+9n^{2}+8n^{3}\right)\left(-\frac{1}{2}\right)_{n}}{n!}~. \notag
\end{gather*}
By following the algorithm detailed in \cite{chebterrab2008hypergeometric}, one can identify a transformation that brings the equation \eqref{eq:new_phi_diffeq} to a form satisfied by the hypergeometric function $_3F_2$.

The coefficients in the series \eqref{eq:d6_alignment_app_series} can be expressed in terms of hypergeometric functions. The result~is
\begin{equation*} \notag
\widehat{y}\=1+\sum_{n=1}^{\infty}\left[\frac{\left(-\frac{1}{2}\right)_{n}}{n!}\threeFtwo{n-\frac12}{n}{n+\frac12}{\frac12}{\frac12}{\widehat{\sigma}^{2}} - \widehat{\sigma}\,\threeFtwo{n}{n+\frac12}{n+1}{1}{\frac32}{\widehat{\sigma}^{2}}\right]\gamma^{2n}~.
\end{equation*}
The hypergeometric functions in the above series can be expanded as a power series in $\widehat{\sigma}$, yielding a double sum. By interchanging the order of summation and applying identities enjoyed by the Pochhammer symbols, we arrive at the expression \eqref{eq:D6_Alignment_Hypergeometric}:
\begin{equation*} \nonumber
\widehat{y}=\sqrt{1-\gamma^{2}}-\frac{\gamma^{2}}{2}\,\sum_{k=1}^{\infty}(k+1)\,\threeFtwo{\frac{k+1}{2}}{\frac{k+2}{2}}{\frac{k+3}{2}}{\frac32}{2}{\gamma^{2}}\widehat{\sigma}^{k}~.
\end{equation*}
In a roundabout way we have arrived at an identity, since $\widehat{y}$ is a solution to \eqref{eq:D6quantumrecast}, and so can also be expressed directly in terms of radicals, as we have done in \eqref{eq:D6_Alignment_Radical}. We do not know if the above expression is an instance of a known identity.

\newpage
\section{Convergence of Solutions}\label{sect:Asymptotics_Convergence}
\vskip-10pt
In this appendix, we discuss convergence of the $N$-expansions, such as those in \eqref{eq:t_solution_D6_special_Equations} that appear as solutions to attractor equations. We concentrate on the solutions \eqref{eq:t_solution_D6_special_Equations} for concreteness, but we expect that a similar argument can be made to apply to all the solutions found in \sref{sect:D4-system} and \sref{sect:D6-system}, even if we do not have a closed-form expressions for the polynomials $P^x_\fp,P^y_\fp$ and $P^t_\fp$ appearing in these solutions.

We begin by making a few comments on the asymptotics of the terms in the sum \eqref{eq:t_solution_D6_special_Equations} as $j \to \infty$. Let us first consider the individual factors that appear in each summand, $N_\fp$,  $\bm{k}_{\length(\fp)-1}$, and $a_\fp$. 

The asymptotics of Gromov-Witten invariants are determined by the radius of convergence of the instanton sum $\cI$ in \eqref{eq:Instanton_Sum_Definition} (see for example \cite{Candelas:1990rm,Bershadsky:1993cx,Klemm:1999gm,Couso-Santamaria:2016vcc}). This leads to a formula for the asymptotic growth of $N^{\mbox{\tiny GW}}_k$, which relates the growth of the Gromov-Witten invariants at large $k$ to the coordinate $y_{\text{sing}}$ of the singularity closest to the large complex structure point.
\begin{align*} \notag
N_k \simeq N_k^{\mbox{\tiny GW}} \simeq \me^{2\pi k y_{\text{sing}}}~,
\end{align*}
where, by the notation $A \simeq B~,$ we mean $\log A \sim \log B~,$ with the asymptotic equality understood in the classical sense. From this, it follows that there is an integer $w$ such that that the quantity
\begin{align*} \notag
\me^{\frac{1}{w} \log N_w}\=N_{w}^{\frac{1}{w}}
\end{align*}
is maximal. For $\fp$ a partition of $j$, this $w$ gives a bound
\begin{align} \label{eq:GW_bound}
N_\fp \= \prod_{k=0}^\infty N_{k}^{\mu_k}\=\prod_{k=0}^\infty \left(N_{k}^{\frac{1}{k}}\right)^{k\mu_k}\leqslant\prod_{k=0}^\infty \left(N_{w}^{\frac{1}{w}}\right)^{k\mu_k} \=N_{w}^{\frac{1}{w}\sum_{k}k\mu_{k}}\= N_{w}^{\frac{j}{w}}\= \me^{\frac{j}{w} \log N_{w}}~.
\end{align}
The asymptotics of the Bessel functions $\bm{k}_{\length(\fp)}(2\pi y_0 |\fp|)$ can be found by noting that $\length(\fp)<|\fp|$ so that we can write $|\fp| = \Ic \left(\length(\fp) - \smallfrac{1}{2} \right)$ for some $\Ic > 1$. Then we use the asymptotic formula for large order \cite{NIST:DLMF}
\begin{align*}
\bm{k}_{\length(\fp)-1}(2\pi y_0 |\fp|) &\= \frac{1}{\pi \sqrt{y_0 |\fp|}} K_{\length(\fp)-\frac{1}{2}}\left(2\pi y_0 \Ic \left(\length(\fp) - \smallfrac{1}{2} \right)  \right) \nonumber \\
&\; \= \; O\left(\;\me^{-\sqrt{1+(2\pi y_0 \Ic)^2}\left(\length(\fp) - \frac{1}{2} \right)} \left(  \frac{1+ \sqrt{1 + (2\pi y_0 \Ic)^2}}{(2 \pi y_0 \Ic)} \right)^{\length(\fp)- \frac{1}{2}}\;\right). \nonumber
\end{align*}
If $y_0>\frac{1}{2\pi}$, then $2\pi y_0 \Ic>1$ for any $\Ic>1$. This means that we can find constants $C_{1}$ and $L$ such that
\begin{align} \label{eq:Bessel_bound}
k_{\length(\fp)-1}(2\pi y_0 |\fp|) < C_1 \me^{-2\pi y_0 \Ic \left( \length(\fp) -\frac{1}{2} \right)} (1 + \sqrt{2})^{\length(\fp)} \= C_1 \me^{-2\pi y_0 |\fp|} (1+ \sqrt{2})^{\length(\fp)}~,
\end{align}
for $\length(\fp)>L$. Later we argue that the coefficients $a_\fp$ satisfy the bound
\begin{align} \label{eq:ap_bound}
a_\fp \; \leqslant \; \left(\frac{\ee\pi^{2}}{6\length(\fp)}\right)^{\length(\fp)}~.
\end{align}
This bound can be improved, but it suffices for our purposes here.

Denoting the sum appearing in the solutions \eqref{eq:t_solution_D6_special_Equations} by $S(y_0)$,
\begin{align*} \notag
S(y_0) \= \sum_{j=1}^\infty \sum_{\fp \in \pt(j)} a_\fp N_\fp \left( \frac{j}{2 \pi y_{0} Y} \right)^{\length(\fp)} \bm{k}_{\length(\fp)-1}(2 \pi j y_{0})~,
\end{align*}
 the bound \eqref{eq:GW_bound}  allows us to write
\begin{align*}
S(y_0) \; \leqslant \; \sum_{j=1}^\infty \sum_{\fp \in \pt(j)} a_\fp \left( \frac{j}{2 \pi y_{0} Y} \right)^{\length(\fp)} \bm{k}_{\length(\fp)-1}(2 \pi j y_{0})\, \me^{\frac{j}{w} \log N_{w} }~. \nonumber
\end{align*}
We can break the finite sum over partitions of $j$ into sums over partitions of different length. This gives a further bound
\begin{align*}
S(y_0) \; \leqslant \; \sum_{j=1}^\infty \sum_{l = 1}^j |\pt_l(j)| \max \left(a_{j,l} \right) \left( \frac{j}{2 \pi y_{0} Y} \right)^{l} \bm{k}_{l-1}(2 \pi j y_{0})\, \me^{ \frac{j}{w} \log N_{w} }~, \nonumber
\end{align*}
where $\max \left(a_{j,l} \right)$ is the maximal value $a_\fp$ takes over $\pt_l(j)$, and $|\pt_l(j)|$ is the number of partitions of $j$ of length $l$. Ramanujan's asymptotic formula for $|\pt(j)|$ tells us that
\begin{align*} \notag
|\pt_l(j)| \; \leqslant |\pt(j)| \; < \me^{\pi \sqrt{\frac{2}{3}j}}~.
\end{align*}
This gives
\begin{align*}
S(y_0) \; \leqslant \;\sum_{j=1}^\infty \sum_{l = 1}^j  \max \left(a_{j,l} \right) \left( \frac{j}{2 \pi y_{0}Y } \right)^{l}\bm{k}_{l-1}(2 \pi j y_0) \me^{\frac{j}{w}\log N_{w}  + \pi \sqrt{\frac{2}{3}j}}~. \nonumber
\end{align*}
For $j>L$, we break the sum over $l$ into two parts, one from $1$ to $L$ and the other from $L+1$ to $j$. We can then apply the bounds \eqref{eq:Bessel_bound} and \eqref{eq:ap_bound}, allowing us to give another upper bound:
\begin{align*}
S(y_0) \; \leqslant \; C_2 + \sum_{j=L+1}^\infty \left(P(j) + C_{1}\sum_{l=L+1}^j \left( \frac{\ee\pi j(1+\sqrt{2})}{12  y_0 Y l} \right)^{l}  \right)\me^{-j\left(2\pi y_0 - \frac{1}{w}\log N_{w} \right) + \pi \sqrt{\frac{2}{3}j}} \;, \nonumber
\end{align*}
where $P(j)$ is a polynomial of degree $L$ in $j$ and $C_2$ is a constant corresponding to a finite sum over $j$ from $1$ to $L$. To understand the appearance of $P(j)$, recall the relation between $\bm{k}$ and the polynomials $f$ as in \eqref{eq:D6-D0_P_s_M}. The summand in the second sum over $l$ is maximised when
\begin{align*} \notag
l \= j\frac{\pi(1+\sqrt{2})}{12 y_0 Y}~.
\end{align*}
At this point the summand takes the value
\begin{align*} \notag
\me^{j\frac{\pi(1+\sqrt{2})}{12 y_0 Y}}~.
\end{align*}
Therefore we have an upper bound
\begin{align*}
S(y_0) \; \leqslant \; C_2 + \sum_{j=1}^\infty \left(P(j) + C_{1} j \me^{j\frac{\pi(1+\sqrt{2})}{12 y_0 Y}} \right)\me^{-j\left(2\pi y_0 - \frac{1}{w}\log N_{w} \right)+ \pi \sqrt{\frac{2}{3}j}}\; , \nonumber
\end{align*}
and the sum converges when
\begin{align*} \notag
2\pi y_0 - \frac{1}{w} \log N_{w} - \frac{\pi(1+\sqrt{2})}{12 Y y_0} \; > \; 0~.
\end{align*}
This gives a sufficient condition for the sums in our solutions, such as \eqref{eq:t_solution_D6_special_Equations}, to converge. However, this bound can be improved.

For the case of AESZ34, the above bound is $y_0 > 0.4442$, while the numerical value of $y_0$ at the rank two attractor point is $y_0 = \frac{\sqrt{15}}{6} \approx 0.64$.
\subsection{Bound for \texorpdfstring{$a_\fp$}{ap}}
\vskip-10pt
We seek a sufficiently strong bound on $a_{\fp}$, for partitions $\fp$ of a fixed length $l$. To this end, we begin with the definition
\begin{align*}\notag
\frac{1}{a_\fp} \= \prod_{j=1}^\infty \mu_j! \; j^{2 \mu_j}~.
\end{align*}
For positive integers $n$, one has that $n!\geq\sqrt{2\pi}n^{n+1/2}\ee^{-n}$. Hence
\begin{align*}\notag
\frac{1}{a_\fp} \;\geqslant\; \prod_{j|\mu_{j}\neq0} \sqrt{2\pi}\mu_{j}^{\mu_{j}+1/2}\ee^{-\mu_{j}} \; j^{2 \mu_j}\=\ee^{-l}\prod_{j|\mu_{j}\neq0} \sqrt{2\pi}\mu_{j}^{\mu_{j}+1/2} \; j^{2 \mu_j}~,
\end{align*}
wherein we have used that $\sum_{j}\mu_{j}=l$. Because the $\mu_{j}$ are integers, we can write
\begin{align*}\notag
\frac{1}{a_\fp} \;\geqslant\; \ee^{-l}\prod_{j=1}^{l} \mu_{j}^{\mu_{j}} \; j^{2 \mu_j}~,
\end{align*}
where $0^0$ is to be interpreted as $1$. We need only concern ourselves with $\mu_{j}$ where $j\in\{1,...,l\}$, since $\mu_{k}$ must vanish for $k>l$. On the space $\{\mu_{j}\geq0,\;\sum_{j}\mu_{j}=l\}$, the above function has a single global minimum. One can determine the minimising $\mu_{j}$ by using a Lagrange multiplier $\lambda$:
\begin{align*}\notag
\partial_{\mu_{i}} \left[\left(\prod_{j=1}^{l}\mu_{j}^{\mu_{j}}\;j^{2\mu_{j}}\right)-\lambda\sum_{j=1}^{l}\mu_{j}\right]\=0~.
\end{align*}
These relations are solved by $\mu_{j}=Cj^{-2}$, where $C$ is a constant. Of course, these extremising $\mu_{j}$ are not necessarily integers, but the result suffices for deriving a bound. Taking a logarithm,
\begin{align*}\notag
\log\frac{1}{a_\fp} \;\geqslant\; -l+\sum_{j=1}^{l}\,\left(\frac{C}{j^{2}}\,\log\frac{C}{j^{2}}+2\,\frac{C}{j^{2}}\,\log j\right)\=-l+(\log C)\sum_{j=1}^{l}\frac{C}{j^{2}}\=-l+l\,\log C~,
\end{align*}
where the relation $\sum_{j}Cj^{-2}=l$ was employed. Upon recognising the Basel sum, we can write $C>\frac{6l}{\pi^{2}}$. Therefore
\begin{align*}\notag
\log\frac{1}{a_\fp} \;\geqslant\; l\left(-1+\log\frac{6}{\pi^{2}}\right)+l\log l~.
\end{align*}
And so, as was required,
\begin{align*}\notag
a_\fp\;\leqslant\; \left(\frac{\,\ee\pi^{2}}{6l}\right)^{l}~.
\end{align*}

%%%%%%%%%%%%%%%%%%%%%%%%%%%%%%%%%%%%%%%%%
%%%%%%%%%%%%%%%%%%%%%%%%%%%%%%%%%%%%%%%%%
\newpage

\bibliographystyle{JHEP}
\bibliography{Attractors}
%%%%%%%%%%%%%%%%%%%%%%%%%%%%%%%%%%%%%%%%%
%%%%%%%%%%%%%%%%%%%%%%%%%%%%%%%%%%%%%%%%%

%%%%%%%%%%%%%%%%%%%%%%%%%%%%%%%%%%%%
%%%%%%%%%%%%%%%%%%%%%%%%%%%%%%%%%%%%
\end{document}